%% file: MHTdiscrete_final_arxiv.tex
\documentclass[12pt]{article}
\usepackage{amssymb}
\usepackage{amsmath}
\usepackage{amsthm}
\usepackage{amsfonts}
\usepackage{graphicx,psfrag}
\usepackage{float}
\usepackage{enumerate}
\usepackage[numbers]{natbib}
\usepackage{url} 
\usepackage{pdflscape}
\usepackage{rotating}
\usepackage{multirow}
\usepackage{booktabs}
\usepackage{color}
\newcommand{\blind}{0}

\newtheorem{theorem}{Theorem}[section]

\newtheorem{definition}{Definition}[section]
\newtheorem{assumption}{Assumption}[section]
\newtheorem{procedure}{Procedure}[section]
\newtheorem{proposition}{Proposition}[section]

\newtheorem{remark}{Remark}[section]

\begin{document}

\bibliographystyle{natbib}

\def\spacingset#1{\renewcommand{\baselinestretch}%
{#1}\small\normalsize} \spacingset{1}


\if0\blind
{
  \title{\bf Familywise Error Rate Controlling Procedures for Discrete Data}
  \author{Yalin Zhu\hspace{.2cm}\\
     Biostatistics and Research Decision Sciences, \\ Merck Research Laboratories, Rahway, NJ, U.S.A. \\
     \\
    Wenge Guo\hspace{.2cm}\\
    Department of Mathematical Sciences, \\ New Jersey Institute of Technology, Newark, NJ, U.S.A. \\
    }
  \maketitle
} \fi

\if1\blind
{
  \bigskip
  \bigskip
  \bigskip
  \begin{center}
    {\LARGE\bf Familywise Error Rate Controlling Procedures for Discrete Data}
\end{center}
  \medskip
} \fi

\bigskip
\begin{abstract}
In applications such as clinical safety analysis, the data of the experiments usually consists of frequency counts. In the analysis of such data, researchers often face the problem of multiple testing based on discrete test statistics, aimed at controlling family-wise error rate (FWER).
Most existing FWER controlling procedures are developed for continuous data, which are often conservative when analyzing discrete data. By using minimal attainable $p$-values, several FWER controlling procedures have been specifically developed for discrete data in the literature. In this paper, by utilizing known marginal distributions of true null $p$-values, three more powerful stepwise procedures are developed, which are modified versions of the conventional Bonferroni, Holm and Hochberg procedures, respectively. It is shown that the first two procedures strongly control the FWER under arbitrary dependence and are more powerful than the existing Tarone-type procedures, while the last one only ensures control of the FWER in special settings. Through extensive simulation studies, we provide numerical evidence of superior performance of the proposed procedures in terms of the FWER control and minimal power. A real clinical safety data is used to demonstrate applications of our proposed procedures. An R package ``MHTdiscrete" and a web application are developed for implementing the proposed procedures.

\end{abstract}

\noindent%
\vfill

\newpage
\spacingset{1.45} 
\section{Introduction}
In the applications of clinical trials, multiple hypotheses testing is a very useful statistical tool to analyze  efficacy or safety data. Simultaneously testing multiple hypotheses is often required in such applications. For single hypothesis testing, a typical error measure which needs to be controlled is type I error rate, the probability of rejecting the hypothesis while the hypothesis is true. There are several possible measures for overall type I error rate while testing multiple hypotheses. A standard error rate for clinical trials is familywise error rate (FWER), which is the probability of making at least one false rejection.

In the existing literature, most FWER controlling procedures are developed for continuous data and some are widely used in practice such as Bonferroni procedure, Holm procedure \cite{holm1979simple}, Hochberg procedure \cite{hochberg1988sharper}, etc. However, these procedures might be conservative when they are used to analyze discrete data. In the literature, several FWER controlling procedures have been specifically developed for discrete data. Tarone \cite{tarone1990modified} proposed a modified Bonferroni procedure for discrete data, which reduces the number of tested hypotheses by  eliminating those hypotheses with relatively large minimal attainable $p$-values. The Tarone procedure is more powerful than the conventional Bonferroni procedure, but it lacks $\alpha$-consistency, that is, a hypothesis which is accepted at a given $\alpha$ level may be rejected at a lower $\alpha$ level. To overcome this issue, Hommel and Krummenauer \cite{hommel1998improvements} and Roth \cite{roth1999multiple} developed two modified versions of the Tarone procedure, which not only control the FWER, but also satisfy the desired property of $\alpha$-consistency. By using Tarone's idea, Hommel and Krummenauer \cite{hommel1998improvements} also developed a step-down procedure for discrete data, which improves the conventional Holm procedure. By using the similar idea, Roth \cite{roth1999multiple} developed  a two-stage step-up procedure for discrete data, which improves the conventional Hochberg procedure by eliminating non-significant tests in the first stage. Westfall and Wolfinger \cite{westfall1997multiple} introduced a resampling based approach by simulating the null distribution of minimal $p$-value, which uses all attainable values for each $p$-value. Gutman and Hochberg \cite{gutman2007improved} developed new stepwise procedures by using the idea of Tarone and the algorithm of Westfall and Wolfinger, but these procedures are computationally intensive and only ensure asymptotic control of the FWER. For references of recent developments in multiple testing for discrete test statistics, see Heyse \cite{heyse2011false}, Chen et al. \cite{chen2014multiple}, D{\"o}hler \cite{dohler2016discrete} and He and Heyse \cite{he2019improved}. For applications of multiple testing procedures in clinical safety studies, see Mehrotra and Heyse \cite{mehrotra2012flagging}, Gould \cite{gould2015statistical}, Jiang and Xia \cite{jiang2014quantitative}, Dimitrienko et al. \cite{dmitrienko2009multiple}, and Goeman and Solari \cite{SIM:SIM6082}.

It is noted that these existing procedures for discrete data are mainly developed based on minimal attainable $p$-values. In practice, if the minimal attainable $p$-values are known, the corresponding true null distributions of the $p$-values are often also known. By fully utilizing the true null distributions rather than the minimal attainable $p$-values, we develop three simple and powerful stepwise procedures for
discrete data. Specifically, we develop new single-step, step-down, and step-up procedures for discrete data, which are modified versions of the conventional Bonferroni, Holm, and Hochberg procedures, respectively. Theoretically, we show that the first two procedures strongly control the FWER under arbitrary dependence, whereas the last one only ensures control of the FWER in special settings. We also show that the proposed procedures have several desired properties: (i) the proposed single-step procedure is more powerful than the existing Tarone and modified Tarone procedures, whereas the proposed step-down procedure is more powerful than the existing Tarone-Holm procedure; (ii) the proposed procedures satisfy the properties of $\alpha$-consistency and $p$-value monotonicity, which are desired for a multiple testing procedure; (iii) simple formulas for adjusted $p$-values are given for these proposed procedures. Through extensive simulation studies, we provide numerical evidence of superior performance of the proposed procedures in terms of the FWER control and minimal power. Even for the proposed step-up procedure, although we cannot provide theoretical guarantee of its FWER control for general cases, we find out numerical validation of its FWER control under various simulation settings. A real data set of clinical safety study is also used to demonstrate applications of our proposed procedures.

The rest of the paper is organized as follows. With notations, assumptions and several existing procedures for discrete data given in Section 2, we present our proposed stepwise procedures and discuss their statistical properties in Section 3. The numerical findings from simulation studies
are given in Section 4 and a real application of clinical safety study is presented in Section 5.
Some concluding remarks are made in Section 6 and all proofs are deferred to the appendix section.

\section{Preliminary}
Consider the problem of simultaneously testing $m$ hypotheses $H_1, \dots, H_m$, among which there are $m_0$ true and $m_1$ false null hypotheses. Suppose the test statistics are discrete. Let $P_i$ denote the $p$-value for testing $H_i$ and $\mathbb{P}_i$ denote the full set of all attainable values for $P_i$  such that $P_i \in \mathbb{P}_i$. Let $F_i$ denote the cumulative distribution function (CDF) of $P_i$ when $H_i$ is true, that is $F_i(u)=Pr(P_i \leq u | H_i \text{ is true})$ for $u \in [0, 1]$. Let $P_{(1)} \leq \dots \leq P_{(m)}$ denote the ordered $p$-values and $H_{(1)}, \ldots, H_{(m)}$ denote the corresponding hypotheses, with $F_{(i)}$ denoting the corresponding CDF of $P_{(i)}$ when $H_{(i)}$ is true and $\mathbb{P}_{(i)}$ the corresponding set of all attainable values of $P_{(i)}$. We make the following assumption regarding $F_i$:
\begin{assumption} \label{uniform}
	The marginal distribution functions $F_i$ of all true null $p$-values $P_i$ are known and satisfy that for any $u \in [0, 1]$,
	$F_i(u) = u, \ \text{if } u \in \mathbb{P}_i$; otherwise,  $F_i(u) < u$.
\end{assumption}
The assumption implies that each true null $p$-values is exactly $U(0, 1)$ distributed  when it takes an attainable $p$-value, and is stochastically larger than $U(0, 1)$ when it takes an unattainable value. For the joint distributions of the $p$-values, throughout the paper we only consider two types of dependence structure, arbitrary dependence, which allows any joint distribution of the $p$-values, and positive regression dependence on subset (PRDS) (Benjamini and Yekutieli \cite{benjamini2001control}; Sarkar \cite{sarkar2002some}), which is often satisfied in many multiple testing situations. The PRDS assumption is defined as follow.
\begin{assumption}[] \label{PRDS}
	A set of $p$-values $ \{P_1, \ldots, P_m\} $ is said to be PRDS, if for any non-decreasing function of the $p$-values $\phi$,  $E\{ \phi(P_1, \dots, P_m)|P_i \leq p \}$ is non-decreasing in $p$ for each true null hypothesis $H_i$.
\end{assumption}

For any multiple testing procedure (MTP), let $V$ denote the number of falsely rejected hypotheses. Then, the FWER of this procedure, defined by $FWER=\Pr(V \ge 1)$ is said to be controlled at level $\alpha$, strongly unless stated otherwise, if it is bounded above by $\alpha$ for any configuration of hypotheses. That is, for any combination of true and false null hypotheses, the FWER of this procedure is less than or equal to $\alpha$. In the literature, there are several popular FWER controlling procedures available for any test statistics, such as Bonferroni, Sidak, Holm, Hochberg, and Hommel procedures (Dimitrienko et al. \cite{dmitrienko2009multiple}). Specifically, for discrete test statistics, Tarone (1990) introduced a modified Bonferroni procedure below by using the smallest attainable $p$-values to eliminate the non-significant tests, which has larger critical constant than the conventional Bonferroni procedure.

\begin{procedure}[Tarone] \label{ProcT}
	Suppose that $ p_i^* $ are the smallest attainable $p$-values for $H_i$. Let $M(\alpha,k)=\sum\limits_{i =1}^{m}I\{p_i^* \le \dfrac{\alpha}{k}\} $ and $K(\alpha)= \min\{1 \leq k\leq m: M(\alpha,k) \leq k\}$. Then, reject $H_i$ if $P_i \le \dfrac{\alpha}{K(\alpha)}$.
\end{procedure}

As pointed out by Hommel and Krummenauer \cite{hommel1998improvements}, the Tarone procedure does not satisfy the desired property of $\alpha$-consistency defined in Section 3.4. In order to overcome this issue,  Hommel and Krummenauer developed a modified Tarone procedure as follows, which is shown satisfying the property of $\alpha$-consistency.	

\begin{procedure}[Modified Tarone] \label{T*}
	Suppose that $ p_i^* $ are the smallest attainable $p$-values for $H_i$. For any $\gamma\in (0,\alpha]$, let $M(\gamma, k)=\sum\limits_{i =1}^{m}I\{p_i^* \le \dfrac{\gamma}{k}\} $ and\\ $K(\gamma)= \min\{1 \leq k\leq m: M(\gamma, k) \leq k\}$. Then reject $H_i$ if there exists an $\gamma\in (0,\alpha]$, such that $P_i \leq \dfrac{\gamma}{K(\gamma)}$.
\end{procedure}	

By incorporating the idea of Tarone \cite{tarone1990modified} into the conventional Holm procedure, Hommel and Krummenauer \cite{hommel1998improvements} also developed a modified Holm procedure as follows for discrete test statistics.

\begin{procedure}[Tarone-Holm]  \label{TH*}
	\
	\begin{enumerate}
		\item Set $I=\{1, \dots, m\}$.
		\item For $k=1, \dots, |I|$, let $ M_I(\gamma, k)= \sum\limits_{i \in I } I\{p_i^* \le \dfrac{\gamma}{k}\}$ and  $K_I(\gamma)= \min\{ k =1, \dots, |I|: M_I(\gamma, k) \le k  \}$.
		\item For $i \in I$, reject $H_i$ if and only if $P_i \le \dfrac{\gamma}{K_I(\gamma)}$ for some $0 < \gamma \le \alpha$. Let $J$ be the index set of the rejected hypotheses.
		\item If $J$ is empty, stop testing; otherwise, set $I=I-J$ and then return to step 2.
	\end{enumerate}
\end{procedure}

In addition, by using the similar idea of Tarone \cite{tarone1990modified}, Roth \cite{roth1999multiple} developed a modified Hochberg procedure for discrete test statistics based on the conventional Hochberg procedure.

\section{Proposed Stepwise Procedures for Discrete Data}

Many existing FWER controlling procedures for discrete data are developed based on the idea of Tarone \cite{tarone1990modified}, which only utilizes partial information of true null $p$-values, so these procedures might be conservative. In this section, we develop more powerful stepwise procedures by fully exploiting known marginal distributions of true null $p$-values.

\subsection{A new single-step procedure}

By using the CDFs of true null $p$-values, we develop a new modified Bonferroni procedure for discrete data as follows.

\begin{procedure}[Modified Bonferroni]  \label{MBonf}
	Let $s^* = \max\{p \in \bigcup\limits_{i=1}^{m} \mathbb{P}_i: \sum\limits_{i=1}^{m} F_i(p) \leq \alpha \}$ and set $s^*  = \dfrac{\alpha}{m}$ if the maximum does not exist. For any hypothesis $H_i$, reject $H_i$ if its corresponding $p$-value $P_i\leq s^*$.
\end{procedure}

It should be noted that the proposed modified Bonferroni procedure for discrete data is a natural extension of the usual Bonferroni method. When all true null $p$-values are $U[0, 1]$,  its critical value $s^*=\max\{p \in [0,1]: mp \leq \alpha \}=\dfrac{\alpha}{m}$,  which is the same as that of the usual Bonferroni procedure. Thus, the modified Bonferroni reduces to the conventional Bonferroni procedure under such setting. For the proposed Procedure \ref{MBonf}, the following result holds. 	
						
\begin{theorem}
	Procedure \ref{MBonf} (Modified Bonferroni) strongly controls the FWER at level $\alpha$ under Assumption \ref{uniform}.
\end{theorem}

Compared to the existing Tarone procedure (Procedure \ref{ProcT}) and modified Tarone procedure (Procedure \ref{T*}) for discrete data, we have
\begin{proposition} \label{MBonf>T}
	Procedure \ref{MBonf} (Modified Bonferroni) is universally more powerful than Procedures \ref{ProcT} (Tarone) and \ref{T*} (Modified Tarone), that is, for any $H_i$, if it is rejected by Procedure \ref{ProcT} or \ref{T*}, it is also rejected by
Procedure \ref{MBonf}.
\end{proposition}

It is useful to calculate its \textit{adjusted $p$-values} for a multiple testing procedure, since one can make decisions of rejection and acceptance as in single hypothesis testing by simply comparing the adjusted $p$-values with the given significance level. By Westfall and Young \cite{westfall1993resampling}, the adjusted $p$-value for a hypothesis in multiple testing is the smallest significance level at which one would reject the hypothesis using the given multiple testing procedure. Thus, the adjusted $p$-values $\tilde{P}_{i,MBonf}$ of Procedure \ref{MBonf} for $H_i$ can be derived as follow:
\begin{equation} \label{MBonf_adjP}
	\tilde{P}_{i,MBonf} = \min \left\{1,\    \sum\limits_{j=1}^{m} F_j(P_i)  \right\}, \quad  for \ i=1, \dots, m.
	\end{equation}  	
It is easy to see that the adjusted $p$-values of Procedure \ref{MBonf} are smaller than or equal to those of the conventional Bonferroni procedure, since for any given $p$-value $P_i$ and $j=1, \dots, m$, $F_j(P_i) \leq P_i$, then $\sum\limits_{j=1}^m F_j(P_i) \leq m  P_i$. Thus, Procedure \ref{MBonf} is uniformly more powerful than the conventional Bonferroni.

\subsection{A new step-down procedure}

By using the similar idea as in Section 3.1, we develop a new modified Holm procedure for discrete data as follows.

\begin{procedure}[Modified Holm] \label{MHolm}
	For $i = 1, \ldots, m$, let $\alpha_i = \max\{ p \in \bigcup\limits_{j=i}^{m} \mathbb{P}_{(j)}: \sum\limits_{j=i}^{m} F_{(j)}(p) \leq \alpha \}$ if the maximum exists; otherwise,
	set $ \alpha_i = \max\left\{\alpha_{i-1},\ \dfrac{\alpha}{m-i+1}\right\}$ with $\alpha_0 = 0$.  Then reject $H_{(1)}, \dots, H_{(i^*)}$ and retain $H_{(i^*+1)}, \dots, H_{(m)}$, where $i^* = \max\{i: P_{(1)} \leq \alpha_1, \dots, P_{(i)} \le \alpha_i \}$, if the maximum exists; otherwise, accepts all the null hypotheses.
\end{procedure}

It should be noted that when all true null $p$-values are $U[0, 1]$, the critical values $$\alpha_i=\max\{p \in [0,1]: (m-i+1)p \leq \alpha \}=\dfrac{\alpha}{m-i+1}.$$ Thus, the proposed modified Holm procedure reduces to the conventional Holm procedure under such case.

\begin{theorem}
	Procedure \ref{MHolm} (Modified Holm) strongly controls the FWER at level $\alpha$ under Assumption \ref{uniform}.
\end{theorem}

Compared to the existing Tarone-Holm procedure for discrete data (Procedure \ref{TH*}), we can show that Procedure \ref{MHolm} is universally more powerful than Procedure \ref{TH*}. That is, for any $H_i$, if it is rejected by Procedure \ref{TH*}, it is also rejected by
Procedure \ref{MHolm}.

\begin{proposition}
	Procedure \ref{MHolm} (Modified Holm) is universally more powerful than Procedure \ref{TH*} (Tarone-Holm).
\end{proposition}

Similar to Procedure \ref{MBonf}, the adjusted $p$-values $\tilde{P}_{(i),MHolm}$ of Procedure \ref{MHolm}  for corresponding hypotheses $H_{(i)}$ can be directly calculated as follows.
\begin{eqnarray} \label{MHolm_adjP}
	\tilde{P}_{(i),MHolm} =\begin{cases}
	\min \left\{1,\    \sum\limits_{j=1}^{m} F_{(j)}(P_{(1)})  \right\}, &i=1, \cr
	\max\left\{\tilde{P}_{(i-1),MHolm} , \min \left\{1,\    \sum\limits_{j=i}^{m} F_{(j)}(P_{(i)})  \right\} \right\}, &i=2, \dots, m.
	\end{cases}
\end{eqnarray}

\subsection{A new step-up procedure}

Similar to Procedures \ref{MBonf} and \ref{MHolm}, by fully exploiting the marginal distributions of true null $p$-values, we can also develop a new modified Hochberg procedure for discrete data as follows, which uses the same critical constants as Procedure \ref{MHolm}.

\begin{procedure}[Modified Hochberg]  \label{MHoch}
	For $i=1, \dots, m$, let $\alpha_i = \max\{ p \in \bigcup\limits_{j=i}^{m} \mathbb{P}_{(j)}: \sum\limits_{j=i}^{m} F_{(j)}(p) \leq \alpha \}$ if the maximum exists; otherwise set $ \alpha_i = \max\left\{\alpha_{i-1},\ \dfrac{\alpha}{m-i+1}\right\}$ with $\alpha_0 = 0$.  Then reject $H_{(1)}, \dots, H_{(i^*)}$ and retain $H_{(i^*+1)}, \dots, H_{(m)}$, where $i^* = \max\{i: P_{(i)} \le \alpha_i \}$, if the maximum exists; otherwise accepts all the null hypotheses.
\end{procedure}

It should be noted that when all true null $p$-values are $U[0, 1]$, the above procedure reduces to the conventional Hochberg procedure.

\begin{proposition}
Suppose that the true null $p$-values are identically distributed, then
\begin{itemize}
  \item[(i)] Procedure \ref{MHoch} (Modified Hochberg) strongly controls the FWER at level $\alpha$ under Assumptions \ref{uniform} and \ref{PRDS}.
  \item[(ii)] Procedure \ref{MHoch} (Modified Hochberg) rejects the same hypotheses as the conventional Hochberg procedure.
\end{itemize}
\end{proposition}

When the true null $p$-values are not identically distributed, let us consider a special case of testing two null hypotheses $H_i, i=1, 2$ for which corresponding $p$-values $P_i$ under $H_i$ only take two attainable values in $[0, 1]$. Denote the support of $P_i$ under $H_i$ as $$\mathbb{P}_i = \{p_i, 1\}, \ \text{where} \ 0< p_i <1.$$  Without loss of generality, assume $p_1 < p_2$ and at least one of two hypotheses is true.

\begin{proposition}
Under the special case of testing two hypotheses described as above, Procedure \ref{MHoch} (Modified Hochberg) strongly controls the FWER under Assumption \ref{uniform}.
\end{proposition}

Similar to Procedures \ref{MBonf} and \ref{MHolm}, the adjusted $p$-values of Procedure \ref{MHoch} for corresponding hypotheses  $H_{(i)}$ can be directly calculated as follows.
\begin{eqnarray} \nonumber
	\tilde{P}_{(i),MHoch} =\begin{cases}
	F_{(m)}(P_{(m)})  , &i=m,
	\cr
	\min\left\{\tilde{P}_{(i+1), MHoch} ,  \sum\limits_{j=i}^{m} F_{(j)}(P_{(i)})  \right\}, &i=m-1, \dots, 1.
	\end{cases}
\end{eqnarray}

\subsection{Statistical property}

In multiple testing, $\alpha$-consistency is a desired statistical property for a multiple testing procedure in terms of the significance level $\alpha$, which is defined as follow:
\begin{definition} [Dimitrienko et al. \cite{dmitrienko2009multiple}] \label{alpha_consist}
	A multiple testing procedure is called to be \emph{$\alpha$-consistent} if any hypothesis that is rejected at a given $\alpha$ level by the procedure is always rejected at a higher $\alpha$ level by the same procedure.
\end{definition}
This property of $\alpha$-consistency  implies that for a given $\alpha' > \alpha$,  the set of rejections determined at $\alpha'$ level will not become smaller than that at $\alpha$ level. This is a desirable property in practice. For single hypothesis testing, it is trivial that this property is always satisfied by any conventional test. However, for multiple hypotheses testing, not all multiple testing procedures satisfy this property. For example, the Tarone procedure (Procedure \ref{ProcT})  does not satisfy this property. For our proposed Procedures \ref{MBonf}, \ref{MHolm} and \ref{MHoch}, it is easy to see that they all satisfy this property.

Another favorable property of a multiple testing procedure is monotonicity in terms of $p$-values, which is defined as follow:
\begin{definition} [Dimitrienko et al. \cite{dmitrienko2009multiple}] \label{p_monotone}
	A multiple testing procedure is called to be \emph{$p$-value monotone}	if one or more $p$-values are made smaller, then at least the same or even more hypotheses would be rejected by this procedure.
\end{definition}
The property of $p$-value monotonicity is always satisfied by conventional $p$-value based stepwise procedures. It is easy to see that it is also satisfied by all of our proposed procedures. This property helps to avoid logical inconsistency of decisions of rejection and acceptance; as such it is an essential requirement for a multiple testing procedure. Summarizing the above discussion, we have

\begin{proposition} \label{MBonf_pmono}
	Procedures \ref{MBonf}, \ref{MHolm} and \ref{MHoch} satisfy the properties of $\alpha$-consistency and $p$-value monotonicity.
\end{proposition}

\begin{remark}\rm
A referee brought our attention to the recently published paper, He and Heyse \cite{he2019improved}. Procedures 2.1 and 3.1 in \cite{he2019improved} are very similar as our proposed single-step Procedures 3.1 and step-down Procedure 3.2. The only difference is that their definitions of the critical values for these two procedures are incomplete. When the maximums do not exist (see the definitions of Procedure 3.1 and 3.2), which often occurs for discrete data, they do not show how to determine the critical values for these procedures. For our proposed step-up Procedure 3.3, they only briefly referred to it, whereas we showed its FWER control in two special settings.

Although He and Heyse \cite{he2019improved} proposed the similar procedures as our Procedures 3.1 and 3.2, they didn't discuss these procedures in details. In \cite{he2019improved}, the main goal is to develop more powerful FWER controlling procedures by utilizing exact pairwise permutation dependence of the $p$-values, however, our goal in this paper is to develop simple and powerful FWER controlling procedures by fully exploiting known marginal distributions of true null $p$-values. Besides showing the FWER control of Procedures 3.1-3.3, we also provide further theoretical discussions to show that the proposed procedures are more powerful than the existing Tarone-type procedures. Finally, we also discuss the statistical properties of the proposed procedures, including $\alpha$-consistency and $p$-value monotone, and provide simple formulas to calculate the adjusted $p$-values of the proposed procedures.
\end{remark}

\section{Simulation studies}

In this section, we perform extensive simulation studies to investigate the performances of the proposed procedures in terms of the FWER control and minimal power, the probability of correctly rejecting at least one false null hypotheses. The simulations are conducted based on two typical discrete tests settings: Fisher's Exact Test (FET) and Binomial Exact Test (BET). Suppose that there are two groups of patients, study group (1) and control group (2).
\begin{enumerate}
	\item FET:  There are $m$ independent binomial responses $X_{ij}$ observed for each of $N$ individuals in each group, such as $X_{i1} \sim Bin(N, p_{i1})$, $X_{i2} \sim Bin(N,p_{i2})$ for $i=1, \ldots, m$. The goal is to simultaneously test $m$ one-sided hypotheses $H_i: p_{i1}=p_{i2}$ vs. $H'_i: p_{i1} < p_{i2}$, where $p_{ij}$ is the success probability for the $i$-th response in group $j$, and $i=1, \ldots, m$, $j=1,2$.
	We conduct the experiment using one-sided FET under $\alpha$ level, then under $H_i$, the test statistics $T_i \sim Hypergeometric (X_{i1}, N, X_{i1}+X_{i2}, 2N)$.
	
	In the simulation with FET, we set the number of hypotheses $m=\{5,10,15\}$, the true null proportion $\pi_0=\{0.2,0.4,0.6,0.8\}$, and the sample size for the binomial response per group used is $N=\{25, 50, 75, 100, 125, 150\}$.   For true null hypotheses, set the success probability parameter of each binomial response in each group as $p_{i1}=p_{i2}=0.1$; for false null hypotheses, set the success probability for study group as $p_{i1}=0.1$, and for control group as $p_{i2}=0.2$. 
	
	\item BET: There are $m$ Poisson responses observed in each group, such as $X_{i1} \sim Poi(\lambda_{i1})$, $X_{i2} \sim Poi(\lambda_{i2})$ for $i=1, \ldots, m$. The goal is to simultaneously test $m$ one-sided hypotheses $H_i: \lambda_{i1}=\lambda_{i2}$ vs. $H'_i: \lambda_{i1} < \lambda_{i2}$, where $\lambda_{ij}$ is the mean parameter for the $i$-th response in group $j$, and $i=1, \ldots, m$, $j=1,2$.  We conduct the experiment using one-sided BET under $\alpha$ level, then  under $H_i$, the test statistics for study group follow binomial distribution  $Bin(X_{i1}+X_{i2}, p_{i})$, where $p_i=\dfrac{\lambda_{i1}}{\lambda_{i1}+\lambda_{i2}}$, i.e., $T_{i} \sim Bin(X_{i1}+X_{i2}, 0.5)$.
	
	In the simulation with BET, we set the number of hypotheses $m=\{5,10,15\}$, with true null proportion $\pi_0=\{0.2,0.4,0.6,0.8\}$, respectively. For true null hypotheses, set the mean parameter of Poisson response in each group as $\lambda_{i1}=\lambda_{i2}=2$,; for false null hypotheses, set the mean parameter for study group as $\lambda_{i1}=2$, and for control group as $\lambda_{i2}=10$. 

\end{enumerate}
By using the true null distribution of the FET or BET statistic, one can calculate the available $p$-values $P_i$ and all attainable $p$-values in the set $\mathbb{P}_i$.  Then the simulated FWER, minimal power, and number of rejections for each multiple testing procedure are calculated by taking average of $B=2000$ iterations.

\subsection{Numerical comparisons for single-step procedures}\label{sim_setting}

\begin{figure}[!htb]
	\centering
	\includegraphics[width=6.4in]{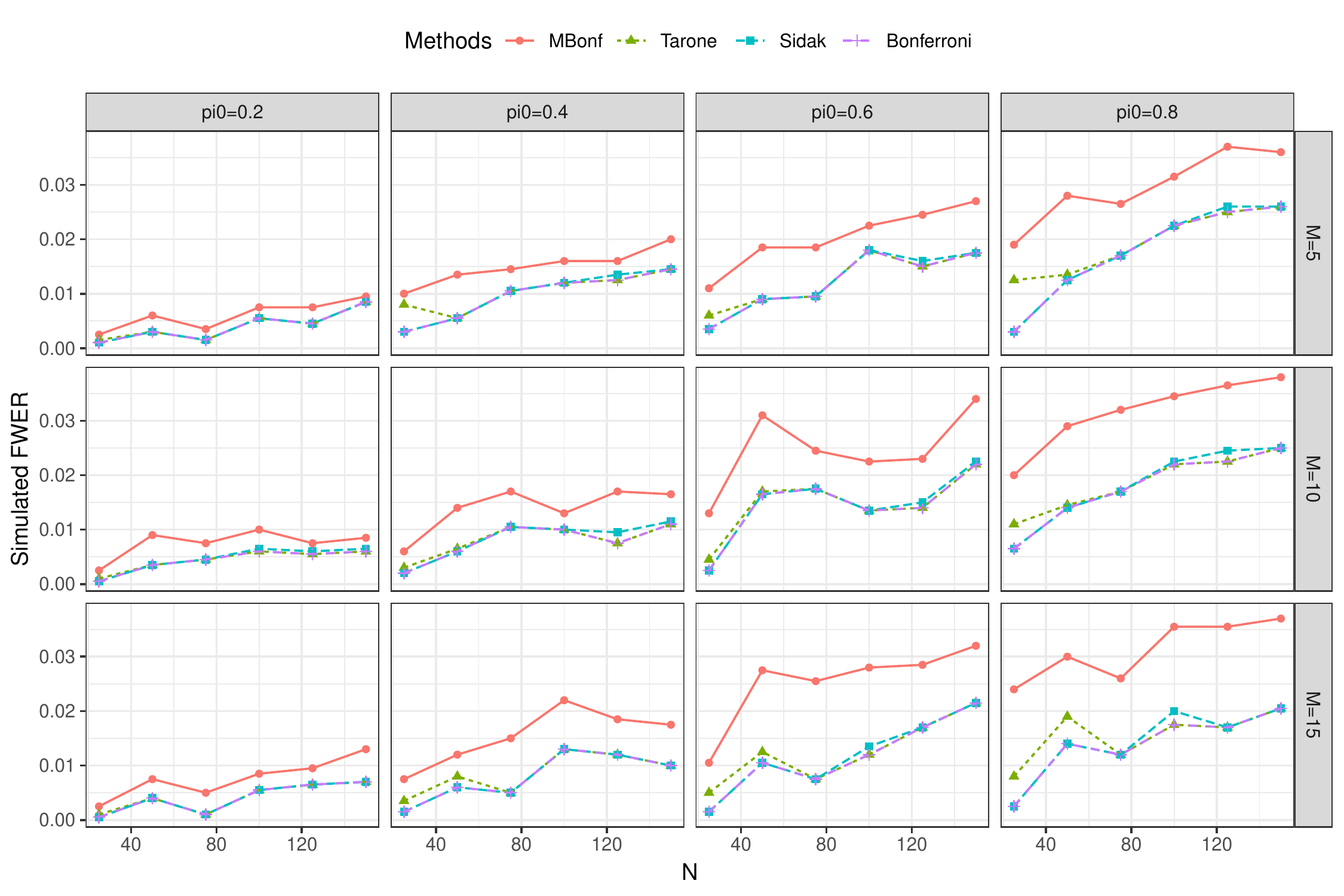}
	\caption{Simulated FWER comparisons for different single-step procedures based on FET, including Procedure \ref{MBonf} (MBonf), Procedure \ref{ProcT} (Tarone), and the conventional Sidak and Bonferroni procedures.}\label{SS_FWER}
\end{figure}

\begin{figure}[!htb]
	\centering
	\includegraphics[width=6.4in]{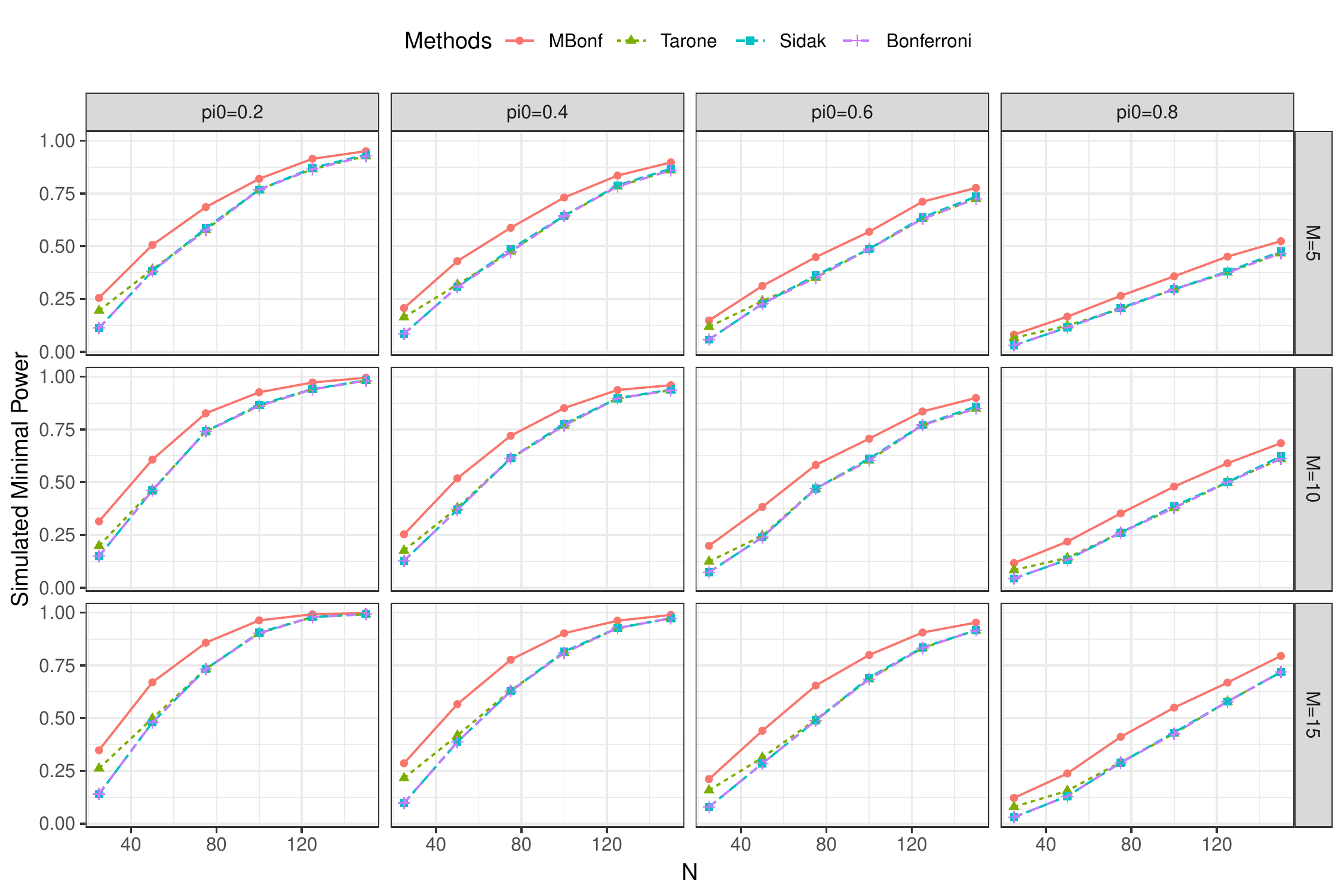}
	\caption{Simulated minimal power comparisons for different single-step procedures based on FET, including Procedure \ref{MBonf} (MBonf), Procedure \ref{ProcT} (Tarone), and the conventional Sidak and Bonferroni procedures.} \label{SS_Pow}
\end{figure}

We now present simulation studies comparing the proposed Procedure \ref{MBonf} with existing single-step procedures, including the Bonferroni procedure, Sidak procedure and Modified Tarone procedure. Figures \ref{SS_FWER} and \ref{SS_Pow} show the simulated FWER levels and minimal powers of all four procedures under the FET setting.
The detailed results can be found in Tables S1 and  S2 in the supplementary materials. From these simulation results one can observe:
\begin{enumerate}[(i)]
	\item The proposed Procedure \ref{MBonf} always controls the FWER at the pre-specified level $\alpha=0.05$ and has higher FWER level and greater power than the existing three procedures for different sample size $N$.
	\item Compared to the existing procedures, the FWER level of Procedure \ref{MBonf} is less conservative and the power advantage is larger for smaller size $N$,  since the data is more discrete under such setting. For example, when testing $m=10$ hypotheses with $\pi_0=0.2$, the FWER level of Procedure \ref{MBonf} (0.0020) is $300\%$ higher than that of the Tarone procedure (0.0005) when the simulated data is generated from binomial distribution with $N=5$, however, when $N=125$, the FWER improvement is only $35.7\%$ (0.0095 versus 0.0070).
	\item As the proportion of true nulls becomes larger, compared to the existing procedures, the FWER level of Procedure \ref{MBonf} is closer to nominal significance level 0.05, but its power performance becomes smaller.
\end{enumerate}

We also conduct simulations by using the BET statistics. The corresponding simulation results are shown in Tables S3 and  S4 in the supplementary materials. One can observe from Tables S3 and  S4 that under the BET setting, the proposed Procedure \ref{MBonf} can also control the FWER  at level  0.05 or 0.1, and is more powerful than the existing three procedures. For other findings, they are similar to the simulation results obtained under the FET setting. 	

\begin{figure}[!htb]
	\centering
\includegraphics[width=6.4in]{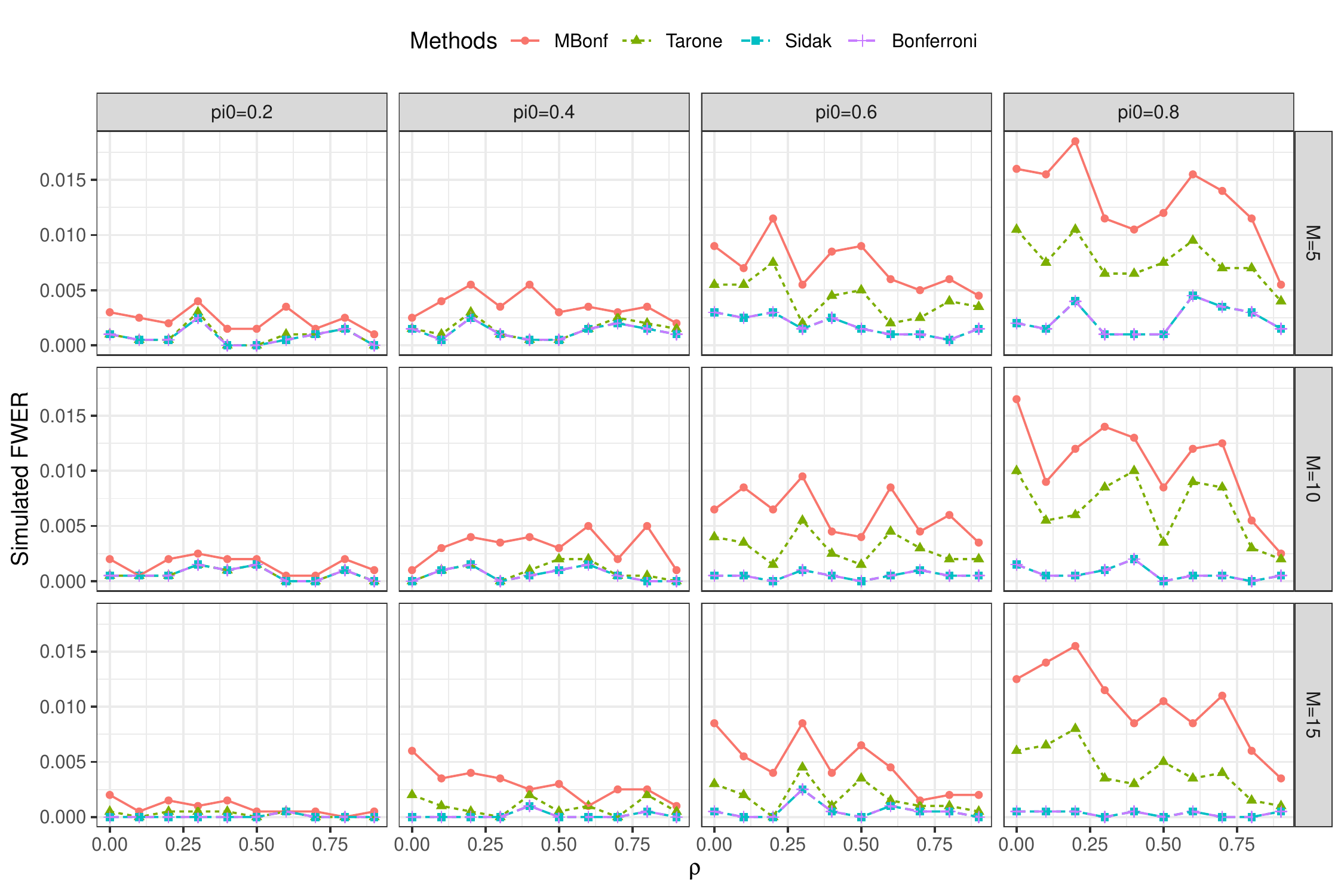}
	\caption{Simulated FWER comparisons for different single-step procedures based on the blocking dependent BET, including Procedure \ref{MBonf} (MBonf), Procedure \ref{ProcT} (Tarone), and the conventional Sidak and Bonferroni procedures.}\label{SSdept_FWER}
\end{figure}

We also perform simulations to evaluate the effect of dependence among the test statistics on the performance of the proposed Procedure \ref{MBonf}. The simulation is conducted under block dependence structure for the BET. Details for generating the dependent simulation data can be found in Section S2 of the supplementary materials. In the simulations, we set the number of hypotheses $m=\{5, 10, 15\}$, the true null proportion $\pi_0=\{0.2, 0.4,0.6,0.8\}$, and the correlation $\rho=\{0, 0.1, \dots, 0.9\}$. The simulation results under such setting are shown in Figures \ref{SSdept_FWER} and \ref{SSdept_Pow}, part of detailed results can be also found in Tables S9 and S10 of the supplementary materials. These simulation results show that the simulated FWERs of all the four procedures are lower than the pre-specified level 0.05, and the minimal powers for these procedures are decreasing as the correlation $\rho$ becomes larger. More importantly, the proposed Procedure \ref{MBonf} is always more powerful than the existing three procedures no matter how the correlation $\rho$ changes.

\begin{figure}[!htb]
	\centering
	\includegraphics[width=6.4in]{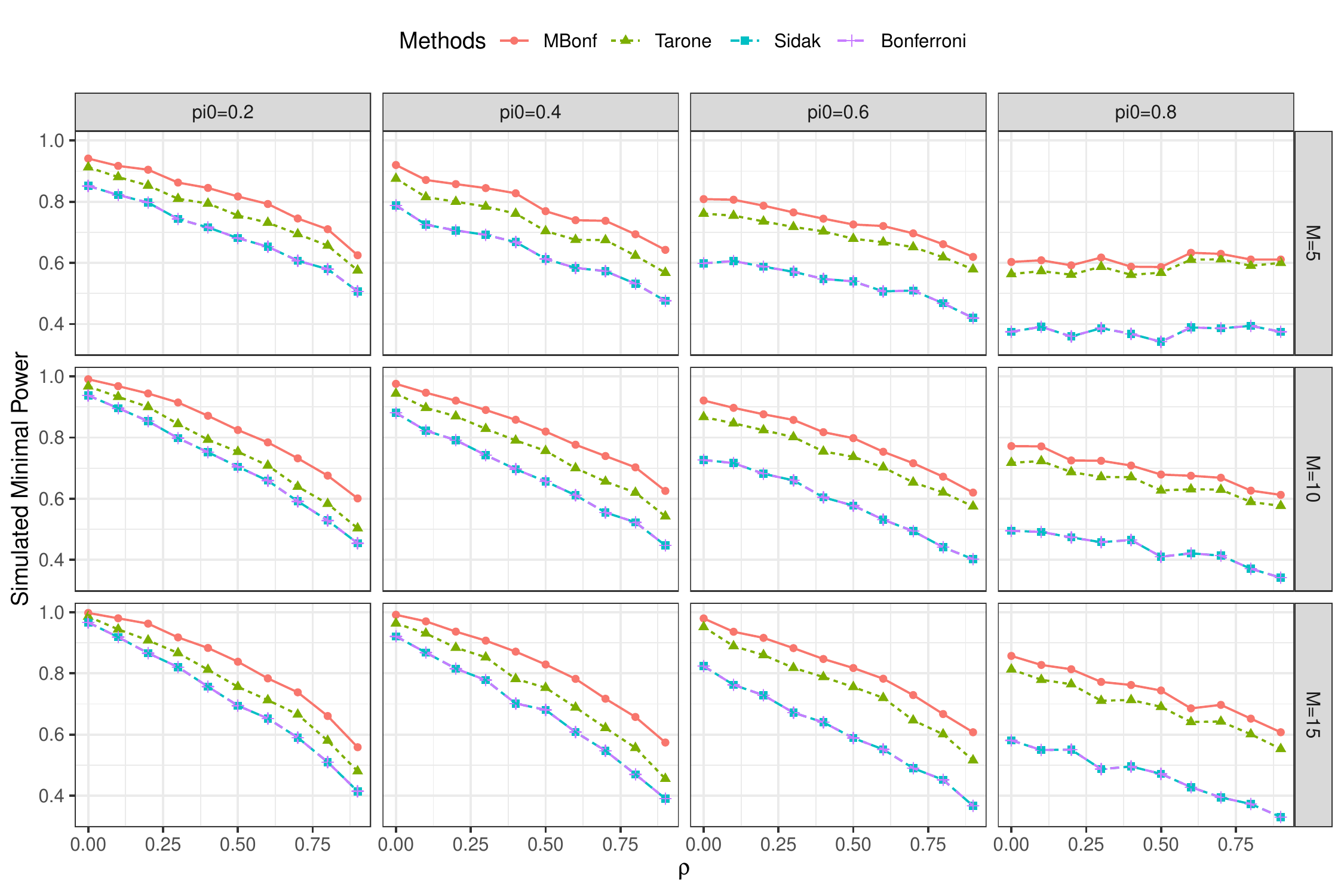}
	\caption{Simulated minimal power comparisons for different single-step procedures based on the blocking dependent BET, including Procedure \ref{MBonf} (MBonf), Procedure \ref{ProcT} (Tarone), and the conventional Sidak and Bonferroni procedures.} \label{SSdept_Pow}
\end{figure}

\subsection{Numerical comparisons for step-down procedures}
Similar to numerical comparisons for the single-step procedures, we also conduct simulation studies to evaluate the proposed step-down Procedure \ref{MHolm} in terms of the FWER control and minimal power compared with two existing step-down procedures: the conventional Holm procedure and the Tarone-Holm procedure in Hommel and Krummenauer \cite{hommel1998improvements}. We only use FET in the simulations for step-down procedures, since using BET produces similar patterns. Figures S1 and S2 in the supplementary materials show the simulation results of these step-down procedures under the FET setting. The detailed results can also be found in Tables S5 and S6 in the supplementary materials. These simulation results show that the proposed Procedure \ref{MHolm} always controls the FWER at the pre-specified level $0.05$ and is more powerful than the existing  step-down procedures. Other numerical findings are similar to those of the single-step procedures. In addition, as seen in Tables S1 and S2, the proposed step-down Procedure \ref{MHolm} is more powerful than the proposed single-step Procedure \ref{MBonf}.

We also conduct simulations for step-down procedures under the block dependence structure for BET. In the simulations, we set the number of hypotheses $m=\{5, 10, 15\}$, the true null proportion $\pi_0=\{0.2, 0.4,0.6,0.8\}$, and the correlation $\rho=\{0, 0.1, \dots, 0.9\}$. The simulation results under such setting are displayed in Figures S5 and S6 in the supplementary materials. The detailed results can be found in Tables S11 and S12 in the supplementary materials. From these simulation results, one can observe that the simulated FWERs of all these three procedures are lower than the pre-specified level 0.05; the simulated power for each procedure is decreasing in terms of the correlation $\rho$, and the proposed Procedure \ref{MHolm} is always more powerful than the two existing  step-down procedures no matter how the correlation $\rho$ changes.

\subsection{Numerical comparisons for step-up procedures}

In this subsection, we conduct simulation studies to evaluate the proposed step-up Procedure \ref{MHoch} in terms of the FWER control and minimal power compared with two existing step-up procedures: the conventional Hochberg procedure and the Roth procedure in Roth (1999). Similar as in Section 4.2, we only use FET in the simulations for step-up procedures, since using BET produces similar patterns. Figures S3 and S4 in the supplementary materials show the simulation results of these step-up procedures under the FET setting.  The detailed results can be found in Tables S7 and S8 in the supplementary materials.
These simulation results show that the proposed Procedure \ref{MHoch} always controls the FWER at the pre-specified level $0.05$ and has greater power than the existing procedures. Other numerical findings are similar to those of the single-step procedures. In addition, as seen in Tables S7 and S8, the proposed step-up Procedure \ref{MHoch} is always more powerful than the proposed single-step and step-down procedures \ref{MBonf} and \ref{MHolm} for different sample size $N$.

We also conduct simulations for step-up procedures under the block dependence structure for BET. In the simulations, we set the number of hypotheses $m=\{5, 10, 15\}$, the true null proportion $\pi_0=\{0.2, 0.4,0.6,0.8\}$, and the correlation $\rho=\{0, 0.1, \dots, 0.9\}$. The simulation results are displayed in Figures S7 and S8 in the supplementary materials.  The detailed results can be found in Tables S13 and S14 in the supplementary materials. From these simulation results, one can observe that the simulated FWERs of all the three step-up procedures are lower than the pre-specified level 0.05; the power for each procedure is decreasing in terms of the correlation $\rho$, and the proposed Procedure \ref{MHoch} is always more powerful than the two existing procedures no matter how the correlation $\rho$ changes.

It should be noted that as the sample size $N$ is larger than some cutoff value, the power improvement for these proposed stepwise procedures could be negligible compared with other existing procedures. Through the simulations studies under the FET setting, we find out that the cutoff value is mainly determined by the sample size $N$ and the number of tested hypotheses $m$; specifically, when the ratio $N/m$ is larger than 50, the power improvement of the proposed procedures can be relatively minimal.

\section{A clinical safety example}

The proposed stepwise procedures can be applied in clinical safety studies to detect significant AEs (so-called ``flagging''), since clinical safety data is usually based on the count of patients of having the adverse events exposures. The following example is modified from Table 1 of Mehrotra and Heyse \cite{mehrotra2004use}, which reports the AE types of two groups of patients.  For illustrative purpose, the AEs of skin body system (BS=10) are only analyzed in this example and the numbers of randomized patients are respectively enlarged to 600 and 650 for two arms. The data is re-ordered based on the rank order of the $p$-values corresponding to the AE types, which are shown in the first three columns of Tables \ref{table:clinsafe_single}-\ref{table:clinsafe_su}.

The skin body system includes nine AE types, which are those have a relatively large number of AEs. Thus, they are possibly detected to be significant at level $\alpha=0.05$ by using Fisher's Exact Test (FET), conditional on the fixed marginal totals. In the data, $ X_{ij}$ is the observed number of the $j$-th group patients experiencing the $i$-th AE for $ i = 1, \ldots, 9$ and $j = 1, 2$ (``1" denotes the study group receiving the candidate treatment and ``2" denotes the control group receiving standard of care), and $ N_{j} $ is the number of patients randomized in group $j$ with $N_1=600$ and $N_2=650$. In Tables \ref{table:clinsafe_single}-\ref{table:clinsafe_su}, the first column shows the indices of the AE types after reordering the data, and the second and third columns show the numbers of toddlers experiencing the corresponding AE in the control and study groups. By using the two-sided FET $T$, the available $p$-value $P_{i}$ and minimal attainable $p$-value $p_i^*$ for $i$-th AE type can be calculated as follows:
\begin{equation}
P_i = \Pr\{T \ge X_{i2} \}=\sum\limits_{k=X_{i2}}^{X_{\cdot i}}\frac{{N_{2}\choose k}{N_{1}\choose {X_{i \cdot }-k}}}{{N_1+N_2\choose k}}
\end{equation}
and
\begin{equation}
p_i^*=\dfrac{{N_{2}\choose X_{i \cdot }}}{{N_1+N_2\choose X_{i \cdot }}},
\end{equation}
where $X_{i\cdot} = X_{i1}+X_{i2}$.

\input{clinsafe_adjP_SS}
Based on the calculated $p$-values, we apply our proposed stepwise procedures to the clinical safety data analysis. We can make decisions of rejection and acceptance by calculating and comparing the adjusted $p$-values for these procedures with the given significance level.
Table \ref{table:clinsafe_single} shows that for the first three AE types, the corresponding adjusted $p$-values of Procedure \ref{MBonf} are smaller than those of the Tarone, Sidak, and Bonferroni procedures, which implies the corresponding hypotheses $H_{(1)}, \dots, H_{(4)} $ are more likely to be rejected by Procedure \ref{MBonf}  than these existing single-step procedures, that is, those AEs are more easily flagged by using Procedure \ref{MBonf}. Given $\alpha=0.05$, Procedure \ref{MBonf} can flag two AEs, Tarone procedure can only flag one AE, and Sidak and Bonferroni procedures cannot flag any AE.


\input{clinsafe_adjP_SD}
Table \ref{table:clinsafe_sd} shows that for the AEs corresponding to hypotheses $H_{(1)}, \dots, H_{(4)} $, the adjusted $p$-values of Procedure \ref{MHolm} are smaller than those of the Tarone-Holm and Holm procedures. It implies that our proposed Procedure \ref{MHolm} has more chances to reject these three hypotheses than these two procedures, which in turn implies Procedure \ref{MHolm} could be more powerful than the existing step-down methods. Given $\alpha=0.05$, Procedure \ref{MHolm} can flag two AEs, Tarone-Holm procedure can only flag one AE, and Holm procedure cannot flag any AE.


\input{clinsafe_adjP_SU}
Table \ref{table:clinsafe_su} shows that for the aforementioned AEs, the adjusted $p$-values of Procedure \ref{MHoch} are smaller than those of Roth and Hochberg procedures. It implies that Procedure \ref{MHoch} has more chances to reject $H_{(1)}, \dots, H_{(4)} $ than these two procedures, which in turn implies our proposed Procedure \ref{MHoch} could be more powerful than the existing step-up methods. Given $\alpha=0.05$, Procedure \ref{MHoch} can flag two AEs, Roth procedure can only flag one AE, and Hochberg procedure cannot flag any AE.

\section{Conclusions}
In this paper, we have developed three new FWER controlling procedures for discrete data by fully utilizing marginal distributions of true null $p$-values rather than minimal attainable $p$-values, which is often used in the developments of existing procedures for discrete data. We have shown that the proposed modified Bonferroni and Holm procedures strongly control the FWER under arbitrary dependence and are more powerful than the existing Tarone-type procedures, whereas the proposed modified Hochberg procedure ensures control of the FWER in special scenarios. Through extensive simulation studies, we have provided numerical evidence of superior performance of the proposed procedures in terms of the FWER control and minimal power, even for the modified Hochberg. We have also developed an R package ``MHTdiscrete" and a web application for implementing the proposed procedures.

A limitation for the proposed methods is when the sample size proportional to the number of tested hypotheses is considerable large, their power improvements are relatively minimal compared to the existing methods.
It should be noted that the proposed procedures are developed for controlling the FWER, they are more appropriate for small-scale multiple testing where the number of tested hypotheses is pretty small. It should be also noted that these proposed procedures are developed under a weak assumption of the marginal distributions (CDF) of true null $p$-values being known. When the marginal distributions are unknown, the proposed procedures are not directly applicable; some known upper bounds or estimates of the marginal distributions are needed.
In practice, some resampling methods such as permutation or bootstrap can be potentially considered to estimate the distributions. We leave it for future work. Another possible future work is to explore optimality of the suggested modified Bonferroni and Holm procedures under arbitrary dependence, in the sense of that one cannot increase even one of the critical constants  while keeping the remaining fixed without losing control of the FWER.


\section{Appendix}
\subsection{Proof of Theorem 3.1.}
\begin{proof}
Let $V$ denote the number of falsely rejected hypotheses and $I_0$ the index set of true null hypotheses, then
\begin{equation}
\begin{aligned}
FWER &=  Pr \{ V \geq 1\} = Pr \left\{ \bigcup\limits_{i \in I_0} \{P_i \leq s^*\} \right\}\\
&\leq \sum\limits_{i \in I_0} Pr\{P_i \leq s^* \} =  \sum\limits_{i \in I_0} F_i(s^*)\\
& \leq \sum\limits_{i=1}^{m} F_i(s^*) \leq \alpha.
\end{aligned}
\end{equation}
The first inequality follows from Bonferroni inequality. The last inequality follows from the following arguments on the definition of $s^*$: (i) if the maximum exists, the inequality automatically holds;
(ii) if the maximum does not exist, $s^* =\dfrac{\alpha}{m}$ and thus by Assumption \ref{uniform},
$$\sum\limits_{i=1}^{m} F_i(s^*) \le \sum\limits_{i=1}^{m} s^* = m \cdot \dfrac{\alpha}{m}= \alpha.$$
The proof is complete.	
\end{proof}

\subsection{Proof of Proposition \ref{MBonf>T}.}
\begin{proof}
Firstly, we prove that Procedure \ref{MBonf} is universally more  powerful than Procedures \ref{ProcT}.

For Procedure \ref{ProcT}, let $R_{K(\alpha)}=\{ i: p_i^* \le \dfrac{\alpha}{K(\alpha)} \}$, then $|R_{K(\alpha)}|=M(\alpha, K(\alpha))\leq K(\alpha)$. Thus,
	
	\begin{equation} \label{eq1}
	\sum\limits_{i=1}^{m} F_i(\dfrac{\alpha}{K(\alpha)}) = \sum\limits_{i \in R_{K(\alpha)}} F_i(\dfrac{\alpha}{K(\alpha)}) \leq |R_{K(\alpha)}| \cdot \dfrac{\alpha}{K(\alpha)} \leq \alpha.
	\end{equation}
	
	Let
	$$t^*=\min\left\{p \in \bigcup\limits_{i=1}^{m} \mathbb{P}_i: p > s^* \right\}.$$
	Then, by the definition of $s^*$ in Procedure \ref{MBonf}, we have
	\begin{equation} \label{eq2}
	\sum\limits_{i=1}^{m} F_i(t^*) > \alpha.
	\end{equation}
	Combining (\ref{eq1}) and (\ref{eq2}), we have $\dfrac{\alpha}{K(\alpha)} < t^* $. Then there are two cases regarding the critical values $s^*$ and $\dfrac{\alpha}{K(\alpha)}$ of Procedures \ref{MBonf} and \ref{ProcT}:
	\begin{itemize}
		\item[(i)] if $ \dfrac{\alpha}{K(\alpha)} \leq s^* $, it is trivial that the set of rejections by Procedures \ref{ProcT} is no larger than that of Procedure \ref{MBonf};
		\item[(ii)]  if $s^* <\dfrac{\alpha}{K(\alpha)} < t^*$, by the definition of $t^*$, it follows that
		$$ \{ H_i: P_i \leq s^* \} = \{ H_i: P_i < t^* \} = \{ H_i: P_i \le \dfrac{\alpha}{K(\alpha)} \},$$
which implies that the rejection sets for these two methods are the same.
	\end{itemize}
	
	Summarizing the above two cases, Procedure \ref{MBonf} always rejects any hypotheses rejected by Procedures \ref{ProcT}. That is, Procedure \ref{MBonf} is universally more  powerful than Procedures \ref{ProcT}.

Secondly, we prove that Procedure \ref{MBonf} is universally more  powerful than Procedure \ref{T*}.

We show that for any $\gamma\in (0,\alpha]$, $\sum\limits_{i=1}^{m} F_i(\dfrac{\gamma}{K(\gamma)}) \leq \alpha $. Let $R_{K(\gamma)}=\{ i: p_i^* \le \dfrac{\gamma}{K(\gamma)} \}$, then $|R_{K(\gamma)}|=M(\gamma, K(\gamma))\leq K(\gamma)$. Thus,
	\begin{equation}
	\begin{aligned}
	& \quad \quad \sum\limits_{i=1}^{m} F_i(\dfrac{\gamma}{K(\gamma)}) =  \sum\limits_{i \in R_{K(\gamma)}} F_i(\dfrac{\gamma}{K(\gamma)}) \\
	&\leq |R_{K(\gamma)}| \cdot \dfrac{\gamma}{K(\gamma)} =M(\gamma, K(\gamma))\cdot \dfrac{\gamma}{K(\gamma)}\\
	&\leq \gamma \le \alpha.\\
	\end{aligned}
	\end{equation}
	For the rest of proof, it is similar to the arguments used in the first part and the conclusion follows.
\end{proof}

\subsection{Proof of Theorem 3.2.}
\begin{proof}
Let $I_0$ be the indices of the true null hypotheses and $V$ the number of falsely rejected hypotheses. If $|I_0|=0$, then $V=0$ and $FWER=0 \leq \alpha$ is trivial. When $|I_0|=m_0 > 0$, let
	$\hat{P}_{(1)} \le \dots \le \hat{P}_{(m_0)}$ denote the $m_0$ ordered true null $p$-values, and 	${P}_{(1)} \le \dots \le {P}_{(m)}$ denote the $m$ ordered $p$-values.
	
	Let $k$ be the smallest (random) index of the minimal true null $p$-values $\hat{P}_{(1)}$,  thus $k$ is the smallest element in the index set $I_0$ of true null $p$-values. Hence,
\begin{equation} \label{sd_eq0}
I_0 \subseteq \left\{(k), \dots, (m)\right\}.
\end{equation}
Define $\alpha_{I_0} = \max\{ p \in \bigcup\limits_{i=1}^{m} \mathbb{P}_i: \sum\limits_{i \in I_0} F_i(p) \leq \alpha \}$ if the maximum exists; otherwise set $\alpha_{I_0} = \dfrac{\alpha}{m_0}$.
Note that $F_i$ are known and $I_0$ is a fixed set, thus $\alpha_{I_0}$ is constant. In the following, we show by using induction that for $i=1, \ldots, k$,
\begin{equation} \label{sd_eq2}
	\alpha_i \le \max\left\{\alpha_{I_0}, \dfrac{\alpha}{m_0}\right\}.
	\end{equation}
When $i=1$, by the definition of $\alpha_1$, if the maximum exists, then by (\ref{sd_eq0}), we have $\alpha_1 \le \alpha_{I_0}$; if the maximum does not exist, then $\alpha_1 = \dfrac{\alpha}{m} \le \dfrac{\alpha}{m_0}$. Therefore, (\ref{sd_eq2}) holds for the case of $i = 1$.

Assume that the inequality (\ref{sd_eq2}) holds for $i = l < k$. In the following, we show that (\ref{sd_eq2}) also holds for $i = l+1$. Note that by (\ref{sd_eq0}) and  $l < k$, we have $m_0 \le m-k +1 \le m-l$ and then
$\dfrac{\alpha}{m-l} \le \dfrac{\alpha}{m_0}$. Thus, if the maximum does not hold, by the induction assumption,
$$\alpha_{l+1} = \max\left\{\alpha_{l}, \dfrac{\alpha}{m-l}\right\} \le \max\left\{\alpha_{I_0}, \dfrac{\alpha}{m_0}\right\}.$$
If the maximum exists, by using the similar argument as in the case of $i = 1$, the inequality (\ref{sd_eq2}) also holds. Therefore, (\ref{sd_eq2}) holds for the case of $i = l+1$. By induction, the inequality (\ref{sd_eq2}) holds for $i = 1, \ldots, k$. Thus,
\begin{equation} \label{sd_eq1}
	\begin{aligned}
	FWER &=  Pr \{ V \geq 1\}  \le Pr \{ \hat{P}_{(1)} \leq \alpha_k \}\\
	& \leq  \sum\limits_{i \in I_0} Pr \left\{ P_i \leq \max\left\{\alpha_{I_0}, \dfrac{\alpha}{m_0}\right\} \right\} \\
&= \max\left\{\sum\limits_{i \in I_0} F_i(\alpha_{I_0}),  \sum\limits_{i \in I_0} F_i\left(\dfrac{\alpha}{m_0}\right) \right\} \\
&\le \alpha.
	\end{aligned}
	\end{equation}
Here, the second inequality follows from (\ref{sd_eq2}) and the Bonferroni inequality, and the third inequality follows from the same argument as in Theorem 3.1 along with the definition of $\alpha_{I_0}$ and Assumption \ref{uniform}. 	The proof is complete.	
\end{proof}

\subsection{Proof of Proposition 3.2.}
\begin{proof}
For $i = 1, \ldots, m$, denote $I_{i}$ as the index set of the ordered $p$-values starting from $P_{(i)}$, i.e., $I_{i} = \{(i), \dots, (m)\}$. Let $R_{K_{I_i}(\gamma)}=\{ j \in I_i : p_j^* \le \dfrac{\gamma}{K_{I_i}(\gamma)} \}$, then $|R_{K_{I_i}(\gamma)}|=M_{I_i}(\gamma, K_{I_i}(\gamma))\leq K_{I_i}(\gamma)$. Thus, by using the similar argument as in the proof of Proposition \ref{MBonf>T}, we have
	\begin{equation}
	\begin{aligned}
	& \quad \quad \sum\limits_{j=i}^{m} F_{(j)}(\dfrac{\gamma}{K_{I_i}(\gamma)})  = \sum\limits_{j \in I_i } F_j(\dfrac{\gamma}{K_{I_i}(\gamma)}) \\
	&=  \sum\limits_{j \in R_{K_{I_i}(\gamma)}} F_j(\dfrac{\gamma}{K_{I_i}(\gamma)}) \leq |R_{K_{I_i}(\gamma)}| \cdot \dfrac{\gamma}{K_{I_i}(\gamma)} \\
	&=M_{I_i}(\gamma, K_{I_i}(\gamma))\cdot \dfrac{\gamma}{K_{I_i}(\gamma)} \leq \gamma \le \alpha.\\
	\end{aligned}
	\end{equation}
	For the rest of proof, it is similar to that of Proposition \ref{MBonf>T} and the conclusion follows.
\end{proof}

\subsection{Proof of Proposition 3.3.}
\begin{proof}
Since the true null $p$-values are identically distributed, let us assume that the true null $p$-values $P_i$ have the same support $\mathbb{P}$ and CDF $F(\cdot)$. Thus, for each $i = 1, \ldots, m$,
	\begin{equation} \label{mhoch_proof}
	\begin{split}
	\alpha_i & = \max\{ p \in \bigcup\limits_{j=i}^{m} \mathbb{P}_{(j)}: \sum\limits_{j=i}^{m} F_{(j)}(p) \leq \alpha \}\\
	& =  \max\{ p \in  \mathbb{P}: (m-i+1) F(p) \leq \alpha \}\\
	&=  \max\left\{ p \in  \mathbb{P}: p \leq \dfrac{\alpha}{m-i+1} \right\}. \\
	\end{split}
	\end{equation}
	The last equality follows from Assumption \ref{uniform}. Obviously, $\alpha_i \le \dfrac{\alpha}{m-i+1}$, that is, $\alpha_i$ is always smaller than or equal to the critical value $\dfrac{\alpha}{m-i+1}$ of the conventional Hochberg. By the FWER control of the Hochberg procedure
	under under Assumption \ref{PRDS}, we have that Procedure \ref{MHoch} also controls the FWER under Assumption \ref{PRDS}.
	
	To prove (ii), let $R = \max\{i: P_{(i)} \le \dfrac{\alpha}{m-i+1} \}$ be the number of rejections by the Hochberg procedure, then for each $H_i$, $H_i$ is rejected by the Hochberg procedure if $P_i \le P_{(R)}$. Thus, by (\ref{mhoch_proof}),
	$$P_{(R)} = \max\{P_{(i)}: P_{(i)} \le \dfrac{\alpha}{m-i+1}\} = \max\{P_i: P_i \le \dfrac{\alpha}{m-R+1}\} \le \alpha_R.$$
We also note that
$$P_{(i)} > \dfrac{\alpha}{m-i+1} \ge \alpha_{i} ~\text{ for } i = R+1, \ldots, m.$$
Thus, the number of rejections by Procedure \ref{MHoch} is $\max\{i: P_{(i)} \le \alpha_{i} \} = R$. Therefore, Procedure \ref{MHoch} rejects the same hypotheses $H_{(1)}, \ldots, H_{(R)}$ as the conventional Hochberg procedure.
\end{proof}

\subsection{Proof of Proposition 3.4.}
\begin{proof}
By the definition of the critical values of Procedure \ref{MHoch}, the critical values $\alpha_i, i=1, 2$ for this procedure under the special case of two null hypotheses are calculated as
	\begin{eqnarray} \nonumber
	\alpha_1 =\begin{cases}
	\alpha/2 , & \alpha < p_1 \cr
	p_1, & p_1 \le \alpha < p_1+p_2 \cr
	p_2, & \alpha \ge p_1+p_2
	\end{cases}
	\end{eqnarray}
	and if $P_1 \le P_2$,
	\begin{eqnarray} \nonumber
	\alpha_2 =\begin{cases}
	\alpha, & \alpha < p_2 \cr
	p_2, & \alpha \ge p_2;
	\end{cases}
	\end{eqnarray}
otherwise,
	\begin{eqnarray} \nonumber
	\alpha_2 =\begin{cases}
	\alpha, & \alpha < p_1 \cr
	p_1, & \alpha \ge p_1.
	\end{cases}
	\end{eqnarray}

In the following, we prove control of the FWER for Procedure \ref{MHoch} for different combinations of true and false null hypotheses.
	
	\vskip 4pt
	\textbf{Case 1. } $H_1$ and $H_2$ are both true.
	\vskip 4pt	
	
	There are three attainable $p$-value settings in which at least one hypothesis is rejected.
	\begin{itemize}
		\item[(i)] $P_1=p_1$ and $P_2=1$. Since $P_2=1> \alpha_2$, accept $H_2$. To reject $H_1$, one needs to check if $P_1\le \alpha_1$, i.e., $p_1 \le \alpha$. Thus, $H_1$ is rejected iff $p_1 \le \alpha$.
		\item[(ii)] $P_1=1$ and $P_2=p_2$. Similarly, $H_1$ is accepted since $P_1 > \alpha_2$. To reject $H_2$, one needs  to check if $P_2 \le \alpha_1$, which is equivalent to $p_1+p_2 \le \alpha$.
		Thus, $H_2$ is rejected iff $p_1+p_2 \le \alpha$.
		\item[(iii)] $P_1 = p_1$ and $P_2 = p_2$. By the definition of step-up procedure, it is easy to check that $H_1$ and $H_2$ are both rejected iff $p_2 \le \alpha$; only $H_1$ is rejected iff $p_1 \le \alpha < p_2$. Thus, by $p_1 < p_2$, we have that at least one hypothesis is rejected iff $p_1 \le \alpha$.
	\end{itemize}
	Therefore, if $p_1 \le \alpha$ but $p_1+p_2 > \alpha$,
	\begin{equation} \label{SUeq1}
	\begin{split}
	FWER & = \Pr\{H_1 \text{ or } H_2 \text{ rejected} \} \\
	& = \Pr(P_1=p_1, P_2=1)+\Pr(P_1=p_1, P_2=p_2) \\
	& = Pr(P_1=p_1) = p_1 \le \alpha.
	\end{split}
	\end{equation}
	If $p_1+p_2 \le \alpha$,
	\begin{equation} \label{SUeq2}
	\begin{split}
	FWER & = \Pr\{H_1 \text{ or } H_2 \text{ rejected} \} \\
	& = \Pr(P_1=p_1, P_2=1) + \Pr(P_1=1, P_2=p_2) + \Pr(P_1=p_1, P_2=p_2) \\
	& \le Pr(P_1=p_1) + \Pr(P_2=p_2) \\
	& = p_1 + p_2 \le \alpha.
	\end{split}
	\end{equation}
	If $p_1 > \alpha$,
	\begin{equation} \label{SUeq3}
	\begin{split}
	FWER & = 0.
	\end{split}
	\end{equation}
	Combining (\ref{SUeq1})-(\ref{SUeq3}), the desired result follows under Case 1.
	
	\vskip 4pt
	\textbf{Case 2. } $H_1$ is true but $H_2$ is false.
	\vskip 4pt	
	
	By the $p$-value monotonicity of Procedure \ref{MHoch}, its FWER is maximized when
	$P_2 = 0$ with probability 1. Thus, $H_1$ is rejected iff $P_1 \le \alpha_2$, which is equivalent to
	$p_1 \le \alpha$. Therefore, if $p_1 \le \alpha$,
	\begin{equation} \label{SUeq4}
	\begin{split}
	FWER & = \Pr\{H_1 \text{ rejected} \} = \Pr(P_1=p_1)  \\
	& = p_1 \le \alpha;
	\end{split}
	\end{equation}
	otherwise,
	\begin{equation} \label{SUeq5}
	\begin{split}
	FWER & = 0.
	\end{split}
	\end{equation}
	Combining (\ref{SUeq4})-(\ref{SUeq5}), the desired result follows under Case 2.
	
	\vskip 4pt
	\textbf{Case 3. } $H_1$ is false but $H_2$ is true.
	\vskip 4pt	
	
	\indent By using the similar arguments as in Case 2, we have
	\begin{equation}
	\begin{split}
	FWER & = \Pr\{H_2 \text{ rejected} \} \le \alpha.
	\end{split}
	\end{equation}

	\vskip 4pt
	
	Summarizing the above discussions under Cases 1-3, we have that
	Procedure \ref{MHoch} strongly controls the FWER, which completes the proof.
\end{proof}

\vskip 14pt
\noindent {\large\bf Supplementary Materials}

The online supplementary materials contain additional simulation results for independence and dependence settings, and a brief description of statistical computing softwares for the proposed methods.

\par
\vskip 14pt
\noindent {\large\bf Acknowledgements}

The research of Wenge Guo was supported in part by NSF Grant DMS-1309162.

\par

\pagebreak

  \begin{center}
	{\LARGE\bf Familywise Error Rate Controlling \vskip -0.15in Procedures for Discrete Data - \vskip 0.1in Supplementary Materials}
\end{center}

\newcommand{\beginsupplement}{%
\setcounter{section}{0}
\renewcommand{\thesection}{S\arabic{section}}
\setcounter{table}{0}
\renewcommand{\thetable}{S\arabic{table}}%
\setcounter{figure}{0}
\renewcommand{\thefigure}{S\arabic{figure}}%
}

\beginsupplement

\vskip 10pt

\section{ Results from Independence Simulation Settings}
The simulation results under the independence setting for stepwise procedures comparisons are shown in this section. Tables \ref{sim_FWER_SS_FET}, \ref{sim_Pow_SS_FET} and Tables \ref{sim_FWER_SS_BT}, \ref{sim_Pow_SS_BT}
respectively provide the results of numerical comparisons of single-step procedures using Fisher and Binomial Exact Tests (as plotted in Figures 1 and 2 in main paper).

\input{simulation_table_FWER_SS_FET}

\input{simulation_table_Pow_SS_FET}

\input{simulation_table_FWER_SS_BT}

\begin{figure}[H]
	\centering
	\includegraphics[width=6.4in]{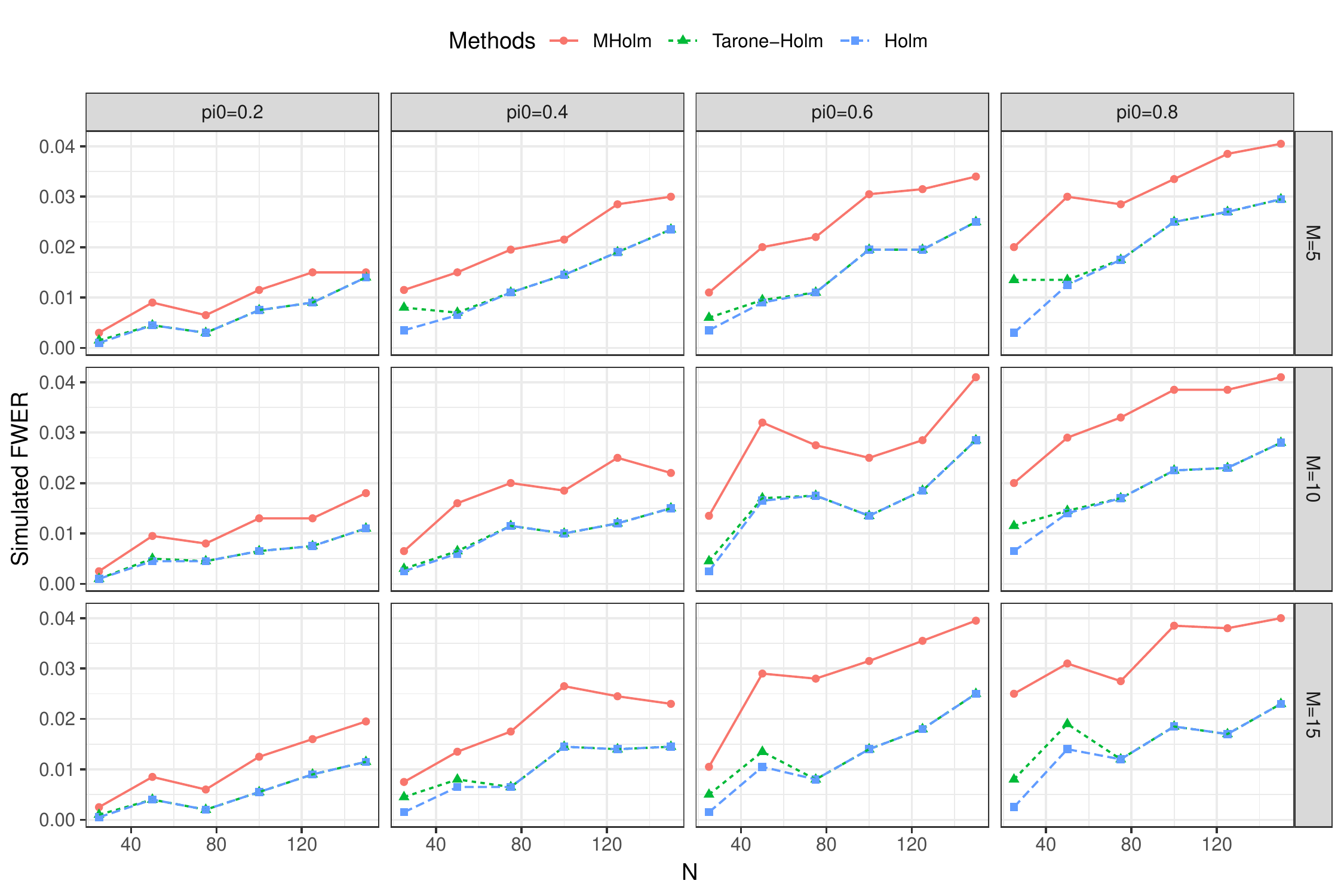}
	\caption{Simulated FWER comparisons for different step-down procedures based on FET, including Procedure 3.2 (MHolm),  Procedure 2.3 (Tarone-Holm), and the conventional Holm procedure (Holm).}\label{SD_FWER}
\end{figure}

\begin{figure}[H]
	\centering
	\includegraphics[width=6.4in]{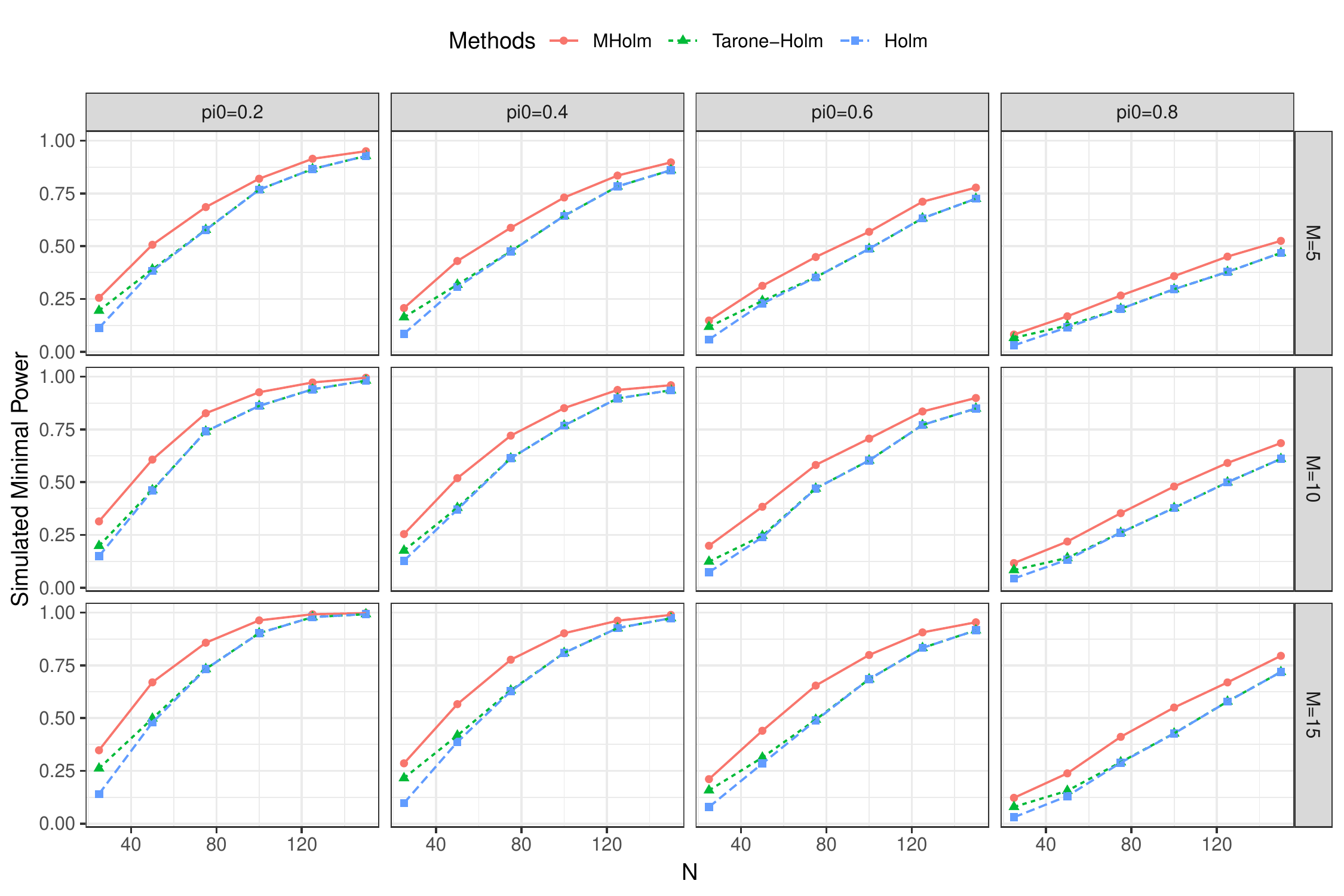}
	\caption{Simulated minimal power comparisons for different step-down procedures based on FET, including Procedure 3.2 (MHolm),  Procedure 2.3 (Tarone-Holm), and the conventional Holm procedure (Holm).}\label{SD_Pow}
\end{figure}

\begin{figure}[H]
	\centering
	\includegraphics[width=6.4in]{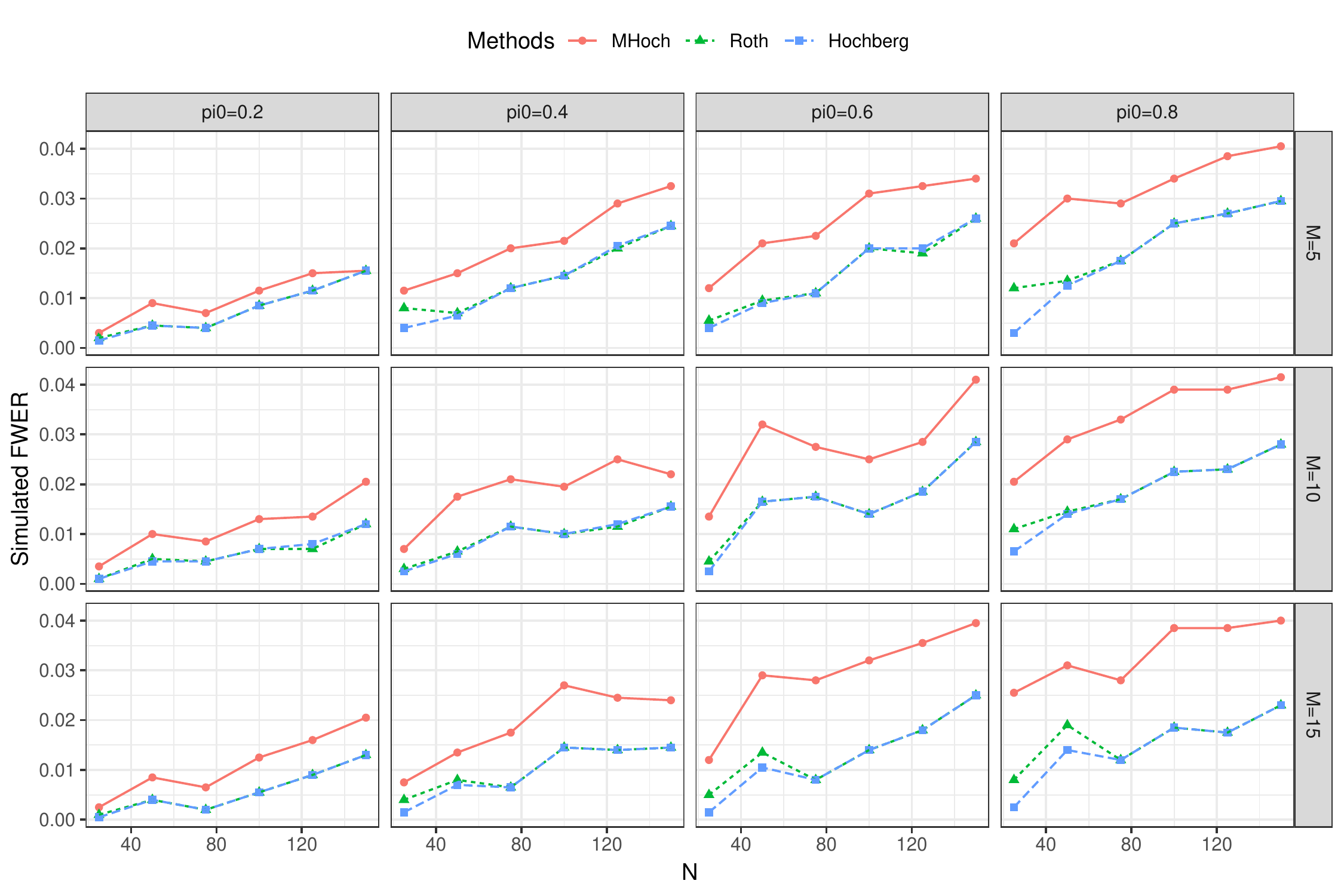}
	\caption{Simulated FWER comparisons for different step-up procedures based on FET, including Procedure 3.3 (MHoch), the Roth procedure (Roth), and the conventional Hochberg procedure (Hochberg).}\label{SU_FWER}
\end{figure}

\begin{figure}[H]
	\centering
	\includegraphics[width=6.4in]{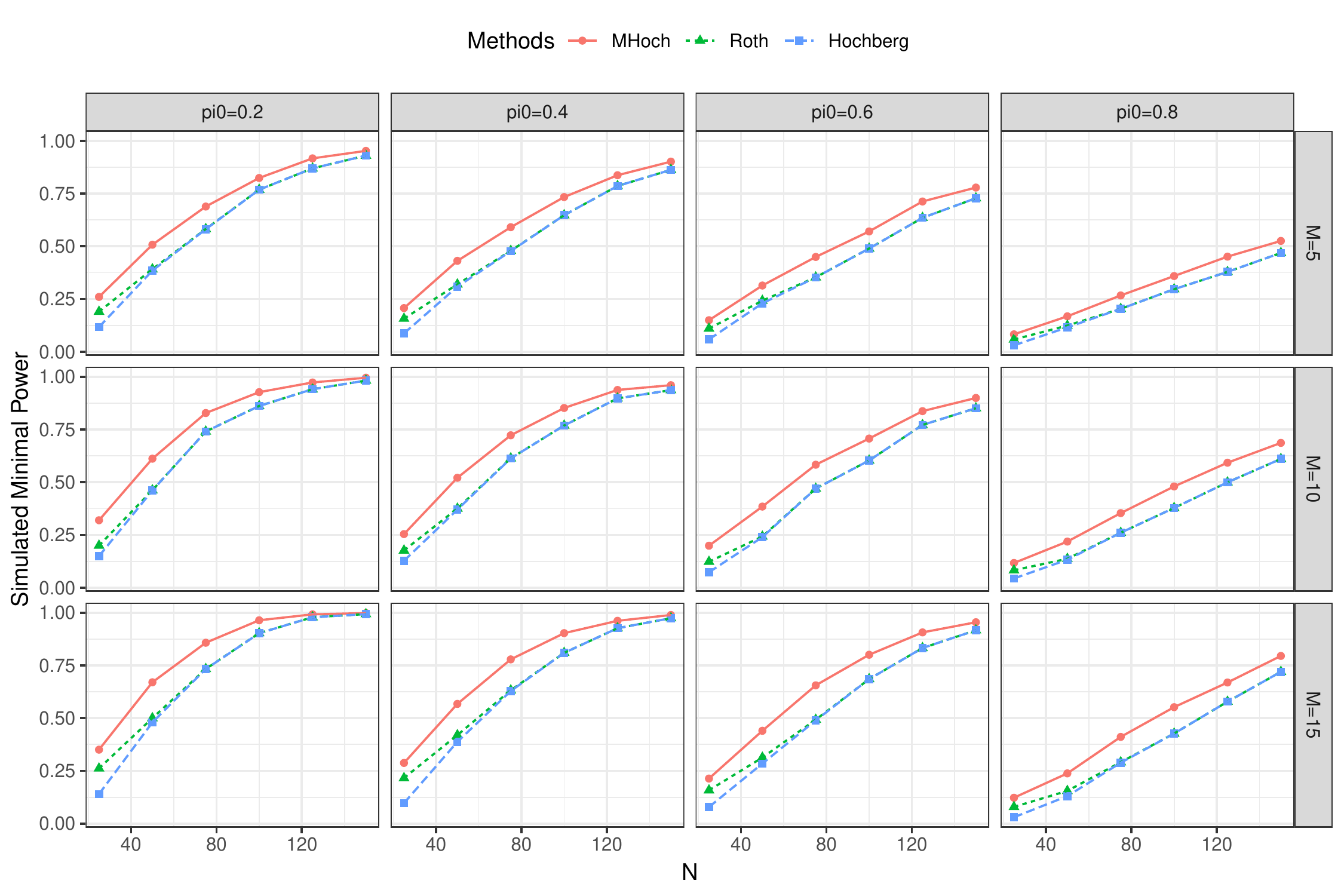}
	\caption{Simulated minimal power comparisons for different step-up procedures based on FET, including Procedure 3.3 (MHoch), the Roth procedure (Roth), and the conventional Hochberg procedure (Hochberg).}\label{SU_Pow}
\end{figure}

\input{simulation_table_Pow_SS_BT}
\vskip 4pt
Tables \ref{sim_FWER_SD_FET} and \ref{sim_Pow_SD_FET} provide numerical results of step-down procedures comparisons using Fisher Exact Test, which are also plotted as Figures \ref{SD_FWER} and \ref{SD_Pow}. Tables \ref{sim_FWER_SU_FET} and \ref{sim_Pow_SU_FET} provide numerical results of step-up procedures comparisons using Fisher Exact Test, which are plotted as Figures \ref{SU_FWER} and \ref{SU_Pow}.

\input{simulation_table_FWER_SD_FET}

\input{simulation_table_Pow_SD_FET}

\newpage


\input{simulation_table_FWER_SU_FET}

\input{simulation_table_Pow_SU_FET}


\section{Results from Dependence Simulation Settings}

In this section, we provide the details for simulating the block dependent binomial exact test (BET) statistics and the simulation results for the stepwise procedures comparisons. The following steps illustrate how to generate the dependent BET statistics and corresponding $p$-values.

\vskip 10pt
\textbf{Step 1. Generate dependent Poisson observed counts for each group}
\vskip 4pt
In order to generate $m$ dependent BET statistics $T_{i}$, we use the following algorithm to generate $m$ dependent Poisson random variables within each group, noting that the Poisson random variables between two groups are independent.

\begin{enumerate}
	\item Let $\lambda_{i1} = 2$ for $i=1, \dots, m$,  generate $m$ independent Poisson random variable $Y_{i1} \sim Poi((1-\rho)\lambda_{i1})$ and one $Y_{01} \sim Poi(2\rho)$.
	
	\item Let $X_{i1}=Y_{i1}+Y_{01}$ for $i=1, \dots, m$, then $X_{i1} \sim Poi(2)$ and the correlation between $X_{i1}$ and $X_{j1}$  is  $\dfrac{Cov(X_{i1},X_{j1})}{\sqrt{Var(X_{i1})}\sqrt{Var(X_{j1})}}=\dfrac{Var(Y_{01})}{\sqrt{2}\sqrt{2}}= \dfrac{2\rho}{2}=\rho$ for $i,j=1,\dots,m$ and $i\neq j$.
	
	\item Let $\lambda_{i2} = 2$ for $i=1, \dots, m_0$ and $\lambda_{i2} = 10$ for $i=m_0+1, \dots, m$,
generate $m$ independent Poisson random variable $Y_{i2} \sim Poi((1-\rho)\lambda_{i2})$ for $i=1, \dots, m$, one $Y_{02} \sim Poi(2\rho)$, and one $Y'_{02} \sim Poi(10\rho)$.
	
\item Let $X_{i2}=Y_{i2}+Y_{02}$ for $i=1, \dots, m_0$ and $X_{i2}=Y_{i2}+Y'_{02}$ for $i=m_0+1, \dots, m$,
then $X_{i2} \sim Poi(2)$ for $i=1, \dots, m_0$ and $X_{i2} \sim Poi(10)$ for $i=m_0+1, \dots, m$.
For $i,j=1,\dots,m_0$ and $i\neq j$, the correlation between $X_{i2}$ and $X_{j2}$  is  $\dfrac{Cov(X_{i2},X_{j2})}{\sqrt{Var(X_{i2})}\sqrt{Var(X_{j2})}}=\dfrac{Var(Y_{02})}{\sqrt{2}\sqrt{2}}= \dfrac{2\rho}{2}=\rho$. Similarly, for $i,j=m_0+1, \dots, m$ and $i\neq j$,  the correlation between $X_{i2}$ and $X_{j2}$ is also equal to $\rho$; for $i=1,\dots,m_0$ and $j=m_0+1, \dots, m$, the correlation between $X_{i2}$ and $X_{j2}$ is equal to zero.
\end{enumerate}

\vskip 4pt
\textbf{Step 2. Obtain the conditional test statistics}
\vskip 4pt
Since the generated Poisson random variables between two groups are independent, we can directly conduct BET for each hypothesis. After generating Poisson observed counts $x_{i1}$ and $x_{i2}$, let $c_i=x_{i1}+x_{i2}$ be the total observed count for two groups. Then the test statistics $T_i$ is conditional test statistics $X_{i1}$ given $X_{i1}+X_{i2}=c_i$ and the critical value is the observed count $x_{i1}$ for Group 1.

\vskip 4pt
\textbf{Step 3. Conditional distribution of the test statistics}
\vskip 4pt
Based on the conditional inference in Lehmann and Romano \cite{lehmann2005testing}, which is the BET in our paper, the conditional distribution of $X_{i1}$ given $X_{i1}+X_{i2}=c_i $ is Binomial, $Bin(c_i, p_i)$, where $p_i = \dfrac{\lambda_{i1}}{\lambda_{i1}+\lambda_{i2}}$.
\vskip 4pt
\textbf{Step 4. Calculate available $p$-value $P_i$ and attainable $p$-values}
\vskip 4pt	
When $H_i$ is true, i.e., $\lambda_{i1}=\lambda_{i2}$, $p_i=0.5$. Thus,  $X_{i1}| X_{i1}+X_{i2}=c_i   \sim Bin(c_i, 0.5)$ under $H_i$. Therefore, the available conditional $p$-value for $H_i$  can be calculated by
\begin{equation} \label{S1}
\begin{split}
P_i &= {\Pr}_{ H_i }\left\{X_{i1} \ge x_{i1} | X_{i1}+X_{i2}=c_i\right\}\\
& = \sum\limits_{j = x_{i1}}^{c_i} {c_i \choose j} 0.5^j (1-0.5)^{c_i-j}\\
& = \sum\limits_{j = x_{i1}}^{c_i} {c_i \choose j} 0.5^{c_i}.
\end{split}
\end{equation}
The corresponding attainable $p$-values can be calculated by
\begin{equation}
{\Pr}_{ H_i }\left\{X_{i1} \ge x | X_{i1}+X_{i2}=c_i\right\} = \sum\limits_{j = x}^{c_i} {c_i \choose j} 0.5^{c_i} \ \text{ for } x=0, 1, \dots, c_i.
\end{equation}

\begin{figure}[h]
	\centering
	\includegraphics[width=6.4in]{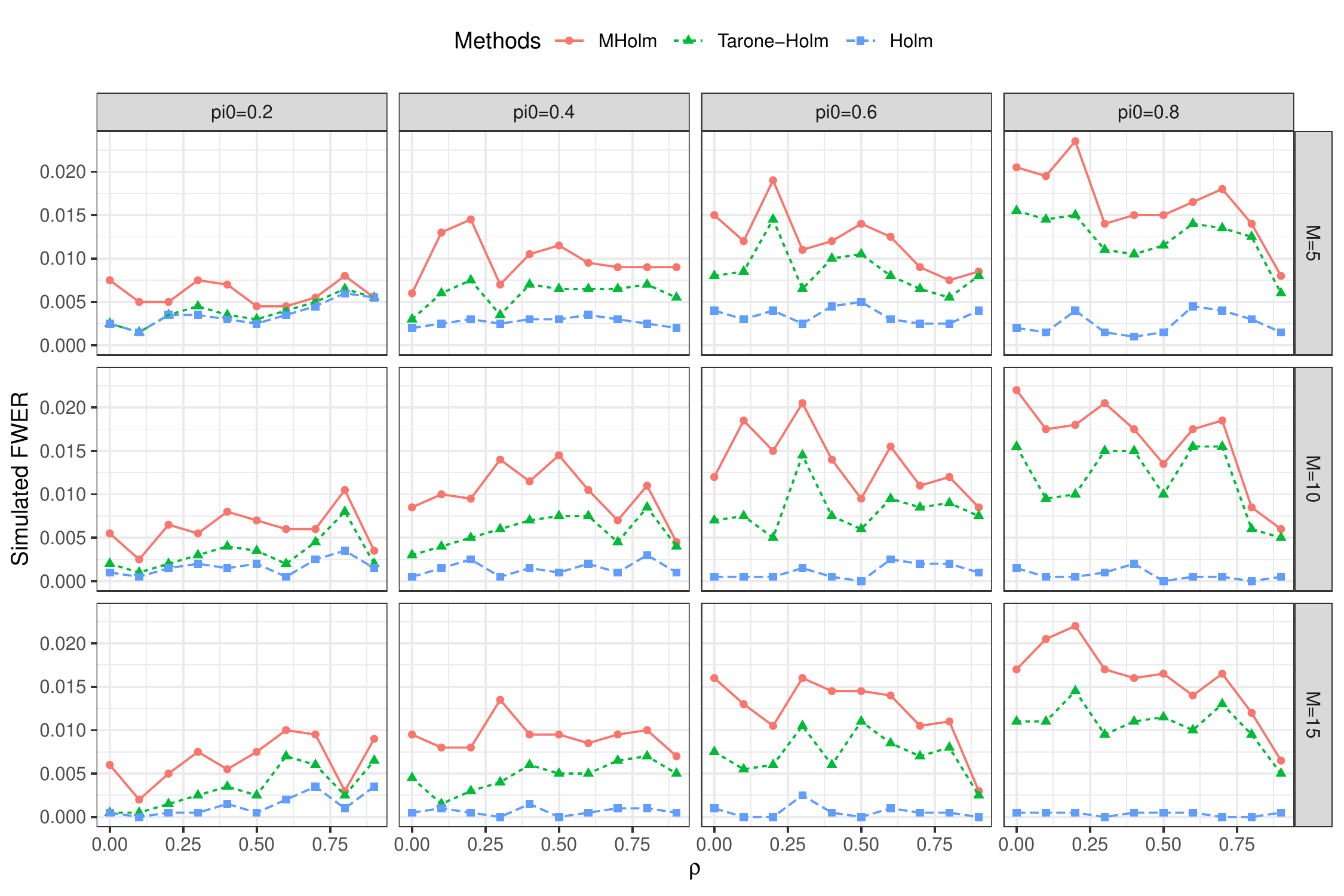}
	\caption{Simulated FWER comparisons for different step-down procedures based on the blocking dependent BET, including Procedure 3.2 (MHolm),  Procedure 2.3 (Tarone-Holm), and the conventional Holm procedure (Holm).}\label{SDdept_FWER}
\end{figure}

\begin{figure}[h]
	\centering
	\includegraphics[width=6.4in]{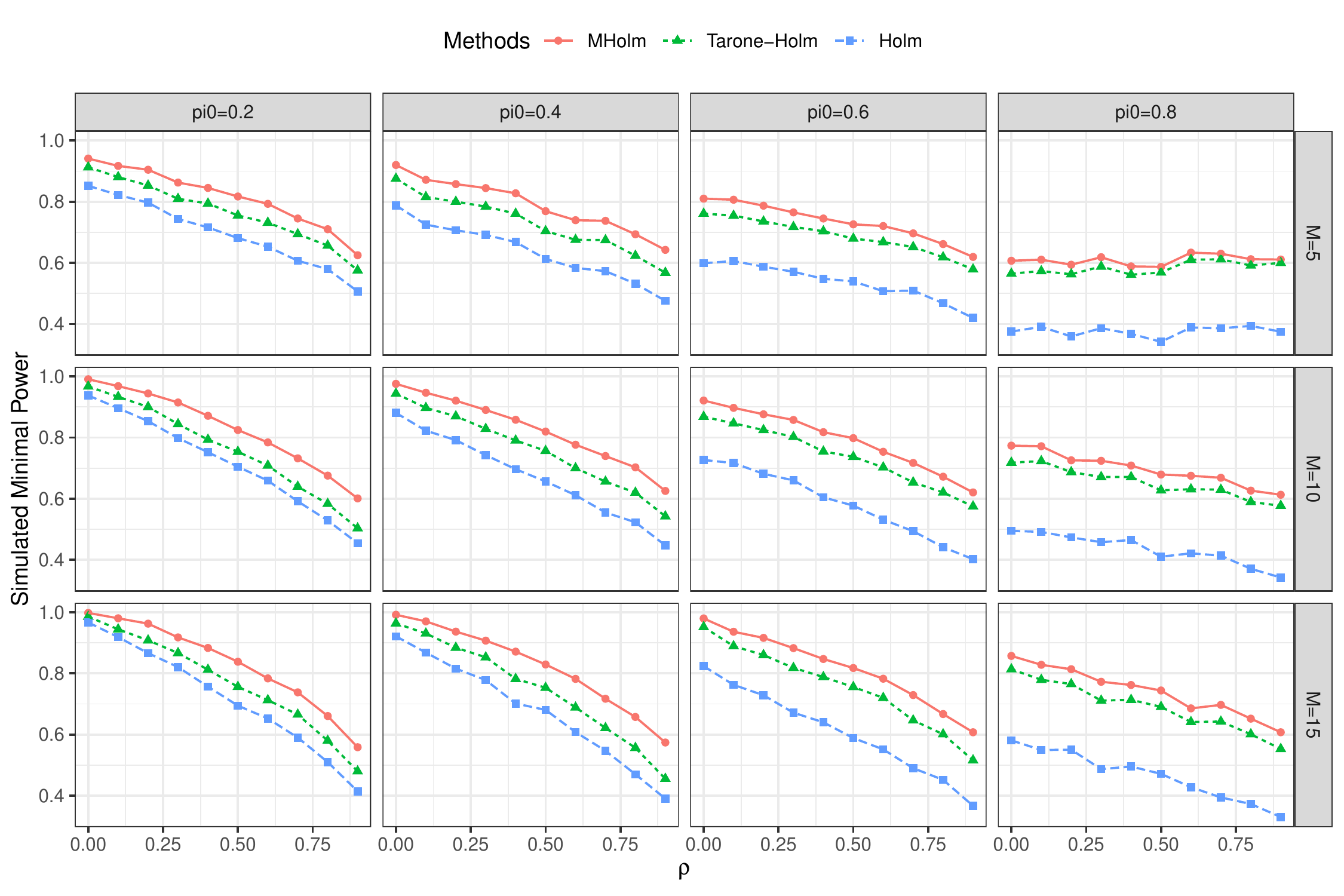}
	\caption{Simulated minimal power comparisons for different step-down procedures based on the blocking dependent BET, including Procedure 3.2 (MHolm),  Procedure 2.3 (Tarone-Holm), and the conventional Holm procedure (Holm).}\label{SDdept_Pow}
\end{figure}

\begin{figure}[h]
	\centering
	\includegraphics[width=6.4in]{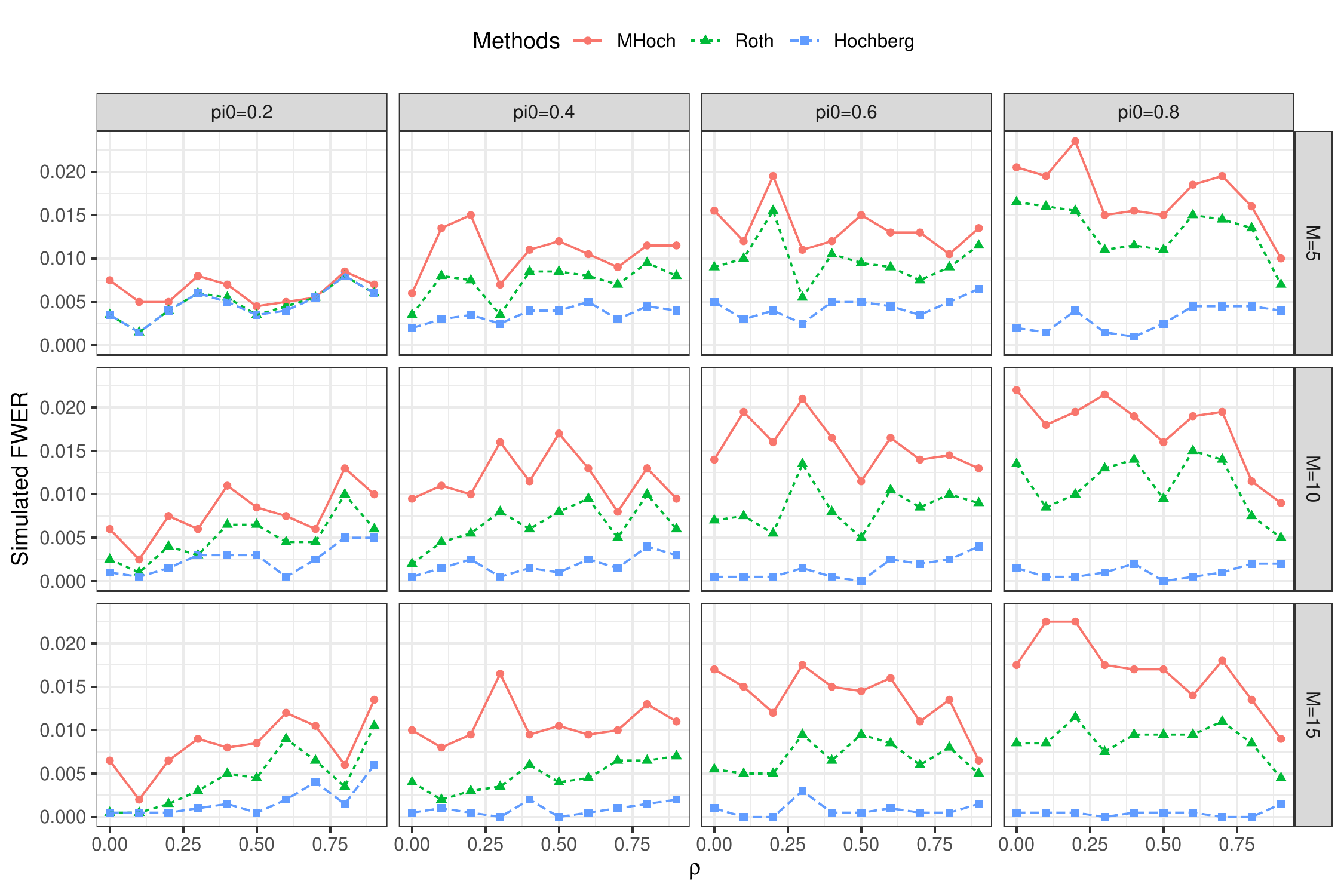}
	\caption{Simulated FWER comparisons for different step-up procedures based on the blocking dependent BET, including Procedure 3.3 (MHoch), the Roth procedure (Roth), and the conventional Hochberg procedure (Hochberg).}\label{SUdept_FWER}
\end{figure}

\begin{figure}[h]
	\centering
	\includegraphics[width=6.4in]{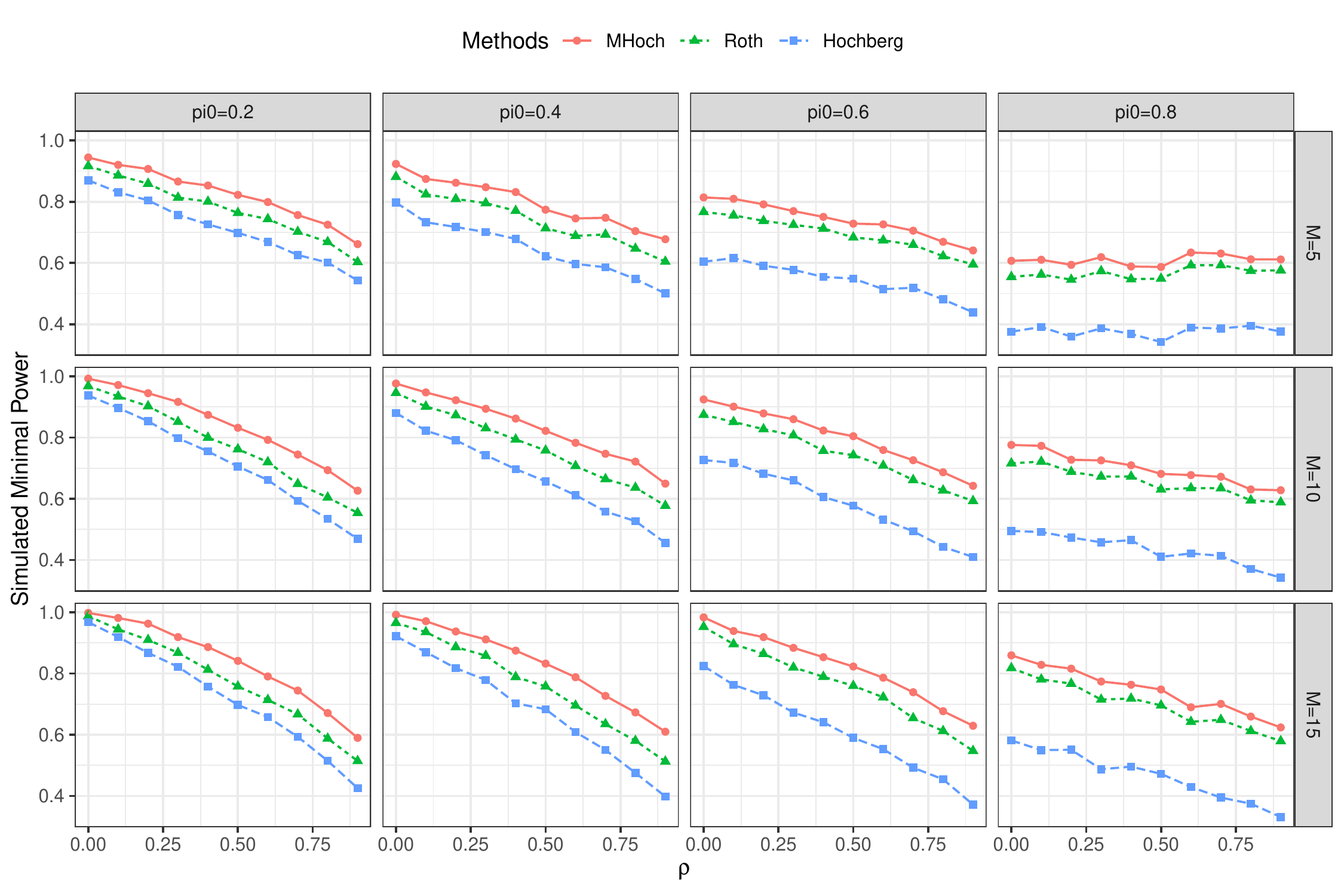}
	\caption{Simulated minimal power comparisons for different step-up procedures based on the blocking dependent BET, including Procedure 3.3 (MHoch), the Roth procedure (Roth), and the conventional Hochberg procedure (Hochberg).}\label{SUdept_Pow}
\end{figure}

The simulation results under the above simulation setting for stepwise procedures comparisons are shown in Tables \ref{sim_FWER_SS_BT_dept} - \ref{sim_Pow_SU_BT_dept} and Figures \ref{SDdept_FWER} - \ref{SUdept_Pow}.
It is easy to see that in such block dependence simulation setting, the $p$-values calculated based on the Poisson outcomes satisfies the PRDS Assumption 2.2, since $\rho \ge 0$ and the tests are one-sided.

\begin{sidewaystable}
	\input{simulation_table_FWER_SS_BT_dept}
\end{sidewaystable}
\begin{sidewaystable}

\input{simulation_table_Pow_SS_BT_dept}
\end{sidewaystable}
\begin{sidewaystable}

\input{simulation_table_FWER_SD_BT_dept}
\end{sidewaystable}
\begin{sidewaystable}

\input{simulation_table_Pow_SD_BT_dept}
\end{sidewaystable}
\begin{sidewaystable}

\input{simulation_table_FWER_SU_BT_dept}
\end{sidewaystable}
\begin{sidewaystable}

\input{simulation_table_Pow_SU_BT_dept}
\end{sidewaystable}

\begin{description}
\item[R-package for  MHTdiscrete:] R-package MHTdiscrete \cite{rpack} contains R code to implement our proposed methods and several existing FWER controlling procedures for discrete data, which are described in this paper. The package can be downloaded from \url{https://cran.r-project.org/web/packages/MHTdiscrete}.

\item[Web Application for  MHTdiscrete:] A web application containing the proposed procedures and several comparable procedures can be accessed at \url{https://allen.shinyapps.io/MTPs}.
\end{description}

\nocite{*}
\bibliography{MHTdiscrete_final_arxiv}{}

\end{document}

%% file: clinsafe_adjP_SS.tex
\begin{table}[t]\small
	\renewcommand\arraystretch{0.8}
\centering
\caption{A comparison of adjusted $p$-values for the Procedure \ref{MBonf} ($\tilde{P}_{i,MBonf}$), Procedure \ref{T*} ($\tilde{P}_{i,T^*}$), Sidak Procedure ($\tilde{P}_{i,Sidak}$) and Bonferroni Procedure ($\tilde{P}_{i,Bonf}$) when testing the hypotheses for nine AE types of Body System 10 in the clinical safety data example modified from Mehrotra and Heyse \cite{mehrotra2004use}, where the numbers of patients for two groups are $N_1=600$ and $N_2=650$.}
\vskip .25in

\label{table:clinsafe_single}

\centering
    \begin{tabular}
	{c|c|c|c|c|c|c|c} \hline
		 $i$  &
		$X_{i1} $  &
		$X_{i2} $  &
		$P_{i}$  &
		$\tilde{P}_{i,MBonf}$ &
		$\tilde{P}_{i,T^*}$ &
		$\tilde{P}_{i,Sidak}$ &
		$\tilde{P}_{i,Bonf}$ 
		\\ \hline 
1 & 13 & 3 & 0.0098 & 0.0218 & 0.0295 & 0.0851 & 0.0885 \\ 
2 & 8 & 1 & 0.0170 & 0.0469 & 0.0679 & 0.1428 & 0.1527 \\ 
3 & 4 & 0 & 0.0528 & 0.1978 & 0.2640 & 0.3863 & 0.4753 \\ 
4 & 6 & 2 & 0.1634 & 0.8467 & 1.0000 & 0.7993 & 1.0000 \\ 
5 & 2 & 0 & 0.2302 & 1.0000 & 1.0000 & 0.9051 & 1.0000 \\ 
6 & 4 & 2 & 0.4353 & 1.0000 & 1.0000 & 0.9942 & 1.0000 \\ 
7 & 0 & 2 & 0.5004 & 1.0000 & 1.0000 & 0.9981 & 1.0000 \\ 
8 & 2 & 1 & 0.6103 & 1.0000 & 1.0000 & 0.9998 & 1.0000 \\ 
9 & 1 & 2 & 1.0000 & 1.0000 & 1.0000 & 1.0000 & 1.0000 \\   \hline
    \end{tabular}
\\

\end{table}

%% file: clinsafe_adjP_SD.tex
\begin{table}[t]\small
	\renewcommand\arraystretch{0.8}
\centering
\caption{A comparison of adjusted $p$-values for the Procedure \ref{MHolm} ($\tilde{P}_{(i),MHolm}$) , Procedure \ref{TH*} ($\tilde{P}_{(i),TH}$) and Holm Procedure ($\tilde{P}_{(i),Holm}$) when testing the hypotheses for nine AE types of Body System 10 in the clinical safety data example modified from Mehrotra and Heyse \cite{mehrotra2004use}, where the numbers of patients for two groups are $N_1=600$ and $N_2=650$.}
\vskip .25in

\label{table:clinsafe_sd}

    \begin{tabular}
	{c|c|c|c|c|c|c} \hline
 $(i)$  &
$X_{i1}$  &
	$X_{i2}$  &
		$P_{(i)}$  &
		$\tilde{P}_{(i),MHolm}$ &
		$\tilde{P}_{(i),TH}$ &
			$\tilde{P}_{(i),Holm}$ 
									\\ \hline 
									
(1) & 13 & 3 & 0.0098 & 0.0218 & 0.0295 & 0.0885 \\ 
(2) & 8 & 1 & 0.0170 & 0.0370 & 0.0509 & 0.1358 \\ 
(3) & 4 & 0 & 0.0528 & 0.1165 & 0.1584 & 0.3697 \\ 
(4) & 6 & 2 & 0.1634 & 0.4948 & 0.6536 & 0.9804 \\ 
(5) & 2 & 0 & 0.2302 & 0.9009 & 1.0000 & 1.0000 \\ 
(6) & 4 & 2 & 0.4353 & 1.0000 & 1.0000 & 1.0000 \\ 
(7) & 0 & 2 & 0.5004 & 1.0000 & 1.0000 & 1.0000 \\ 
(8) & 2 & 1 & 0.6103 & 1.0000 & 1.0000 & 1.0000 \\ 
(9) & 1 & 2 & 1.0000 & 1.0000 & 1.0000 & 1.0000 \\ 
									\hline
									\end{tabular}
										\\

\end{table}

%% file: clinsafe_adjP_SU.tex
\begin{table}[t]\small
	\renewcommand\arraystretch{0.8}
\centering
\caption{A comparison of adjusted $p$-values for the Procedure \ref{MHoch} ($\tilde{P}_{(i),MHoch}$) , Roth Procedure ($\tilde{P}_{(i),Roth}$) and Hochberg Procedure ($\tilde{P}_{(i),Hochberg}$)  when testing the hypotheses for nine AE types of Body System 10 in the clinical safety data example modified from Mehrotra and Heyse \cite{mehrotra2004use}, where the numbers of patients for two groups are $N_1=600$ and $N_2=650$.}
\vskip .25in

\label{table:clinsafe_su}

    \begin{tabular}
	{c|c|c|c|c|c|c} \hline
 $(i)$  &
$X_{i1}$  &
	$X_{i2}$  &
		$P_{(i)}$  &
		$\tilde{P}_{(i),MHoch}$ &
		$\tilde{P}_{(i),Roth}$ &
			$\tilde{P}_{(i),Hochberg}$ 
									\\ \hline 
(1) & 13 & 3 & 0.0098 & 0.0218 & 0.0296 & 0.0885 \\ 
(2) & 8 & 1 & 0.0170 & 0.0370 & 0.0510 & 0.1358 \\ 
(3) & 4 & 0 & 0.0528 & 0.1165 & 0.1585 & 0.3697 \\ 
(4) & 6 & 2 & 0.1634 & 0.4948 & 0.7722 & 0.9804 \\ 
(5) & 2 & 0 & 0.2302 & 0.9009 & 1.0000 & 1.0000 \\ 
(6) & 4 & 2 & 0.4353 & 1.0000 & 1.0000 & 1.0000 \\ 
(7) & 0 & 2 & 0.5004 & 1.0000 & 1.0000 & 1.0000 \\ 
(8) & 2 & 1 & 0.6103 & 1.0000 & 1.0000 & 1.0000 \\ 
(9) & 1 & 2 & 1.0000 & 1.0000 & 1.0000 & 1.0000 \\ 
									\hline
									\end{tabular}
										\\

\end{table}

%% file: simulation_table_FWER_SS_FET.tex
\renewcommand{\baselinestretch}{1}

\begin{table}[H] \small
	\renewcommand\arraystretch{0.5}
	\centering
	\caption{Simulated FWER comparisons for single-step procedures with independent $p$-values generated from Fisher's Exact Test statistics, including Procedure 3.1 (MBonf), Procedure 2.2 (Tarone), and the conventional Sidak (Sidak) and Bonferroni (Bonf) procedures.} 
	\label{sim_FWER_SS_FET}
	\vspace{0.15 in}
	\begin{tabular}{clcccccc}
		\toprule
	&	& $N=25$ & $N=50$ & $N=75$ & $N=100$ & $N=125$ & $N=150$ \\ 
		\midrule
		\multirow{4}{2cm}{$m=5$\\$\pi_0 = 0.2$ }
& MBonf & 0.0025 & 0.0060 & 0.0035 & 0.0075 & 0.0075 & 0.0095 \\ 
  & Tarone & 0.0015 & 0.0030 & 0.0015 & 0.0055 & 0.0045 & 0.0085 \\ 
  & Sidak & 0.0010 & 0.0030 & 0.0015 & 0.0055 & 0.0045 & 0.0085 \\ 
  & Bonf & 0.0010 & 0.0030 & 0.0015 & 0.0055 & 0.0045 & 0.0085 \\ 
		\midrule
		\multirow{4}{2cm}{$m=5$\\$\pi_0 = 0.4$ }
& MBonf & 0.0100 & 0.0135 & 0.0145 & 0.0160 & 0.0160 & 0.0200 \\ 
  & Tarone & 0.0080 & 0.0055 & 0.0105 & 0.0120 & 0.0125 & 0.0145 \\ 
  & Sidak & 0.0030 & 0.0055 & 0.0105 & 0.0120 & 0.0135 & 0.0145 \\ 
  & Bonf & 0.0030 & 0.0055 & 0.0105 & 0.0120 & 0.0125 & 0.0145 \\ 
		\midrule
		\multirow{4}{2cm}{$m=5$\\$\pi_0 = 0.6$ }
& MBonf & 0.0110 & 0.0185 & 0.0185 & 0.0225 & 0.0245 & 0.0270 \\ 
  & Tarone & 0.0060 & 0.0090 & 0.0095 & 0.0180 & 0.0150 & 0.0175 \\ 
  & Sidak & 0.0035 & 0.0090 & 0.0095 & 0.0180 & 0.0160 & 0.0175 \\ 
  & Bonf & 0.0035 & 0.0090 & 0.0095 & 0.0180 & 0.0150 & 0.0175 \\  
  \midrule
  \multirow{4}{2cm}{$m=5$\\$\pi_0 = 0.8$ }
& MBonf & 0.0190 & 0.0280 & 0.0265 & 0.0315 & 0.0370 & 0.0360 \\ 
  & Tarone & 0.0125 & 0.0135 & 0.0170 & 0.0225 & 0.0250 & 0.0260 \\ 
  & Sidak & 0.0030 & 0.0125 & 0.0170 & 0.0225 & 0.0260 & 0.0260 \\ 
  & Bonf & 0.0030 & 0.0125 & 0.0170 & 0.0225 & 0.0250 & 0.0260 \\ 
  \bottomrule
		\toprule
		\multirow{4}{2cm}{$m=10$\\$\pi_0 = 0.2$ }
& MBonf & 0.0025 & 0.0090 & 0.0075 & 0.0100 & 0.0075 & 0.0085 \\ 
  & Tarone & 0.0010 & 0.0035 & 0.0045 & 0.0060 & 0.0055 & 0.0060 \\ 
  & Sidak & 0.0005 & 0.0035 & 0.0045 & 0.0065 & 0.0060 & 0.0065 \\ 
  & Bonf & 0.0005 & 0.0035 & 0.0045 & 0.0060 & 0.0055 & 0.0060 \\ 
		\midrule
		\multirow{4}{2cm}{$m=10$\\$\pi_0 = 0.4$ }
& MBonf & 0.0060 & 0.0140 & 0.0170 & 0.0130 & 0.0170 & 0.0165 \\ 
  & Tarone & 0.0030 & 0.0065 & 0.0105 & 0.0100 & 0.0075 & 0.0110 \\ 
  & Sidak & 0.0020 & 0.0060 & 0.0105 & 0.0100 & 0.0095 & 0.0115 \\ 
  & Bonf & 0.0020 & 0.0060 & 0.0105 & 0.0100 & 0.0075 & 0.0110 \\ 
		\midrule
		\multirow{4}{2cm}{$m=10$\\$\pi_0 = 0.6$ }
& MBonf & 0.0130 & 0.0310 & 0.0245 & 0.0225 & 0.0230 & 0.0340 \\ 
  & Tarone & 0.0045 & 0.0170 & 0.0175 & 0.0135 & 0.0140 & 0.0220 \\ 
  & Sidak & 0.0025 & 0.0165 & 0.0175 & 0.0135 & 0.0150 & 0.0225 \\ 
  & Bonf & 0.0025 & 0.0165 & 0.0175 & 0.0135 & 0.0140 & 0.0220 \\
		\midrule
		\multirow{4}{2cm}{$m=10$\\$\pi_0 = 0.8$ }
& MBonf & 0.0200 & 0.0290 & 0.0320 & 0.0345 & 0.0365 & 0.0380 \\ 
  & Tarone & 0.0110 & 0.0145 & 0.0170 & 0.0220 & 0.0225 & 0.0250 \\ 
  & Sidak & 0.0065 & 0.0140 & 0.0170 & 0.0225 & 0.0245 & 0.0250 \\ 
  & Bonf & 0.0065 & 0.0140 & 0.0170 & 0.0220 & 0.0225 & 0.0250 \\ 
		\bottomrule		
		\toprule	
		\multirow{4}{2cm}{$m=15$\\$\pi_0 = 0.2$ }
 & MBonf & 0.0025 & 0.0075 & 0.0050 & 0.0085 & 0.0095 & 0.0130 \\ 
  & Tarone & 0.0010 & 0.0040 & 0.0010 & 0.0055 & 0.0065 & 0.0070 \\ 
  & Sidak & 0.0005 & 0.0040 & 0.0010 & 0.0055 & 0.0065 & 0.0070 \\ 
  & Bonf & 0.0005 & 0.0040 & 0.0010 & 0.0055 & 0.0065 & 0.0070 \\ 
		\midrule
		\multirow{4}{2cm}{$m=15$\\$\pi_0 = 0.4$ }
& MBonf & 0.0075 & 0.0120 & 0.0150 & 0.0220 & 0.0185 & 0.0175 \\ 
  & Tarone & 0.0035 & 0.0080 & 0.0050 & 0.0130 & 0.0120 & 0.0100 \\ 
  & Sidak & 0.0015 & 0.0060 & 0.0050 & 0.0130 & 0.0120 & 0.0100 \\ 
  & Bonf & 0.0015 & 0.0060 & 0.0050 & 0.0130 & 0.0120 & 0.0100 \\ 
		\midrule
		\multirow{4}{2cm}{$m=15$\\$\pi_0 = 0.6$ }
 & MBonf & 0.0105 & 0.0275 & 0.0255 & 0.0280 & 0.0285 & 0.0320 \\ 
  & Tarone & 0.0050 & 0.0125 & 0.0075 & 0.0120 & 0.0170 & 0.0215 \\ 
  & Sidak & 0.0015 & 0.0105 & 0.0075 & 0.0135 & 0.0170 & 0.0215 \\ 
  & Bonf & 0.0015 & 0.0105 & 0.0075 & 0.0120 & 0.0170 & 0.0215 \\ 
		\midrule
		\multirow{4}{2cm}{$m=15$\\$\pi_0 = 0.8$ }
& MBonf & 0.0240 & 0.0300 & 0.0260 & 0.0355 & 0.0355 & 0.0370 \\ 
  & Tarone & 0.0080 & 0.0190 & 0.0120 & 0.0175 & 0.0170 & 0.0205 \\ 
  & Sidak & 0.0025 & 0.0140 & 0.0120 & 0.0200 & 0.0170 & 0.0205 \\ 
  & Bonf & 0.0025 & 0.0140 & 0.0120 & 0.0175 & 0.0170 & 0.0205 \\ 
		\bottomrule
	\end{tabular}	
\end{table}

\renewcommand{\baselinestretch}{1}

%% file: simulation_table_Pow_SS_FET.tex
\renewcommand{\baselinestretch}{1}

\begin{table}[H]\small
	\renewcommand\arraystretch{0.5}
	\centering
	\caption{Simulated minimal power comparisons for single-step procedures with independent $p$-values generated from Fisher's exact test statistics, including Procedure 3.1 (MBonf), Procedure 2.2 (Tarone), and the conventional Sidak (Sidak) and Bonferroni (Bonf) procedures.} 
	
	\label{sim_Pow_SS_FET}
	\vspace{0.15 in}
	\begin{tabular}{clcccccc}
		\toprule
	&	& $N=25$ & $N=50$ & $N=75$ & $N=100$ & $N=125$ & $N=150$\\
		\midrule
		\multirow{4}{2cm}{$m=5$\\$\pi_0 = 0.2$ }
& MBonf & 0.2550 & 0.5060 & 0.6855 & 0.8195 & 0.9145 & 0.9505 \\ 
  & Tarone & 0.1945 & 0.3900 & 0.5775 & 0.7680 & 0.8655 & 0.9275 \\ 
  & Sidak & 0.1125 & 0.3825 & 0.5850 & 0.7680 & 0.8710 & 0.9340 \\ 
  & Bonf & 0.1125 & 0.3825 & 0.5765 & 0.7680 & 0.8655 & 0.9275 \\ 
  \midrule
  \multirow{4}{2cm}{$m=5$\\$\pi_0 = 0.4$ }
& MBonf & 0.2070 & 0.4295 & 0.5875 & 0.7310 & 0.8350 & 0.8975 \\ 
  & Tarone & 0.1635 & 0.3180 & 0.4750 & 0.6440 & 0.7835 & 0.8600 \\ 
  & Sidak & 0.0850 & 0.3065 & 0.4865 & 0.6440 & 0.7885 & 0.8675 \\ 
  & Bonf & 0.0850 & 0.3065 & 0.4735 & 0.6440 & 0.7835 & 0.8600 \\  
  \midrule
  \multirow{4}{2cm}{$m=5$\\$\pi_0 = 0.6$ }
& MBonf & 0.1480 & 0.3120 & 0.4485 & 0.5685 & 0.7110 & 0.7765 \\ 
  & Tarone & 0.1180 & 0.2400 & 0.3515 & 0.4870 & 0.6300 & 0.7250 \\ 
  & Sidak & 0.0580 & 0.2275 & 0.3615 & 0.4870 & 0.6370 & 0.7350 \\ 
  & Bonf & 0.0580 & 0.2275 & 0.3510 & 0.4870 & 0.6300 & 0.7250 \\ 
  \midrule
  \multirow{4}{2cm}{$m=5$\\$\pi_0 = 0.8$ }
& MBonf & 0.0810 & 0.1665 & 0.2650 & 0.3575 & 0.4510 & 0.5240 \\ 
  & Tarone & 0.0660 & 0.1235 & 0.2035 & 0.2965 & 0.3755 & 0.4665 \\ 
  & Sidak & 0.0305 & 0.1155 & 0.2080 & 0.2965 & 0.3800 & 0.4750 \\ 
  & Bonf & 0.0305 & 0.1155 & 0.2035 & 0.2965 & 0.3755 & 0.4665 \\ 
  \bottomrule  
		\toprule
		\multirow{4}{2cm}{$m=10$\\$\pi_0 = 0.2$ }
& MBonf & 0.3140 & 0.6070 & 0.8265 & 0.9260 & 0.9725 & 0.9950 \\ 
  & Tarone & 0.1980 & 0.4625 & 0.7405 & 0.8615 & 0.9400 & 0.9810 \\ 
  & Sidak & 0.1490 & 0.4605 & 0.7405 & 0.8665 & 0.9410 & 0.9830 \\ 
  & Bonf & 0.1490 & 0.4605 & 0.7405 & 0.8615 & 0.9400 & 0.9810 \\ 
		\midrule
		\multirow{4}{2cm}{$m=10$\\$\pi_0 = 0.4$ }
& MBonf & 0.2525 & 0.5180 & 0.7200 & 0.8510 & 0.9370 & 0.9595 \\ 
  & Tarone & 0.1760 & 0.3785 & 0.6130 & 0.7685 & 0.8970 & 0.9355 \\ 
  & Sidak & 0.1270 & 0.3700 & 0.6130 & 0.7760 & 0.8975 & 0.9390 \\ 
  & Bonf & 0.1270 & 0.3700 & 0.6130 & 0.7685 & 0.8970 & 0.9355 \\ 
		\midrule
		\multirow{4}{2cm}{$m=10$\\$\pi_0 = 0.6$ }
& MBonf & 0.1980 & 0.3825 & 0.5815 & 0.7060 & 0.8350 & 0.8990 \\ 
  & Tarone & 0.1235 & 0.2460 & 0.4700 & 0.6030 & 0.7705 & 0.8485 \\ 
  & Sidak & 0.0730 & 0.2390 & 0.4695 & 0.6110 & 0.7715 & 0.8585 \\ 
  & Bonf & 0.0730 & 0.2390 & 0.4695 & 0.6030 & 0.7705 & 0.8485 \\ 
		\midrule
		\multirow{4}{2cm}{$m=10$\\$\pi_0 = 0.8$ }
 & MBonf & 0.1165 & 0.2180 & 0.3520 & 0.4790 & 0.5895 & 0.6850 \\ 
  & Tarone & 0.0835 & 0.1410 & 0.2605 & 0.3780 & 0.4995 & 0.6105 \\ 
  & Sidak & 0.0435 & 0.1325 & 0.2605 & 0.3865 & 0.5000 & 0.6225 \\ 
  & Bonf & 0.0435 & 0.1325 & 0.2605 & 0.3780 & 0.4995 & 0.6105 \\ 
		\bottomrule  	
		\toprule
		\multirow{4}{2cm}{$m=15$\\$\pi_0 = 0.2$ }
& MBonf & 0.3475 & 0.6695 & 0.8570 & 0.9630 & 0.9920 & 0.9975 \\ 
  & Tarone & 0.2615 & 0.4980 & 0.7325 & 0.9025 & 0.9785 & 0.9925 \\ 
  & Sidak & 0.1400 & 0.4790 & 0.7320 & 0.9055 & 0.9785 & 0.9925 \\ 
  & Bonf & 0.1400 & 0.4790 & 0.7320 & 0.9020 & 0.9785 & 0.9925 \\  
		\midrule
		\multirow{4}{2cm}{$m=15$\\$\pi_0 = 0.4$ }
& MBonf & 0.2855 & 0.5660 & 0.7765 & 0.9020 & 0.9615 & 0.9890 \\ 
  & Tarone & 0.2155 & 0.4175 & 0.6310 & 0.8090 & 0.9265 & 0.9730 \\ 
  & Sidak & 0.0970 & 0.3865 & 0.6275 & 0.8150 & 0.9270 & 0.9730 \\ 
  & Bonf & 0.0970 & 0.3865 & 0.6275 & 0.8090 & 0.9265 & 0.9730 \\ 
		\midrule
		\multirow{4}{2cm}{$m=15$\\$\pi_0 = 0.6$ }
 & MBonf & 0.2105 & 0.4400 & 0.6540 & 0.7990 & 0.9055 & 0.9525 \\ 
  & Tarone & 0.1575 & 0.3125 & 0.4925 & 0.6845 & 0.8320 & 0.9160 \\ 
  & Sidak & 0.0785 & 0.2845 & 0.4885 & 0.6915 & 0.8350 & 0.9160 \\ 
  & Bonf & 0.0785 & 0.2845 & 0.4885 & 0.6845 & 0.8320 & 0.9160 \\ 
		\midrule
		\multirow{4}{2cm}{$m=15$\\$\pi_0 = 0.8$ }
& MBonf & 0.1215 & 0.2375 & 0.4110 & 0.5495 & 0.6680 & 0.7945 \\ 
  & Tarone & 0.0790 & 0.1555 & 0.2910 & 0.4265 & 0.5785 & 0.7180 \\ 
  & Sidak & 0.0300 & 0.1300 & 0.2885 & 0.4315 & 0.5780 & 0.7180 \\ 
  & Bonf & 0.0300 & 0.1300 & 0.2885 & 0.4260 & 0.5780 & 0.7180 \\ 
		\bottomrule  
	\end{tabular}
\end{table}

\renewcommand{\baselinestretch}{1}

%% file: simulation_table_FWER_SS_BT.tex
\begin{table}[H]\small
	\renewcommand\arraystretch{1}
	\centering
	\caption{Simulated FWER comparisons for single-step procedures with independent $p$-values generated from Binomial Exact Test statistics, including Procedure 3.1 (MBonf), Procedure 2.2 (Tarone), and the conventional Sidak (Sidak) and Bonferroni (Bonf) procedures.} 
	\label{sim_FWER_SS_BT}
	\vspace{0.15 in}
	\begin{tabular}{clcccc}
		\toprule
	&	 & $\pi_0=0.2$ & $\pi_0=0.4$ & $\pi_0=0.6$ & $\pi_0=0.8$  \\ 
		\midrule
		\multirow{4}{2cm}{$m=5$\\$\alpha=0.05$ }
& MBonf & 0.0020 & 0.0060 & 0.0075 & 0.0165 \\ 
& Tarone & 0.0010 & 0.0030 & 0.0055 & 0.0105 \\ 
& Sidak & 0.0010 & 0.0020 & 0.0025 & 0.0030 \\ 
& Bonf & 0.0010 & 0.0020 & 0.0025 & 0.0030 \\ 
		\midrule
		\multirow{4}{2cm}{$m=10$\\$\alpha=0.05$ }
& MBonf & 0.0010 & 0.0045 & 0.0130 & 0.0160 \\ 
& Tarone & 0.0000 & 0.0010 & 0.0050 & 0.0115 \\ 
& Sidak & 0.0000 & 0.0005 & 0.0025 & 0.0025 \\ 
& Bonf & 0.0000 & 0.0005 & 0.0025 & 0.0025 \\ 
		\midrule
		\multirow{4}{2cm}{$m=15$\\$\alpha=0.05$ }
& MBonf & 0.0010 & 0.0065 & 0.0045 & 0.0150 \\ 
& Tarone & 0.0000 & 0.0010 & 0.0020 & 0.0070 \\ 
& Sidak & 0.0000 & 0.0005 & 0.0000 & 0.0000 \\ 
& Bonf & 0.0000 & 0.0005 & 0.0000 & 0.0000 \\ 
	\bottomrule	
  \toprule
  \multirow{4}{2cm}{$m=5$\\$\alpha = 0.1$ }
& MBonf & 0.0070 & 0.0125 & 0.0200 & 0.0365 \\ 
& Tarone & 0.0020 & 0.0065 & 0.0110 & 0.0285 \\ 
& Sidak & 0.0020 & 0.0055 & 0.0065 & 0.0130 \\ 
& Bonf & 0.0020 & 0.0055 & 0.0065 & 0.0130 \\ 
		\midrule
		\multirow{4}{2cm}{$m=10$\\$\alpha = 0.1$ }
& MBonf & 0.0040 & 0.0080 & 0.0275 & 0.0350 \\ 
& Tarone & 0.0000 & 0.0030 & 0.0165 & 0.0195 \\ 
& Sidak & 0.0000 & 0.0015 & 0.0055 & 0.0060 \\ 
& Bonf & 0.0000 & 0.0015 & 0.0055 & 0.0060 \\ 
		\midrule
		\multirow{4}{2cm}{$m=15$\\$\alpha = 0.1$ }
& MBonf & 0.0060 & 0.0155 & 0.0185 & 0.0315 \\ 
& Tarone & 0.0005 & 0.0060 & 0.0045 & 0.0200 \\ 
& Sidak & 0.0000 & 0.0010 & 0.0020 & 0.0025 \\ 
& Bonf & 0.0000 & 0.0010 & 0.0020 & 0.0025 \\ 
		\bottomrule	
	\end{tabular}

\end{table}

\renewcommand{\baselinestretch}{1}

%% file: simulation_table_Pow_SS_BT.tex
\renewcommand{\baselinestretch}{1}
\begin{table}[H]\small
	\renewcommand\arraystretch{1}
	\centering
	\caption{Simulated minimal power comparisons for single-step procedures with independent $p$-values generated from Binomial Exact Test statistics, including Procedure 3.1 (MBonf), Procedure 2.2 (Tarone), and the conventional Sidak (Sidak) and Bonferroni (Bonf) procedures.} 	
	\label{sim_Pow_SS_BT}	
	\vspace{0.15 in}
	\begin{tabular}{clcccc}
		\toprule
&	 & $\pi_0=0.2$ & $\pi_0=0.4$ & $\pi_0=0.6$ & $\pi_0=0.8$  \\ 
		\midrule
		\multirow{4}{2cm}{$m=5$\\$\alpha = 0.05$ }
& MBonf & 0.9205 & 0.8805 & 0.7845 & 0.5565 \\ 
& Tarone & 0.8815 & 0.8240 & 0.7395 & 0.5235 \\ 
& Sidak & 0.8735 & 0.8055 & 0.6610 & 0.4045 \\ 
& Bonf & 0.8735 & 0.8055 & 0.6610 & 0.4045 \\ 
  \midrule
  \multirow{4}{2cm}{$m=10$\\$\alpha = 0.05$ }
& MBonf & 0.9850 & 0.9635 & 0.9035 & 0.7390 \\ 
& Tarone & 0.9470 & 0.9240 & 0.8630 & 0.6855 \\ 
& Sidak & 0.9315 & 0.8635 & 0.7050 & 0.4775 \\ 
& Bonf & 0.9315 & 0.8635 & 0.7050 & 0.4775 \\ 
  \midrule
  \multirow{4}{2cm}{$m=15$\\$\alpha = 0.05$ }
& MBonf & 0.9925 & 0.9810 & 0.9555 & 0.8210 \\ 
& Tarone & 0.9825 & 0.9500 & 0.9095 & 0.7845 \\ 
& Sidak & 0.9820 & 0.9475 & 0.8560 & 0.6135 \\ 
& Bonf & 0.9820 & 0.9475 & 0.8560 & 0.6135 \\ 
  \bottomrule
  \toprule
  \multirow{4}{2cm}{$m=5$\\$\alpha = 0.1$ }
& MBonf & 0.9680 & 0.9415 & 0.8615 & 0.6330 \\ 
& Tarone & 0.9410 & 0.9140 & 0.8240 & 0.5920 \\ 
& Sidak & 0.9050 & 0.8375 & 0.7040 & 0.4520 \\ 
& Bonf & 0.9050 & 0.8375 & 0.7040 & 0.4520 \\ 
  \midrule
  \multirow{4}{2cm}{$m=10$\\$\alpha = 0.1$ }
& MBonf & 0.9965 & 0.9875 & 0.9620 & 0.8315 \\ 
& Tarone & 0.9885 & 0.9660 & 0.9170 & 0.7835 \\ 
& Sidak & 0.9870 & 0.9565 & 0.8690 & 0.6600 \\ 
& Bonf & 0.9870 & 0.9565 & 0.8690 & 0.6600 \\ 
  \midrule
  \multirow{4}{2cm}{$m=15$\\$\alpha = 0.1$ }
& MBonf & 0.9995 & 0.9970 & 0.9830 & 0.9030 \\ 
& Tarone & 0.9960 & 0.9930 & 0.9605 & 0.8400 \\ 
& Sidak & 0.9880 & 0.9615 & 0.8830 & 0.6515 \\ 
& Bonf & 0.9895 & 0.9635 & 0.8880 & 0.6590 \\ 
  \bottomrule	
	\end{tabular}
\end{table}

\renewcommand{\baselinestretch}{1}

%% file: simulation_table_FWER_SD_FET.tex
\renewcommand{\baselinestretch}{1}

\begin{table}[H] \small
	\renewcommand\arraystretch{1}
	\centering
	
	 \caption{Simulated FWER comparisons for step-down procedures with independent $p$-values generated from Fisher's Exact Test statistics, including Procedure 3.2 (MHolm),  Procedure 2.3 (TH), and the conventional Holm procedure (Holm).} 
	\label{sim_FWER_SD_FET}
	\vspace{0.15 in}
	\begin{tabular}{clcccccc}
		\toprule
	&	& $N=25$ & $N=50$ & $N=75$ & $N=100$ & $N=125$ & $N=150$ \\ 
		\midrule
		\multirow{3}{2cm}{  $m=5$\\$\pi_0 = 0.2$ }
	    & MHolm & 0.0030 & 0.0090 & 0.0065 & 0.0115 & 0.0150 & 0.0150 \\
		& TH & 0.0015 & 0.0045 & 0.0030 & 0.0075 & 0.0090 & 0.0140 \\  
		& Holm & 0.0010 & 0.0045 & 0.0030 & 0.0075 & 0.0090 & 0.0140 \\ 
		\midrule
		\multirow{3}{2cm}{ $m=5$\\$\pi_0 = 0.4$ }
& MHolm & 0.0115 & 0.0150 & 0.0195 & 0.0215 & 0.0285 & 0.0300 \\ 
  & TH & 0.0080 & 0.0070 & 0.0110 & 0.0145 & 0.0190 & 0.0235 \\ 
  & Holm & 0.0035 & 0.0065 & 0.0110 & 0.0145 & 0.0190 & 0.0235 \\ 
		\midrule
		\multirow{3}{2cm}{ $m=5$\\$\pi_0 = 0.6$ }
& MHolm & 0.0110 & 0.0200 & 0.0220 & 0.0305 & 0.0315 & 0.0340 \\ 
  & TH & 0.0060 & 0.0095 & 0.0110 & 0.0195 & 0.0195 & 0.0250 \\ 
  & Holm & 0.0035 & 0.0090 & 0.0110 & 0.0195 & 0.0195 & 0.0250 \\ 
  \midrule
  \multirow{3}{2cm}{ $m=5$\\$\pi_0 = 0.8$ }
& MHolm & 0.0200 & 0.0300 & 0.0285 & 0.0335 & 0.0385 & 0.0405 \\ 
  & TH & 0.0135 & 0.0135 & 0.0175 & 0.0250 & 0.0270 & 0.0295 \\ 
  & Holm & 0.0030 & 0.0125 & 0.0175 & 0.0250 & 0.0270 & 0.0295 \\ 
  \bottomrule
		\toprule
		\multirow{3}{2cm}{ $m=10$\\$\pi_0 = 0.2$ }
& MHolm & 0.0025 & 0.0095 & 0.0080 & 0.0130 & 0.0130 & 0.0180 \\ 
  & TH & 0.0010 & 0.0050 & 0.0045 & 0.0065 & 0.0075 & 0.0110 \\ 
  & Holm & 0.0010 & 0.0045 & 0.0045 & 0.0065 & 0.0075 & 0.0110 \\ 
		\midrule
		\multirow{3}{2cm}{ $m=10$\\$\pi_0 = 0.4$ }
 & MHolm & 0.0065 & 0.0160 & 0.0200 & 0.0185 & 0.0250 & 0.0220 \\ 
  & TH & 0.0030 & 0.0065 & 0.0115 & 0.0100 & 0.0120 & 0.0150 \\ 
  & Holm & 0.0025 & 0.0060 & 0.0115 & 0.0100 & 0.0120 & 0.0150 \\ 
		\midrule
		\multirow{3}{2cm}{ $m=10$\\$\pi_0 = 0.6$ }
& MHolm & 0.0135 & 0.0320 & 0.0275 & 0.0250 & 0.0285 & 0.0410 \\ 
  & TH & 0.0045 & 0.0170 & 0.0175 & 0.0135 & 0.0185 & 0.0285 \\ 
  & Holm & 0.0025 & 0.0165 & 0.0175 & 0.0135 & 0.0185 & 0.0285 \\ 
		\midrule
		\multirow{3}{2cm}{ $m=10$\\$\pi_0 = 0.8$ }
& MHolm & 0.0200 & 0.0290 & 0.0330 & 0.0385 & 0.0385 & 0.0410 \\ 
  & TH & 0.0115 & 0.0145 & 0.0170 & 0.0225 & 0.0230 & 0.0280 \\ 
  & Holm & 0.0065 & 0.0140 & 0.0170 & 0.0225 & 0.0230 & 0.0280 \\ 
		\bottomrule		
		\toprule	
		\multirow{3}{2cm}{  $m=15$\\$\pi_0 = 0.2$ }
 & MHolm & 0.0025 & 0.0085 & 0.0060 & 0.0125 & 0.0160 & 0.0195 \\ 
  & TH & 0.0010 & 0.0040 & 0.0020 & 0.0055 & 0.0090 & 0.0115 \\ 
  & Holm & 0.0005 & 0.0040 & 0.0020 & 0.0055 & 0.0090 & 0.0115 \\ 
		\midrule
		\multirow{3}{2cm}{ $m=15$\\$\pi_0 = 0.4$ }
& MHolm & 0.0075 & 0.0135 & 0.0175 & 0.0265 & 0.0245 & 0.0230 \\ 
  & TH & 0.0045 & 0.0080 & 0.0065 & 0.0145 & 0.0140 & 0.0145 \\ 
  & Holm & 0.0015 & 0.0065 & 0.0065 & 0.0145 & 0.0140 & 0.0145 \\ 
		\midrule
		\multirow{3}{2cm}{ $m=15$\\$\pi_0 = 0.6$ }
& MHolm & 0.0105 & 0.0290 & 0.0280 & 0.0315 & 0.0355 & 0.0395 \\ 
  & TH & 0.0050 & 0.0135 & 0.0080 & 0.0140 & 0.0180 & 0.0250 \\ 
  & Holm & 0.0015 & 0.0105 & 0.0080 & 0.0140 & 0.0180 & 0.0250 \\ 
		\midrule
		\multirow{3}{2cm}{ $m=15$\\$\pi_0 = 0.8$ }
& MHolm & 0.0250 & 0.0310 & 0.0275 & 0.0385 & 0.0380 & 0.0400 \\ 
  & TH & 0.0080 & 0.0190 & 0.0120 & 0.0185 & 0.0170 & 0.0230 \\ 
  & Holm & 0.0025 & 0.0140 & 0.0120 & 0.0185 & 0.0170 & 0.0230 \\ 
		\bottomrule
	\end{tabular}	
\end{table}

\renewcommand{\baselinestretch}{1}

%% file: simulation_table_Pow_SD_FET.tex
\renewcommand{\baselinestretch}{1}
\begin{table}[H]\small
	\renewcommand\arraystretch{1}
	\centering
	\caption{Simulated minimal power comparisons for step-down procedures with independent $p$-values generated from Fisher's Exact Test statistics, including Procedure 3.2 (MHolm),  Procedure 2.3 (TH), and the conventional Holm procedure (Holm).} 
	\label{sim_Pow_SD_FET}
	\vspace{0.15 in}
	\begin{tabular}{clcccccc}
		\toprule
	&	& $N=25$ & $N=50$ & $N=75$ & $N=100$ & $N=125$ & $N=150$\\
		\midrule
		\multirow{3}{2cm}{$m=5$\\$\pi_0 = 0.2$ }
& MHolm & 0.2555 & 0.5070 & 0.6855 & 0.8200 & 0.9145 & 0.9505 \\ 
  & TH & 0.1945 & 0.3905 & 0.5780 & 0.7680 & 0.8660 & 0.9280 \\ 
  & Holm & 0.1130 & 0.3830 & 0.5770 & 0.7680 & 0.8660 & 0.9280 \\ 
  \midrule
  \multirow{3}{2cm}{$m=5$\\$\pi_0 = 0.4$ }
& MHolm & 0.2070 & 0.4300 & 0.5875 & 0.7310 & 0.8350 & 0.8975 \\ 
  & TH & 0.1635 & 0.3180 & 0.4760 & 0.6440 & 0.7835 & 0.8605 \\ 
  & Holm & 0.0850 & 0.3065 & 0.4745 & 0.6440 & 0.7835 & 0.8605 \\ 
  \midrule
  \multirow{3}{2cm}{$m=5$\\$\pi_0 = 0.6$ }
& MHolm & 0.1480 & 0.3125 & 0.4490 & 0.5685 & 0.7110 & 0.7780 \\ 
  & TH & 0.1180 & 0.2410 & 0.3530 & 0.4880 & 0.6325 & 0.7255 \\ 
  & Holm & 0.0580 & 0.2285 & 0.3525 & 0.4880 & 0.6325 & 0.7255 \\ 
  \midrule
  \multirow{3}{2cm}{$m=5$\\$\pi_0 = 0.8$ }
& MHolm & 0.0815 & 0.1680 & 0.2665 & 0.3585 & 0.4510 & 0.5255 \\ 
  & TH & 0.0660 & 0.1240 & 0.2035 & 0.2965 & 0.3785 & 0.4685 \\ 
  & Holm & 0.0305 & 0.1155 & 0.2035 & 0.2965 & 0.3785 & 0.4685 \\ 
  \bottomrule  
		\toprule
		\multirow{3}{2cm}{$m=10$\\$\pi_0 = 0.2$ }
& MHolm & 0.3140 & 0.6070 & 0.8265 & 0.9260 & 0.9725 & 0.9950 \\ 
  & TH & 0.1980 & 0.4625 & 0.7405 & 0.8615 & 0.9400 & 0.9815 \\ 
  & Holm & 0.1490 & 0.4605 & 0.7405 & 0.8615 & 0.9400 & 0.9815 \\ 
		\midrule
		\multirow{3}{2cm}{$m=10$\\$\pi_0 = 0.4$ }
& MHolm & 0.2540 & 0.5190 & 0.7205 & 0.8510 & 0.9370 & 0.9595 \\ 
  & TH & 0.1760 & 0.3785 & 0.6130 & 0.7685 & 0.8970 & 0.9355 \\ 
  & Holm & 0.1275 & 0.3700 & 0.6130 & 0.7685 & 0.8970 & 0.9355 \\ 
		\midrule
		\multirow{3}{2cm}{$m=10$\\$\pi_0 = 0.6$ }
& MHolm & 0.1985 & 0.3835 & 0.5815 & 0.7065 & 0.8350 & 0.8990 \\ 
  & TH & 0.1235 & 0.2460 & 0.4700 & 0.6030 & 0.7710 & 0.8495 \\ 
  & Holm & 0.0730 & 0.2390 & 0.4695 & 0.6030 & 0.7710 & 0.8495 \\ 
		\midrule
		\multirow{3}{2cm}{$m=10$\\$\pi_0 = 0.8$ }
& MHolm & 0.1165 & 0.2185 & 0.3530 & 0.4795 & 0.5910 & 0.6850 \\ 
  & TH & 0.0835 & 0.1410 & 0.2605 & 0.3780 & 0.4995 & 0.6105 \\ 
  & Holm & 0.0435 & 0.1325 & 0.2605 & 0.3780 & 0.4995 & 0.6105 \\ 
		\bottomrule  	
		\toprule
		\multirow{3}{2cm}{$m=15$\\$\pi_0 = 0.2$ }
& MHolm & 0.3475 & 0.6695 & 0.8570 & 0.9630 & 0.9920 & 0.9975 \\ 
  & TH & 0.2615 & 0.4980 & 0.7325 & 0.9025 & 0.9785 & 0.9925 \\ 
  & Holm & 0.1400 & 0.4790 & 0.7320 & 0.9020 & 0.9785 & 0.9925 \\ 
		\midrule
		\multirow{3}{2cm}{$m=15$\\$\pi_0 = 0.4$ } 
& MHolm & 0.2855 & 0.5660 & 0.7765 & 0.9020 & 0.9615 & 0.9890 \\ 
  & TH & 0.2155 & 0.4175 & 0.6310 & 0.8090 & 0.9265 & 0.9735 \\ 
  & Holm & 0.0970 & 0.3865 & 0.6275 & 0.8090 & 0.9265 & 0.9735 \\ 
		\midrule
		\multirow{3}{2cm}{$m=15$\\$\pi_0 = 0.6$ }
& MHolm & 0.2105 & 0.4400 & 0.6540 & 0.7990 & 0.9060 & 0.9540 \\ 
  & TH & 0.1575 & 0.3130 & 0.4925 & 0.6845 & 0.8325 & 0.9160 \\ 
  & Holm & 0.0785 & 0.2845 & 0.4885 & 0.6845 & 0.8325 & 0.9160 \\ 
		\midrule
		\multirow{3}{2cm}{$m=15$\\$\pi_0 = 0.8$ }
& MHolm & 0.1220 & 0.2380 & 0.4110 & 0.5500 & 0.6690 & 0.7950 \\ 
  & TH & 0.0790 & 0.1555 & 0.2910 & 0.4275 & 0.5785 & 0.7185 \\ 
  & Holm & 0.0300 & 0.1300 & 0.2885 & 0.4270 & 0.5780 & 0.7185 \\ 
		\bottomrule  
	\end{tabular}
\end{table}

\renewcommand{\baselinestretch}{1}

%% file: simulation_table_FWER_SU_FET.tex
\renewcommand{\baselinestretch}{1}
\begin{table}[H]\small
	\renewcommand\arraystretch{1}
	\centering
	\caption{Simulated FWER comparisons for step-up procedures with independent $p$-values generated from Fisher's Exact Test statistics, including Procedure 3.3 (MHoch), the Roth procedure (Roth), and the conventional Hochberg procedure (Hoch).} 
	
	\label{sim_FWER_SU_FET}	
	\vspace{0.15 in}
	\begin{tabular}{rlcccccc}
		\toprule
	&	& $N=25$ & $N=50$ & $N=75$ & $N=100$ & $N=125$ & $N=150$ \\ 
		\midrule
		\multirow{3}{2cm}{$m=5$\\$\pi_0 = 0.2$}
& MHoch & 0.0030 & 0.0090 & 0.0070 & 0.0115 & 0.0150 & 0.0155 \\ 
  & Roth & 0.0020 & 0.0045 & 0.0040 & 0.0085 & 0.0115 & 0.0155 \\ 
  & Hoch & 0.0015 & 0.0045 & 0.0040 & 0.0085 & 0.0115 & 0.0155 \\ 
		\midrule
		\multirow{3}{2cm}{$m=5$\\$\pi_0 = 0.4$}
& MHoch & 0.0115 & 0.0150 & 0.0200 & 0.0215 & 0.0290 & 0.0325 \\ 
  & Roth & 0.0080 & 0.0070 & 0.0120 & 0.0145 & 0.0200 & 0.0245 \\ 
  & Hoch & 0.0040 & 0.0065 & 0.0120 & 0.0145 & 0.0205 & 0.0245 \\ 
		\midrule
		\multirow{3}{2cm}{$m=5$\\$\pi_0 = 0.6$}
& MHoch & 0.0120 & 0.0210 & 0.0225 & 0.0310 & 0.0325 & 0.0340 \\ 
  & Roth & 0.0055 & 0.0095 & 0.0110 & 0.0200 & 0.0190 & 0.0260 \\ 
  & Hoch & 0.0040 & 0.0090 & 0.0110 & 0.0200 & 0.0200 & 0.0260 \\  

  \midrule
  \multirow{3}{2cm}{$m=5$\\$\pi_0 = 0.8$}
& MHoch & 0.0210 & 0.0300 & 0.0290 & 0.0340 & 0.0385 & 0.0405 \\ 
  & Roth & 0.0120 & 0.0135 & 0.0175 & 0.0250 & 0.0270 & 0.0295 \\ 
  & Hoch & 0.0030 & 0.0125 & 0.0175 & 0.0250 & 0.0270 & 0.0295 \\
  \bottomrule
		\toprule
		\multirow{3}{2cm}{$m=10$\\$\pi_0 = 0.2$}
& MHoch & 0.0035 & 0.0100 & 0.0085 & 0.0130 & 0.0135 & 0.0205 \\ 
  & Roth & 0.0010 & 0.0050 & 0.0045 & 0.0070 & 0.0070 & 0.0120 \\ 
  & Hoch & 0.0010 & 0.0045 & 0.0045 & 0.0070 & 0.0080 & 0.0120 \\ 
		\midrule
		\multirow{3}{2cm}{$m=10$\\$\pi_0 = 0.4$}
& MHoch & 0.0070 & 0.0175 & 0.0210 & 0.0195 & 0.0250 & 0.0220 \\ 
  & Roth & 0.0030 & 0.0065 & 0.0115 & 0.0100 & 0.0115 & 0.0155 \\ 
  & Hoch & 0.0025 & 0.0060 & 0.0115 & 0.0100 & 0.0120 & 0.0155 \\ 
		\midrule
		\multirow{3}{2cm}{$m=10$\\$\pi_0 = 0.6$ }
& MHoch & 0.0135 & 0.0320 & 0.0275 & 0.0250 & 0.0285 & 0.0410 \\ 
  & Roth & 0.0045 & 0.0165 & 0.0175 & 0.0140 & 0.0185 & 0.0285 \\ 
  & Hoch & 0.0025 & 0.0165 & 0.0175 & 0.0140 & 0.0185 & 0.0285 \\ 
		\midrule
		\multirow{3}{2cm}{$m=10$ \\ $\pi_0 = 0.8$ }
& MHoch & 0.0205 & 0.0290 & 0.0330 & 0.0390 & 0.0390 & 0.0415 \\ 
  & Roth & 0.0110 & 0.0145 & 0.0170 & 0.0225 & 0.0230 & 0.0280 \\ 
  & Hoch & 0.0065 & 0.0140 & 0.0170 & 0.0225 & 0.0230 & 0.0280 \\ 
		\bottomrule		
		\toprule	
		\multirow{3}{2cm}{$m=15$\\$\pi_0 = 0.2$ }
 & MHoch & 0.0025 & 0.0085 & 0.0065 & 0.0125 & 0.0160 & 0.0205 \\ 
  & Roth & 0.0010 & 0.0040 & 0.0020 & 0.0055 & 0.0090 & 0.0130 \\ 
  & Hoch & 0.0005 & 0.0040 & 0.0020 & 0.0055 & 0.0090 & 0.0130 \\ 
		\midrule
		\multirow{3}{2cm}{$m=15$\\$\pi_0 = 0.4$ }
& MHoch & 0.0075 & 0.0135 & 0.0175 & 0.0270 & 0.0245 & 0.0240 \\ 
  & Roth & 0.0040 & 0.0080 & 0.0065 & 0.0145 & 0.0140 & 0.0145 \\ 
  & Hoch & 0.0015 & 0.0070 & 0.0065 & 0.0145 & 0.0140 & 0.0145 \\
		\midrule
		\multirow{3}{2cm}{$m=15$\\$\pi_0 = 0.6$ }
& MHoch & 0.0120 & 0.0290 & 0.0280 & 0.0320 & 0.0355 & 0.0395 \\ 
  & Roth & 0.0050 & 0.0135 & 0.0080 & 0.0140 & 0.0180 & 0.0250 \\ 
  & Hoch & 0.0015 & 0.0105 & 0.0080 & 0.0140 & 0.0180 & 0.0250 \\ 
		\midrule
		\multirow{3}{2cm}{$m=15$\\$\pi_0 = 0.8$ }
& MHoch & 0.0255 & 0.0310 & 0.0280 & 0.0385 & 0.0385 & 0.0400 \\ 
  & Roth & 0.0080 & 0.0190 & 0.0120 & 0.0185 & 0.0175 & 0.0230 \\ 
  & Hoch & 0.0025 & 0.0140 & 0.0120 & 0.0185 & 0.0175 & 0.0230 \\ 
		\bottomrule
	\end{tabular}

\end{table}

\renewcommand{\baselinestretch}{1}

%% file: simulation_table_Pow_SU_FET.tex
\renewcommand{\baselinestretch}{1}

\begin{table}[H]\small
	\renewcommand\arraystretch{1}
	\centering
	\caption{Simulated minimal power comparisons for step-up procedures with independent $p$-values generated from Fisher's Exact Test statistics, including Procedure 3.1 (MHoch), the Roth procedure (Roth), and the conventional Hochberg procedure (Hoch).} 
	
	\label{sim_Pow_SU_FET}	
	\vspace{0.15 in}
	\begin{tabular}{clcccccc}
		\toprule
	&	& $N=25$ & $N=50$ & $N=75$ & $N=100$ & $N=125$ & $N=150$\\
		\midrule
		\multirow{3}{2cm}{$m=5$\\$\pi_0 = 0.2$ }
& MHoch & 0.2600 & 0.5075 & 0.6885 & 0.8240 & 0.9170 & 0.9525 \\ 
  & Roth & 0.1900 & 0.3915 & 0.5820 & 0.7685 & 0.8695 & 0.9300 \\ 
  & Hoch & 0.1170 & 0.3845 & 0.5810 & 0.7685 & 0.8695 & 0.9300 \\ 
  \midrule
  \multirow{3}{2cm}{$m=5$\\$\pi_0 = 0.4$ }
& MHoch & 0.2070 & 0.4310 & 0.5905 & 0.7335 & 0.8370 & 0.9015 \\ 
  & Roth & 0.1570 & 0.3195 & 0.4785 & 0.6475 & 0.7860 & 0.8625 \\ 
  & Hoch & 0.0870 & 0.3085 & 0.4770 & 0.6475 & 0.7860 & 0.8625 \\ 
  \midrule
  \multirow{3}{2cm}{$m=5$\\$\pi_0 = 0.6$ }
& MHoch & 0.1495 & 0.3140 & 0.4495 & 0.5705 & 0.7125 & 0.7785 \\ 
  & Roth & 0.1100 & 0.2420 & 0.3530 & 0.4895 & 0.6350 & 0.7280 \\ 
  & Hoch & 0.0585 & 0.2295 & 0.3525 & 0.4895 & 0.6350 & 0.7280 \\ 
  \midrule
  \multirow{3}{2cm}{$m=5$\\$\pi_0 = 0.8$ }
& MHoch & 0.0825 & 0.1680 & 0.2670 & 0.3590 & 0.4510 & 0.5255 \\ 
  & Roth & 0.0580 & 0.1240 & 0.2035 & 0.2965 & 0.3785 & 0.4685 \\ 
  & Hoch & 0.0305 & 0.1155 & 0.2035 & 0.2965 & 0.3785 & 0.4685 \\ 
  \bottomrule  
		\toprule
		\multirow{3}{2cm}{$m=10$\\$\pi_0 = 0.2$ }
& MHoch & 0.3195 & 0.6115 & 0.8280 & 0.9270 & 0.9730 & 0.9955 \\ 
  & Roth & 0.1995 & 0.4620 & 0.7405 & 0.8615 & 0.9410 & 0.9820 \\ 
  & Hoch & 0.1495 & 0.4605 & 0.7405 & 0.8615 & 0.9410 & 0.9820 \\ 
		\midrule
		\multirow{3}{2cm}{$m=10$\\$\pi_0 = 0.4$ }
& MHoch & 0.2540 & 0.5210 & 0.7225 & 0.8520 & 0.9375 & 0.9600 \\ 
  & Roth & 0.1765 & 0.3750 & 0.6135 & 0.7690 & 0.8975 & 0.9365 \\ 
  & Hoch & 0.1275 & 0.3700 & 0.6130 & 0.7690 & 0.8975 & 0.9365 \\ 
		\midrule
		\multirow{3}{2cm}{$m=10$\\$\pi_0 = 0.6$ }
& MHoch & 0.1990 & 0.3845 & 0.5830 & 0.7070 & 0.8370 & 0.8995 \\ 
  & Roth & 0.1230 & 0.2425 & 0.4700 & 0.6030 & 0.7715 & 0.8510 \\ 
  & Hoch & 0.0730 & 0.2395 & 0.4695 & 0.6030 & 0.7715 & 0.8510 \\ 
		\midrule
		\multirow{3}{2cm}{$m=10$\\$\pi_0 = 0.8$ }
& MHoch & 0.1170 & 0.2185 & 0.3535 & 0.4800 & 0.5925 & 0.6865 \\ 
  & Roth & 0.0825 & 0.1370 & 0.2605 & 0.3780 & 0.4995 & 0.6105 \\ 
  & Hoch & 0.0435 & 0.1325 & 0.2605 & 0.3780 & 0.4995 & 0.6105 \\ 
		\bottomrule  	
		\toprule
		\multirow{3}{2cm}{$m=15$\\$\pi_0 = 0.2$ }
& MHoch & 0.3505 & 0.6700 & 0.8575 & 0.9640 & 0.9920 & 0.9980 \\ 
  & Roth & 0.2615 & 0.5000 & 0.7330 & 0.9030 & 0.9785 & 0.9930 \\ 
  & Hoch & 0.1400 & 0.4795 & 0.7325 & 0.9025 & 0.9785 & 0.9930 \\ 
		\midrule
		\multirow{3}{2cm}{$m=15$\\$\pi_0 = 0.4$ } 
& MHoch & 0.2875 & 0.5675 & 0.7785 & 0.9030 & 0.9615 & 0.9890 \\ 
  & Roth & 0.2160 & 0.4195 & 0.6320 & 0.8095 & 0.9265 & 0.9740 \\ 
  & Hoch & 0.0970 & 0.3875 & 0.6285 & 0.8095 & 0.9265 & 0.9740 \\ 
		\midrule
		\multirow{3}{2cm}{$m=15$\\$\pi_0 = 0.6$ }
& MHoch & 0.2135 & 0.4400 & 0.6555 & 0.8005 & 0.9065 & 0.9545 \\ 
  & Roth & 0.1580 & 0.3130 & 0.4925 & 0.6850 & 0.8325 & 0.9165 \\ 
  & Hoch & 0.0785 & 0.2845 & 0.4890 & 0.6850 & 0.8325 & 0.9165 \\ 
		\midrule
		\multirow{3}{2cm}{$m=15$\\$\pi_0 = 0.8$ }
& MHoch & 0.1225 & 0.2380 & 0.4110 & 0.5520 & 0.6690 & 0.7950 \\ 
  & Roth & 0.0790 & 0.1550 & 0.2910 & 0.4270 & 0.5780 & 0.7195 \\ 
  & Hoch & 0.0300 & 0.1300 & 0.2885 & 0.4270 & 0.5780 & 0.7195 \\ 
		\bottomrule  
	\end{tabular}
\end{table}

\renewcommand{\baselinestretch}{1.5}

%% file: simulation_table_FWER_SS_BT_dept.tex
\begin{table}[H]\small
	\renewcommand\arraystretch{0.6}
	\centering
	\caption{Simulated FWER comparisons for single-step procedures with dependent $p$-values generated from Binomial Exact Test statistics, including Procedure 3.1 (MBonf), Procedure 2.1 (Tarone), and the conventional Sidak (Sidak) and Bonferroni (Bonf) procedures.} 
	\label{sim_FWER_SS_BT_dept}
	\vspace{0.15 in}
	\begin{tabular}{clcccccccccc}
		\toprule
	& $\rho$	  & 0 & 0.1 & 0.2 & 0.3 & 0.4 & 0.5 & 0.6 & 0.7 & 0.8 & 0.9 \\ 
		\midrule
		\multirow{4}{2cm}{$m=5$\\$\pi_0=0.4$ }
& MBonf & 0.0025 & 0.0040 & 0.0055 & 0.0035 & 0.0055 & 0.0030 & 0.0035 & 0.0030 & 0.0035 & 0.0020 \\ 
  & Tarone & 0.0015 & 0.0010 & 0.0030 & 0.0010 & 0.0005 & 0.0005 & 0.0015 & 0.0025 & 0.0020 & 0.0015 \\ 
  & Sidak & 0.0015 & 0.0005 & 0.0025 & 0.0010 & 0.0005 & 0.0005 & 0.0015 & 0.0020 & 0.0015 & 0.0010 \\ 
  & Bonf & 0.0015 & 0.0005 & 0.0025 & 0.0010 & 0.0005 & 0.0005 & 0.0015 & 0.0020 & 0.0015 & 0.0010 \\ 
		\midrule
		\multirow{4}{2cm}{$m=5$\\$\pi_0=0.6$ }
& MBonf & 0.0090 & 0.0070 & 0.0115 & 0.0055 & 0.0085 & 0.0090 & 0.0060 & 0.0050 & 0.0060 & 0.0045 \\ 
  & Tarone & 0.0055 & 0.0055 & 0.0075 & 0.0020 & 0.0045 & 0.0050 & 0.0020 & 0.0025 & 0.0040 & 0.0035 \\ 
  & Sidak & 0.0030 & 0.0025 & 0.0030 & 0.0015 & 0.0025 & 0.0015 & 0.0010 & 0.0010 & 0.0005 & 0.0015 \\ 
  & Bonf & 0.0030 & 0.0025 & 0.0030 & 0.0015 & 0.0025 & 0.0015 & 0.0010 & 0.0010 & 0.0005 & 0.0015 \\ 
		\midrule
		\multirow{4}{2cm}{$m=5$\\$\pi_0=0.8$ }
& MBonf & 0.0160 & 0.0155 & 0.0185 & 0.0115 & 0.0105 & 0.0120 & 0.0155 & 0.0140 & 0.0115 & 0.0055 \\ 
  & Tarone & 0.0105 & 0.0075 & 0.0105 & 0.0065 & 0.0065 & 0.0075 & 0.0095 & 0.0070 & 0.0070 & 0.0040 \\ 
  & Sidak & 0.0020 & 0.0015 & 0.0040 & 0.0010 & 0.0010 & 0.0010 & 0.0045 & 0.0035 & 0.0030 & 0.0015 \\ 
  & Bonf & 0.0020 & 0.0015 & 0.0040 & 0.0010 & 0.0010 & 0.0010 & 0.0045 & 0.0035 & 0.0030 & 0.0015 \\ 
		\bottomrule	
  \toprule
  \multirow{4}{2cm}{$m=10$\\$\pi_0=0.4$}
& MBonf & 0.0010 & 0.0030 & 0.0040 & 0.0035 & 0.0040 & 0.0030 & 0.0050 & 0.0020 & 0.0050 & 0.0010 \\ 
  & Tarone & 0.0000 & 0.0010 & 0.0015 & 0.0000 & 0.0010 & 0.0020 & 0.0020 & 0.0005 & 0.0005 & 0.0000 \\ 
  & Sidak & 0.0000 & 0.0010 & 0.0015 & 0.0000 & 0.0005 & 0.0010 & 0.0015 & 0.0005 & 0.0000 & 0.0000 \\ 
  & Bonf & 0.0000 & 0.0010 & 0.0015 & 0.0000 & 0.0005 & 0.0010 & 0.0015 & 0.0005 & 0.0000 & 0.0000 \\ 
		\midrule
		\multirow{4}{2cm}{$m=10$\\$\pi_0=0.6$ }
& MBonf & 0.0065 & 0.0085 & 0.0065 & 0.0095 & 0.0045 & 0.0040 & 0.0085 & 0.0045 & 0.0060 & 0.0035 \\ 
  & Tarone & 0.0040 & 0.0035 & 0.0015 & 0.0055 & 0.0025 & 0.0015 & 0.0045 & 0.0030 & 0.0020 & 0.0020 \\ 
  & Sidak & 0.0005 & 0.0005 & 0.0000 & 0.0010 & 0.0005 & 0.0000 & 0.0005 & 0.0010 & 0.0005 & 0.0005 \\ 
  & Bonf & 0.0005 & 0.0005 & 0.0000 & 0.0010 & 0.0005 & 0.0000 & 0.0005 & 0.0010 & 0.0005 & 0.0005 \\ 
		\midrule
		\multirow{4}{2cm}{$m=10$\\$\pi_0=0.8$}
& MBonf & 0.0165 & 0.0090 & 0.0120 & 0.0140 & 0.0130 & 0.0085 & 0.0120 & 0.0125 & 0.0055 & 0.0025 \\ 
  & Tarone & 0.0100 & 0.0055 & 0.0060 & 0.0085 & 0.0100 & 0.0035 & 0.0090 & 0.0085 & 0.0030 & 0.0020 \\ 
  & Sidak & 0.0015 & 0.0005 & 0.0005 & 0.0010 & 0.0020 & 0.0000 & 0.0005 & 0.0005 & 0.0000 & 0.0005 \\ 
  & Bonf & 0.0015 & 0.0005 & 0.0005 & 0.0010 & 0.0020 & 0.0000 & 0.0005 & 0.0005 & 0.0000 & 0.0005 \\ 
		\bottomrule	
	\end{tabular}

\end{table}

\renewcommand{\baselinestretch}{1.65}

%% file: simulation_table_Pow_SS_BT_dept.tex
\begin{table}[H]\small
	\renewcommand\arraystretch{0.6}
	\centering
	\caption{Simulated minimal power comparisons for single-step procedures with dependent $p$-values generated from Binomial Exact Test statistics, including Procedure 3.1 (MBonf), Procedure 2.1 (Tarone), and the conventional Sidak (Sidak) and Bonferroni (Bonf) procedures.} 
	
	\label{sim_Pow_SS_BT_dept}
	\vspace{0.15 in}
	\begin{tabular}{clcccccccccc}
	\toprule
	& $\rho$	  & 0 & 0.1 & 0.2 & 0.3 & 0.4 & 0.5 & 0.6 & 0.7 & 0.8 & 0.9 \\ 
	\midrule
	\multirow{4}{2cm}{$m=5$\\$\pi_0=0.4$ }
& MBonf & 0.9200 & 0.8710 & 0.8575 & 0.8445 & 0.8275 & 0.7690 & 0.7395 & 0.7375 & 0.6935 & 0.6425 \\ 
  & Tarone & 0.8755 & 0.8150 & 0.8000 & 0.7840 & 0.7610 & 0.7040 & 0.6755 & 0.6745 & 0.6235 & 0.5680 \\ 
  & Sidak & 0.7880 & 0.7250 & 0.7060 & 0.6915 & 0.6680 & 0.6130 & 0.5835 & 0.5730 & 0.5320 & 0.4760 \\ 
  & Bonf & 0.7880 & 0.7250 & 0.7060 & 0.6915 & 0.6680 & 0.6130 & 0.5835 & 0.5730 & 0.5320 & 0.4760 \\ 
	\midrule
	\multirow{4}{2cm}{$m=5$\\$\pi_0=0.6$ }
& MBonf & 0.8085 & 0.8065 & 0.7870 & 0.7650 & 0.7445 & 0.7255 & 0.7205 & 0.6965 & 0.6610 & 0.6195 \\ 
  & Tarone & 0.7610 & 0.7545 & 0.7355 & 0.7175 & 0.7030 & 0.6790 & 0.6675 & 0.6515 & 0.6185 & 0.5790 \\ 
  & Sidak & 0.5985 & 0.6060 & 0.5875 & 0.5710 & 0.5475 & 0.5395 & 0.5070 & 0.5095 & 0.4680 & 0.4200 \\ 
  & Bonf & 0.5985 & 0.6060 & 0.5875 & 0.5710 & 0.5475 & 0.5395 & 0.5070 & 0.5095 & 0.4680 & 0.4200 \\ 
	\midrule
	\multirow{4}{2cm}{$m=5$\\$\pi_0=0.8$ }
& MBonf & 0.6030 & 0.6085 & 0.5920 & 0.6175 & 0.5875 & 0.5865 & 0.6330 & 0.6295 & 0.6110 & 0.6110 \\ 
  & Tarone & 0.5635 & 0.5730 & 0.5615 & 0.5875 & 0.5610 & 0.5680 & 0.6105 & 0.6115 & 0.5905 & 0.6000 \\ 
  & Sidak & 0.3750 & 0.3910 & 0.3595 & 0.3865 & 0.3680 & 0.3420 & 0.3890 & 0.3860 & 0.3940 & 0.3750 \\ 
  & Bonf & 0.3750 & 0.3910 & 0.3595 & 0.3865 & 0.3680 & 0.3420 & 0.3890 & 0.3860 & 0.3940 & 0.3750 \\
	\bottomrule	
	\toprule
	\multirow{4}{2cm}{$m=10$\\$\pi_0=0.4$}
& MBonf & 0.9755 & 0.9465 & 0.9205 & 0.8900 & 0.8580 & 0.8195 & 0.7765 & 0.7395 & 0.7025 & 0.6255 \\ 
  & Tarone & 0.9435 & 0.8970 & 0.8695 & 0.8285 & 0.7905 & 0.7565 & 0.7000 & 0.6560 & 0.6200 & 0.5425 \\ 
  & Sidak & 0.8805 & 0.8225 & 0.7915 & 0.7425 & 0.6965 & 0.6565 & 0.6115 & 0.5550 & 0.5230 & 0.4475 \\ 
  & Bonf & 0.8805 & 0.8225 & 0.7915 & 0.7425 & 0.6965 & 0.6565 & 0.6115 & 0.5550 & 0.5230 & 0.4475 \\ 
	\midrule
	\multirow{4}{2cm}{$m=10$\\$\pi_0=0.6$ }
& MBonf & 0.9210 & 0.8970 & 0.8760 & 0.8575 & 0.8175 & 0.7980 & 0.7535 & 0.7160 & 0.6720 & 0.6200 \\ 
  & Tarone & 0.8670 & 0.8465 & 0.8240 & 0.8015 & 0.7545 & 0.7370 & 0.7025 & 0.6535 & 0.6205 & 0.5750 \\ 
  & Sidak & 0.7265 & 0.7165 & 0.6815 & 0.6600 & 0.6045 & 0.5775 & 0.5315 & 0.4940 & 0.4415 & 0.4025 \\ 
  & Bonf & 0.7265 & 0.7165 & 0.6815 & 0.6600 & 0.6045 & 0.5775 & 0.5315 & 0.4940 & 0.4415 & 0.4025 \\ 
	\midrule
	\multirow{4}{2cm}{$m=10$\\$\pi_0=0.8$}
& MBonf & 0.7720 & 0.7710 & 0.7250 & 0.7240 & 0.7085 & 0.6790 & 0.6750 & 0.6685 & 0.6265 & 0.6125 \\ 
  & Tarone & 0.7175 & 0.7230 & 0.6870 & 0.6710 & 0.6705 & 0.6270 & 0.6310 & 0.6295 & 0.5895 & 0.5765 \\ 
  & Sidak & 0.4950 & 0.4915 & 0.4735 & 0.4580 & 0.4650 & 0.4105 & 0.4215 & 0.4140 & 0.3710 & 0.3420 \\ 
  & Bonf & 0.4950 & 0.4915 & 0.4735 & 0.4580 & 0.4650 & 0.4105 & 0.4215 & 0.4140 & 0.3710 & 0.3420 \\ 
	\bottomrule	
\end{tabular}

\end{table}

\renewcommand{\baselinestretch}{1.65}

%% file: simulation_table_FWER_SD_BT_dept.tex
\begin{table}[H]\small
	\renewcommand\arraystretch{0.8}
	\centering
	\caption{Simulated FWER comparisons for step-down procedures with dependent $p$-values generated from Binomial Exact Test statistics, including Procedure 3.2 (MHolm),  Procedure 2.3 (TH), and the conventional Holm procedure (Holm).} 
	\label{sim_FWER_SD_BT_dept}
	\vspace{0.15 in}
	\begin{tabular}{clcccccccccc}
		\toprule
	& $\rho$	  & 0 & 0.1 & 0.2 & 0.3 & 0.4 & 0.5 & 0.6 & 0.7 & 0.8 & 0.9 \\ 
		\midrule
		\multirow{4}{2cm}{$m=5$\\$\pi_0=0.4$ }
& MHolm & 0.0060 & 0.0130 & 0.0145 & 0.0070 & 0.0105 & 0.0115 & 0.0095 & 0.0090 & 0.0090 & 0.0090 \\ 
  & TH & 0.0030 & 0.0060 & 0.0075 & 0.0035 & 0.0070 & 0.0065 & 0.0065 & 0.0065 & 0.0070 & 0.0055 \\ 
  & Holm & 0.0020 & 0.0025 & 0.0030 & 0.0025 & 0.0030 & 0.0030 & 0.0035 & 0.0030 & 0.0025 & 0.0020 \\ 

		\midrule
		\multirow{4}{2cm}{$m=5$\\$\pi_0=0.6$ }
& MHolm & 0.0150 & 0.0120 & 0.0190 & 0.0110 & 0.0120 & 0.0140 & 0.0125 & 0.0090 & 0.0075 & 0.0085 \\ 
  & TH & 0.0080 & 0.0085 & 0.0145 & 0.0065 & 0.0100 & 0.0105 & 0.0080 & 0.0065 & 0.0055 & 0.0080 \\ 
  & Holm & 0.0040 & 0.0030 & 0.0040 & 0.0025 & 0.0045 & 0.0050 & 0.0030 & 0.0025 & 0.0025 & 0.0040 \\ 
		\midrule
		\multirow{4}{2cm}{$m=5$\\$\pi_0=0.8$ }
& MHolm & 0.0205 & 0.0195 & 0.0235 & 0.0140 & 0.0150 & 0.0150 & 0.0165 & 0.0180 & 0.0140 & 0.0080 \\ 
  & TH & 0.0155 & 0.0145 & 0.0150 & 0.0110 & 0.0105 & 0.0115 & 0.0140 & 0.0135 & 0.0125 & 0.0060 \\ 
  & Holm & 0.0020 & 0.0015 & 0.0040 & 0.0015 & 0.0010 & 0.0015 & 0.0045 & 0.0040 & 0.0030 & 0.0015 \\ 

		\bottomrule	
  \toprule
  \multirow{4}{2cm}{$m=10$\\$\pi_0=0.4$}
& MHolm & 0.0085 & 0.0100 & 0.0095 & 0.0140 & 0.0115 & 0.0145 & 0.0105 & 0.0070 & 0.0110 & 0.0045 \\ 
  & TH & 0.0030 & 0.0040 & 0.0050 & 0.0060 & 0.0070 & 0.0075 & 0.0075 & 0.0045 & 0.0085 & 0.0040 \\ 
  & Holm & 0.0005 & 0.0015 & 0.0025 & 0.0005 & 0.0015 & 0.0010 & 0.0020 & 0.0010 & 0.0030 & 0.0010 \\ 
		\midrule
		\multirow{4}{2cm}{$m=10$\\$\pi_0=0.6$ }
& MHolm & 0.0120 & 0.0185 & 0.0150 & 0.0205 & 0.0140 & 0.0095 & 0.0155 & 0.0110 & 0.0120 & 0.0085 \\ 
  & TH & 0.0070 & 0.0075 & 0.0050 & 0.0145 & 0.0075 & 0.0060 & 0.0095 & 0.0085 & 0.0090 & 0.0075 \\ 
  & Holm & 0.0005 & 0.0005 & 0.0005 & 0.0015 & 0.0005 & 0.0000 & 0.0025 & 0.0020 & 0.0020 & 0.0010 \\ 
		\midrule
		\multirow{4}{2cm}{$m=10$\\$\pi_0=0.8$}
& MHolm & 0.0220 & 0.0175 & 0.0180 & 0.0205 & 0.0175 & 0.0135 & 0.0175 & 0.0185 & 0.0085 & 0.0060 \\ 
  & TH & 0.0155 & 0.0095 & 0.0100 & 0.0150 & 0.0150 & 0.0100 & 0.0155 & 0.0155 & 0.0060 & 0.0050 \\ 
  & Holm & 0.0015 & 0.0005 & 0.0005 & 0.0010 & 0.0020 & 0.0000 & 0.0005 & 0.0005 & 0.0000 & 0.0005 \\ 
		\bottomrule	
	\end{tabular}

\end{table}

\renewcommand{\baselinestretch}{1.65}

%% file: simulation_table_Pow_SD_BT_dept.tex
\begin{table}[H]\small
	\renewcommand\arraystretch{0.8}
	\centering
	\caption{Simulated minimal power comparisons for step-down procedures with dependent $p$-values generated from Binomial Exact Test statistics, including Procedure 3.2 (MHolm),  Procedure 2.3 (TH), and the conventional Holm procedure (Holm).} 
	
	\label{sim_Pow_SD_BT_dept}
	\vspace{0.15 in}
	\begin{tabular}{clcccccccccc}
	\toprule
	& $\rho$	  & 0 & 0.1 & 0.2 & 0.3 & 0.4 & 0.5 & 0.6 & 0.7 & 0.8 & 0.9 \\ 
	\midrule
	\multirow{4}{2cm}{$m=5$\\$\pi_0=0.4$ }
& MHolm & 0.9200 & 0.8715 & 0.8575 & 0.8445 & 0.8275 & 0.7690 & 0.7395 & 0.7375 & 0.6935 & 0.6425 \\ 
  & TH & 0.8755 & 0.8155 & 0.8000 & 0.7840 & 0.7610 & 0.7040 & 0.6755 & 0.6745 & 0.6235 & 0.5680 \\ 
  & Holm & 0.7880 & 0.7250 & 0.7065 & 0.6915 & 0.6680 & 0.6130 & 0.5835 & 0.5730 & 0.5320 & 0.4760 \\ 
	\midrule
	\multirow{4}{2cm}{$m=5$\\$\pi_0=0.6$ }
& MHolm & 0.8100 & 0.8065 & 0.7870 & 0.7650 & 0.7450 & 0.7260 & 0.7205 & 0.6965 & 0.6615 & 0.6195 \\ 
  & TH & 0.7610 & 0.7545 & 0.7355 & 0.7175 & 0.7035 & 0.6795 & 0.6680 & 0.6520 & 0.6185 & 0.5790 \\ 
  & Holm & 0.5990 & 0.6060 & 0.5875 & 0.5710 & 0.5480 & 0.5395 & 0.5075 & 0.5095 & 0.4680 & 0.4200 \\ 
	\midrule
	\multirow{4}{2cm}{$m=5$\\$\pi_0=0.8$ }
& MHolm & 0.6070 & 0.6105 & 0.5940 & 0.6185 & 0.5885 & 0.5870 & 0.6335 & 0.6300 & 0.6120 & 0.6110 \\ 
  & TH & 0.5650 & 0.5730 & 0.5630 & 0.5880 & 0.5615 & 0.5685 & 0.6105 & 0.6115 & 0.5915 & 0.6000 \\ 
  & Holm & 0.3755 & 0.3910 & 0.3595 & 0.3865 & 0.3680 & 0.3420 & 0.3890 & 0.3860 & 0.3940 & 0.3750 \\ 
	\bottomrule	
	\toprule
	\multirow{4}{2cm}{$m=10$\\$\pi_0=0.4$}
& MHolm & 0.9755 & 0.9465 & 0.9205 & 0.8900 & 0.8580 & 0.8195 & 0.7765 & 0.7395 & 0.7025 & 0.6255 \\ 
  & TH & 0.9435 & 0.8970 & 0.8695 & 0.8285 & 0.7905 & 0.7565 & 0.7000 & 0.6560 & 0.6200 & 0.5425 \\ 
  & Holm & 0.8805 & 0.8225 & 0.7915 & 0.7425 & 0.6965 & 0.6565 & 0.6115 & 0.5550 & 0.5230 & 0.4475 \\ 
	\midrule
	\multirow{4}{2cm}{$m=10$\\$\pi_0=0.6$ }
& MHolm & 0.9210 & 0.8970 & 0.8760 & 0.8575 & 0.8175 & 0.7980 & 0.7535 & 0.7170 & 0.6720 & 0.6205 \\ 
  & TH & 0.8680 & 0.8465 & 0.8245 & 0.8020 & 0.7545 & 0.7370 & 0.7025 & 0.6535 & 0.6205 & 0.5750 \\ 
  & Holm & 0.7265 & 0.7165 & 0.6815 & 0.6600 & 0.6045 & 0.5775 & 0.5315 & 0.4940 & 0.4415 & 0.4025 \\ 
\midrule
	\multirow{4}{2cm}{$m=10$\\$\pi_0=0.8$}
& MHolm & 0.7735 & 0.7715 & 0.7255 & 0.7240 & 0.7085 & 0.6790 & 0.6750 & 0.6685 & 0.6265 & 0.6130 \\ 
  & TH & 0.7180 & 0.7230 & 0.6870 & 0.6710 & 0.6710 & 0.6275 & 0.6310 & 0.6295 & 0.5895 & 0.5770 \\ 
  & Holm & 0.4950 & 0.4915 & 0.4735 & 0.4580 & 0.4650 & 0.4105 & 0.4215 & 0.4140 & 0.3710 & 0.3425 \\ 
	\bottomrule	
\end{tabular}

\end{table}

\renewcommand{\baselinestretch}{1.65}

%% file: simulation_table_FWER_SU_BT_dept.tex
\begin{table}[H]\small
	\renewcommand\arraystretch{0.8}
	\centering
	\caption{Simulated FWER comparisons for step-up procedures with dependent $p$-values generated from Binomial Exact Test statistics, including Procedure 3.3 (MHoch), the Roth procedure (Roth), and the conventional Hochberg procedure (Hoch).} 
	\label{sim_FWER_SU_BT_dept}
	\vspace{0.15 in}
	\begin{tabular}{clcccccccccc}
		\toprule
	& $\rho$	  & 0 & 0.1 & 0.2 & 0.3 & 0.4 & 0.5 & 0.6 & 0.7 & 0.8 & 0.9 \\ 
		\midrule
		\multirow{4}{2cm}{$m=5$\\$\pi_0=0.4$ }
  & MHoch & 0.0060 & 0.0135 & 0.0150 & 0.0070 & 0.0110 & 0.0120 & 0.0105 & 0.0090 & 0.0115 & 0.0115 \\ 
  & Roth & 0.0035 & 0.0080 & 0.0075 & 0.0035 & 0.0085 & 0.0085 & 0.0080 & 0.0070 & 0.0095 & 0.0080 \\ 
  & Hoch & 0.0020 & 0.0030 & 0.0035 & 0.0025 & 0.0040 & 0.0040 & 0.0050 & 0.0030 & 0.0045 & 0.0040 \\ 
		\midrule
		\multirow{4}{2cm}{$m=5$\\$\pi_0=0.6$ }
& MHoch & 0.0155 & 0.0120 & 0.0195 & 0.0110 & 0.0120 & 0.0150 & 0.0130 & 0.0130 & 0.0105 & 0.0135 \\ 
  & Roth & 0.0090 & 0.0100 & 0.0155 & 0.0055 & 0.0105 & 0.0095 & 0.0090 & 0.0075 & 0.0090 & 0.0115 \\ 
  & Hoch & 0.0050 & 0.0030 & 0.0040 & 0.0025 & 0.0050 & 0.0050 & 0.0045 & 0.0035 & 0.0050 & 0.0065 \\ 

		\midrule
		\multirow{4}{2cm}{$m=5$\\$\pi_0=0.8$ }
& MHoch & 0.0205 & 0.0195 & 0.0235 & 0.0150 & 0.0155 & 0.0150 & 0.0185 & 0.0195 & 0.0160 & 0.0100 \\ 
  & Roth & 0.0165 & 0.0160 & 0.0155 & 0.0110 & 0.0115 & 0.0110 & 0.0150 & 0.0145 & 0.0135 & 0.0070 \\ 
  & Hoch & 0.0020 & 0.0015 & 0.0040 & 0.0015 & 0.0010 & 0.0025 & 0.0045 & 0.0045 & 0.0045 & 0.0040 \\ 
		\bottomrule	
  \toprule
  \multirow{4}{2cm}{$m=10$\\$\pi_0=0.4$}
& MHoch & 0.0095 & 0.0110 & 0.0100 & 0.0160 & 0.0115 & 0.0170 & 0.0130 & 0.0080 & 0.0130 & 0.0095 \\ 
  & Roth & 0.0020 & 0.0045 & 0.0055 & 0.0080 & 0.0060 & 0.0080 & 0.0095 & 0.0050 & 0.0100 & 0.0060 \\ 
  & Hoch & 0.0005 & 0.0015 & 0.0025 & 0.0005 & 0.0015 & 0.0010 & 0.0025 & 0.0015 & 0.0040 & 0.0030 \\ 
		\midrule
		\multirow{4}{2cm}{$m=10$\\$\pi_0=0.6$ }
& MHoch & 0.0140 & 0.0195 & 0.0160 & 0.0210 & 0.0165 & 0.0115 & 0.0165 & 0.0140 & 0.0145 & 0.0130 \\ 
  & Roth & 0.0070 & 0.0075 & 0.0055 & 0.0135 & 0.0080 & 0.0050 & 0.0105 & 0.0085 & 0.0100 & 0.0090 \\ 
  & Hoch & 0.0005 & 0.0005 & 0.0005 & 0.0015 & 0.0005 & 0.0000 & 0.0025 & 0.0020 & 0.0025 & 0.0040 \\
		\midrule
		\multirow{4}{2cm}{$m=10$\\$\pi_0=0.8$}
& MHoch & 0.0220 & 0.0180 & 0.0195 & 0.0215 & 0.0190 & 0.0160 & 0.0190 & 0.0195 & 0.0115 & 0.0090 \\ 
  & Roth & 0.0135 & 0.0085 & 0.0100 & 0.0130 & 0.0140 & 0.0095 & 0.0150 & 0.0140 & 0.0075 & 0.0050 \\ 
  & Hoch & 0.0015 & 0.0005 & 0.0005 & 0.0010 & 0.0020 & 0.0000 & 0.0005 & 0.0010 & 0.0020 & 0.0020 \\ 
	\bottomrule	
	\end{tabular}

\end{table}

\renewcommand{\baselinestretch}{1.65}

%% file: simulation_table_Pow_SU_BT_dept.tex
\begin{table}[H]\small
	\renewcommand\arraystretch{0.8}
	\centering
	\caption{Simulated minimal power comparisons for step-up procedures with dependent $p$-values generated from Binomial Exact Test statistics, including Procedure 3.3 (MHoch), the Roth procedure (Roth), and the conventional Hochberg procedure (Hoch).} 
	
	\label{sim_Pow_SU_BT_dept}
	\vspace{0.15 in}
	\begin{tabular}{clcccccccccc}
	\toprule
	& $\rho$	  & 0 & 0.1 & 0.2 & 0.3 & 0.4 & 0.5 & 0.6 & 0.7 & 0.8 & 0.9 \\ 
	\midrule
	\multirow{4}{2cm}{$m=5$\\$\pi_0=0.4$ }
& MHoch & 0.9235 & 0.8745 & 0.8620 & 0.8475 & 0.8315 & 0.7740 & 0.7455 & 0.7475 & 0.7040 & 0.6775 \\ 
  & Roth & 0.8815 & 0.8240 & 0.8090 & 0.7955 & 0.7705 & 0.7135 & 0.6885 & 0.6930 & 0.6470 & 0.6045 \\ 
  & Hoch & 0.7975 & 0.7330 & 0.7175 & 0.7010 & 0.6785 & 0.6215 & 0.5970 & 0.5860 & 0.5475 & 0.5005 \\ 
	\midrule
	\multirow{4}{2cm}{$m=5$\\$\pi_0=0.6$ }
& MHoch & 0.8140 & 0.8095 & 0.7915 & 0.7695 & 0.7505 & 0.7285 & 0.7260 & 0.7055 & 0.6690 & 0.6410 \\ 
  & Roth & 0.7665 & 0.7555 & 0.7375 & 0.7245 & 0.7125 & 0.6835 & 0.6740 & 0.6595 & 0.6220 & 0.5955 \\ 
  & Hoch & 0.6040 & 0.6160 & 0.5910 & 0.5775 & 0.5545 & 0.5490 & 0.5145 & 0.5190 & 0.4810 & 0.4390 \\ 
	\midrule
	\multirow{4}{2cm}{$m=5$\\$\pi_0=0.8$ }
& MHoch & 0.6070 & 0.6105 & 0.5940 & 0.6190 & 0.5885 & 0.5870 & 0.6340 & 0.6310 & 0.6120 & 0.6115 \\ 
  & Roth & 0.5545 & 0.5625 & 0.5455 & 0.5735 & 0.5470 & 0.5490 & 0.5925 & 0.5930 & 0.5745 & 0.5760 \\ 
  & Hoch & 0.3755 & 0.3910 & 0.3595 & 0.3865 & 0.3680 & 0.3420 & 0.3890 & 0.3860 & 0.3950 & 0.3760 \\ 
	\bottomrule	
	\toprule
	\multirow{4}{2cm}{$m=10$\\$\pi_0=0.4$}
& MHoch & 0.9765 & 0.9475 & 0.9220 & 0.8940 & 0.8620 & 0.8220 & 0.7830 & 0.7470 & 0.7215 & 0.6495 \\ 
  & Roth & 0.9460 & 0.9015 & 0.8730 & 0.8305 & 0.7940 & 0.7580 & 0.7070 & 0.6645 & 0.6365 & 0.5775 \\ 
  & Hoch & 0.8805 & 0.8225 & 0.7915 & 0.7425 & 0.6970 & 0.6570 & 0.6120 & 0.5585 & 0.5270 & 0.4560 \\ 
	\midrule
	\multirow{4}{2cm}{$m=10$\\$\pi_0=0.6$ }
& MHoch & 0.9245 & 0.9010 & 0.8790 & 0.8600 & 0.8230 & 0.8045 & 0.7595 & 0.7260 & 0.6865 & 0.6425 \\ 
  & Roth & 0.8745 & 0.8510 & 0.8275 & 0.8075 & 0.7570 & 0.7425 & 0.7085 & 0.6615 & 0.6275 & 0.5930 \\ 
  & Hoch & 0.7265 & 0.7170 & 0.6820 & 0.6600 & 0.6055 & 0.5775 & 0.5320 & 0.4940 & 0.4430 & 0.4100 \\ 
\midrule
	\multirow{4}{2cm}{$m=10$\\$\pi_0=0.8$}
& MHoch & 0.7760 & 0.7730 & 0.7275 & 0.7255 & 0.7095 & 0.6815 & 0.6775 & 0.6720 & 0.6305 & 0.6280 \\ 
  & Roth & 0.7160 & 0.7215 & 0.6880 & 0.6725 & 0.6730 & 0.6305 & 0.6350 & 0.6345 & 0.5950 & 0.5885 \\ 
  & Hoch & 0.4950 & 0.4915 & 0.4735 & 0.4580 & 0.4650 & 0.4105 & 0.4215 & 0.4140 & 0.3710 & 0.3425 \\ 
	\bottomrule	
\end{tabular}

\end{table}

\renewcommand{\baselinestretch}{1.65}

%% file: MHTdiscrete_final_arxiv.bbl
\begin{thebibliography}{}

\bibitem[Benjamini and Yekutieli(2001)Benjamini and
  Yekutieli]{benjamini2001control}
Benjamini, Y. and Yekutieli, D. (2001).
\newblock The control of the false discovery rate in multiple testing under
  dependency.
\newblock {\em Annals of Statistics\/}, {\bf 29}, 1165--1188.

\bibitem[Block {\em et~al.}(1985)Block, Savits, and Shaked]{block1985concept}
Block, H.~W., Savits, T.~H., and Shaked, M. (1985).
\newblock A concept of negative dependence using stochastic ordering.
\newblock {\em Statistics \& Probability Letters\/}, {\bf 3}, 81--86.

\bibitem[Chen {\em et~al.}(2018)Chen, Doerge, and Heyse]{chen2014multiple}
Chen, X., Doerge, R.~W., and Heyse, J.~F. (2018).
\newblock Multiple testing with discrete data: proportion of true null
  hypotheses and two adaptive {FDR} procedures.
\newblock {\em Biometrical Journal\/}, {\bf 60}, 761--779.

\bibitem[Dmitrienko {\em et~al.}(2009)Dmitrienko, Tamhane, and
  Bretz]{dmitrienko2009multiple}
Dmitrienko, A., Tamhane, A.~C., and Bretz, F. (2009).
\newblock {\em Multiple Testing Problems in Pharmaceutical Statistics\/}.
\newblock CRC Press.

\bibitem[D{\"o}hler(2018)D{\"o}hler]{dohler2016discrete}
D{\"o}hler, S. (2018).
\newblock A discrete modification of the {Benjamini--Yekutieli} procedure.
\newblock {\em Econometrics and Statistics\/}, {\bf 5}, 137--147.

\bibitem[Goeman and Solari(2014)Goeman and Solari]{SIM:SIM6082}
Goeman, J.~J. and Solari, A. (2014).
\newblock Multiple hypothesis testing in genomics.
\newblock {\em Statistics in Medicine\/}, {\bf 33}, 1946--1978.

\bibitem[Gould(2015)Gould]{gould2015statistical}
Gould, A.~L. (2015).
\newblock {\em Statistical Methods for Evaluating Safety in Medical Product
  Development\/}.
\newblock John Wiley \& Sons.

\bibitem[Gutman and Hochberg(2007)Gutman and Hochberg]{gutman2007improved}
Gutman, R. and Hochberg, Y. (2007).
\newblock Improved multiple test procedures for discrete distributions: {N}ew
  ideas and analytical review.
\newblock {\em Journal of Statistical Planning and Inference\/}, {\bf 137},
  2380--2393.

\bibitem[He and Heyse(2019)He and Heyse]{he2019improved}
He, L. and Heyse, J.~F. (2019).
\newblock Improved power of familywise error rate procedures for discrete data
  under dependency.
\newblock {\em Biometrical Journal\/}, {\bf 61}(1), 101--114.

\bibitem[Heyse(2011)Heyse]{heyse2011false}
Heyse, J.~F. (2011).
\newblock A false discovery rate procedure for categorical data.
\newblock {\em Resent Advances in Biostatistics: False Discovery Rates,
  Survival Analysis, and Related Topics\/}, {\bf 4}, 43--58.

\bibitem[Hochberg(1988)Hochberg]{hochberg1988sharper}
Hochberg, Y. (1988).
\newblock A sharper {Bonferroni} procedure for multiple tests of significance.
\newblock {\em Biometrika\/}, {\bf 75}, 800--802.

\bibitem[Holm(1979)Holm]{holm1979simple}
Holm, S. (1979).
\newblock A simple sequentially rejective multiple test procedure.
\newblock {\em Scandinavian Journal of Statistics\/}, {\bf 6}, 65--70.

\bibitem[Hommel and Krummenauer(1998)Hommel and
  Krummenauer]{hommel1998improvements}
Hommel, G. and Krummenauer, F. (1998).
\newblock Improvements and modifications of {T}arone's multiple test procedure
  for discrete data.
\newblock {\em Biometrics\/}, {\bf 54}, 673--681.

\bibitem[Jiang and Xia(2014)Jiang and Xia]{jiang2014quantitative}
Jiang, Q. and Xia, H.~A. (2014).
\newblock {\em Quantitative Evaluation of Safety in Drug Development: Design,
  Analysis and Reporting\/}.
\newblock Chapman and Hall/CRC.

\bibitem[Lehmann and Romano(2005)Lehmann and Romano]{lehmann2005testing}
Lehmann, E.~L. and Romano, J.~P. (2005).
\newblock {\em Testing Statistical Hypotheses, 3rd Edition\/}.
\newblock Springer.

\bibitem[Mehrotra and Adewale(2012)Mehrotra and Adewale]{mehrotra2012flagging}
Mehrotra, D.~V. and Adewale, A.~J. (2012).
\newblock Flagging clinical adverse experiences: reducing false discoveries
  without materially compromising power for detecting true signals.
\newblock {\em Statistics in Medicine\/}, {\bf 31}, 1918--1930.

\bibitem[Mehrotra and Heyse(2004)Mehrotra and Heyse]{mehrotra2004use}
Mehrotra, D.~V. and Heyse, J.~F. (2004).
\newblock Use of the false discovery rate for evaluating clinical safety data.
\newblock {\em Statistical Methods in Medical Research\/}, {\bf 13}, 227--238.

\bibitem[{R Development Core Team}(2018){R Development Core Team}]{R:2017}
{R Development Core Team} (2018).
\newblock {\em R: A Language and Environment for Statistical Computing\/}.
\newblock Vienna, Austria: R Foundation for Statistical Computing.

\bibitem[Roth(1999)Roth]{roth1999multiple}
Roth, A.~J. (1999).
\newblock Multiple comparison procedures for discrete test statistics.
\newblock {\em Journal of Statistical Planning and Inference\/}, {\bf 82},
  101--117.

\bibitem[Sarkar(2002)Sarkar]{sarkar2002some}
Sarkar, S.~K. (2002).
\newblock Some results on false discovery rate in stepwise multiple testing
  procedures.
\newblock {\em Annals of Statistics\/}, {\bf 30}, 239--257.

\bibitem[Sarkar(2008)Sarkar]{sarkar2008simes}
Sarkar, S.~K. (2008).
\newblock On the simes inequality and its generalization.
\newblock In {\em Beyond parametrics in interdisciplinary research: Festschrift
  in honor of Professor Pranab K. Sen\/}, pages 231--242. Institute of
  Mathematical Statistics.

\bibitem[Tarone(1990)Tarone]{tarone1990modified}
Tarone, R.~E. (1990).
\newblock A modified {Bonferroni} method for discrete data.
\newblock {\em Biometrics\/}, {\bf 46}, 515--522.

\bibitem[Westfall and Wolfinger(1997)Westfall and
  Wolfinger]{westfall1997multiple}
Westfall, P.~H. and Wolfinger, R.~D. (1997).
\newblock Multiple tests with discrete distributions.
\newblock {\em The American Statistician\/}, {\bf 51}, 3--8.

\bibitem[Westfall and Young(1993)Westfall and Young]{westfall1993resampling}
Westfall, P.~H. and Young, S.~S. (1993).
\newblock {\em Resampling-based Multiple Testing: {E}xamples and Methods for
  P-value Adjustment\/}.
\newblock John Wiley \& Sons.

\bibitem[Zhu and Guo(2018)Zhu and Guo]{rpack}
Zhu, Y. and Guo, W. (2018).
\newblock {\em MHTdiscrete: Multiple Hypotheses Testing for Discrete Data\/}.
\newblock R package version 1.0.1.

\end{thebibliography}
